\documentclass[letterpaper,runningheads,envcountsame,orivec]{llncs}
\usepackage[headings]{fullpage}
\usepackage[T1]{fontenc}
\usepackage{lmodern}
\usepackage{amsmath,amssymb,mathtools,mathrsfs}
\usepackage{mathpartir}
\usepackage{tikz-cd}
\usepackage{array}
\usepackage{enumitem}
\usepackage{microtype}
\usepackage[hidelinks,unicode,hypertexnames=false,bookmarksdepth=2,
  bookmarksnumbered=true,bookmarksopen=true,bookmarksopenlevel=1]{hyperref}
\makeatletter
\renewcommand*{\@spthm}[4]{%
  \topsep 7\p@ \@plus2\p@ \@minus4\p@
  \H@refstepcounter{#1}%
  \hyper@makecurrent{#1}%
  \def\hott@theoremheading{%
    \Hy@raisedlink{\hyper@anchorstart{\@currentHref}\hyper@anchorend}#2}%
  \@ifnextchar[{\@spythm{#1}{\hott@theoremheading}{#3}{#4}}%
               {\@spxthm{#1}{\hott@theoremheading}{#3}{#4}}}
\makeatother
\allowdisplaybreaks[2]
\newcommand{\U}{\mathbb U}
\newcommand{\V}{\mathbb V}
\newcommand{\Sps}{\mathcal S}
\newcommand{\G}{\mathcal G}
\newcommand{\B}{\mathcal B}
\newcommand{\C}{\mathcal C}
\newcommand{\D}{\mathcal D}
\newcommand{\X}{\mathcal X}
\newcommand{\T}{\mathbb T}
\newcommand{\id}{\mathrm{id}}
\newcommand{\op}{\mathrm{op}}
\newcommand{\lex}{\mathrm{lex}}
\newcommand{\geo}{\mathrm{geo}}
\newcommand{\ctx}{\mathsf{ctx}}
\newcommand{\type}{\mathsf{type}}
\newcommand{\Disp}{\mathsf{Disp}}
\newcommand{\Term}{\mathsf{Terminal}}
\newcommand{\PB}{\mathsf{PB}}
\newcommand{\HC}{\mathsf{HC}}
\newcommand{\Eq}{\mathsf{Eq}}
\newcommand{\Lim}{\mathsf{Lim}}
\newcommand{\Col}{\mathsf{Col}}
\newcommand{\Id}{\mathsf{Id}}
\newcommand{\refl}{\mathsf{refl}}
\newcommand{\glue}{\mathsf{glue}}
\newcommand{\pack}{\mathsf{pack}}
\newcommand{\hcolim}{\mathop{\mathsf{hocolim}}}
\newcommand{\colim}{\mathop{\mathrm{colim}}}
\newcommand{\Fun}{\operatorname{Fun}}
\newcommand{\Map}{\operatorname{Map}}
\newcommand{\Mod}{\operatorname{Mod}}
\newcommand{\Geom}{\operatorname{Geom}}
\newcommand{\Fam}{\operatorname{Fam}}
\newcommand{\Cart}{\operatorname{Cart}}
\newcommand{\Tm}{\operatorname{Tm}}
\newcommand{\Rec}{\mathsf{Rec}}
\newcommand{\Ar}{\operatorname{Ar}}
\newcommand{\Fib}{\mathsf{Fib}}
\newcommand{\pr}{\operatorname{pr}}
\newcommand{\Lan}{\operatorname{Lan}}
\newcommand{\Psh}{\mathcal P_{\U}}
\newcommand{\cmark}[2]{\langle #1\mid #2\rangle}
\newcommand{\squ}{\mathbin{\square_v}}
\title{Higher Duality Theory:\texorpdfstring{\\}{ }Geometric HoTT and Classifying \texorpdfstring{$\infty$}{∞}-Toposes}
\titlerunning{Geometric HoTT and Classifying $\infty$-Toposes}
\author{Yoshihiro Maruyama}
\authorrunning{Y. Maruyama}
\institute{Kyoto University\\
\email{maruyama@i.h.kyoto-u.ac.jp}}
\begin{document}
\maketitle
\raggedbottom
\begin{abstract}
We develop geometric homotopy type theory (geometric HoTT), which gives a dependent language for constructions preserved by inverse images of geometric morphisms. We give a homotopy-coherent calculus for geometric HoTT and extend the classical duality between geometric logic and classifying toposes to Grothendieck $\infty$-toposes. For each geometric type theory, the $\infty$-category of contexts generated by its rules is equivalent to its classifying $\infty$-topos. As the theory varies, the classifier gives a duality of $(\infty,2)$-categories between geometric HoTT theories with interpretations and Grothendieck $\infty$-toposes with geometric morphisms. We then construct a universal family of classifying $\infty$-toposes whose fiber at each presentation of a theory is the classifying $\infty$-topos of that theory. We also construct a universal family of classifying $\infty$-toposes with coherent transport along Morita equivalences of theories. Geometric HoTT thus serves as an internal language for Grothendieck $\infty$-toposes, extending the role of geometric logic in classical topos theory to the realm of $\infty$-topos theory.
\end{abstract}

\section{Introduction}
Geometric logic describes constructions preserved by inverse images of geometric morphisms. Its finite conjunctions, small disjunctions, and existential quantifiers are interpreted in the classifying topos of each theory. Let $T$ be an ordinary geometric theory with classifying topos $\mathbf{Set}[T]$, and let $\mathcal E$ be a Grothendieck topos. Write $\Mod_T(\mathcal E)$ for the category of $T$-models in $\mathcal E$ and their homomorphisms, and $\Geom$ for the category of geometric morphisms between two toposes. The classical correspondence gives an equivalence of categories~\cite{Johnstone,MakkaiReyes}
\begin{equation}\label{eq:classical-introduction}
\Mod_T(\mathcal E)\simeq\Geom(\mathcal E,\mathbf{Set}[T]).
\end{equation}

We extend this correspondence to Grothendieck $\infty$-toposes by introducing geometric homotopy type theory (geometric HoTT). In this language, a type over a context describes a family, and a dependent term describes a section of that family. The language has a unit type, dependent sums ($\Sigma$), identity types ($\Id$), and small homotopy colimits ($\hcolim$) with dependent elimination. Over a homotopy colimit, descent glues compatible local families into a family on the whole context, and dependent elimination glues compatible local sections of that family into a global section.

Let $W=\hcolim_i D_i$ be such a context. A family over $W$ restricts to a family $A_i$ on each $D_i$. For an edge $\alpha:i\to j$ and a point $x:D_i$, the images of $x$ and $D_\alpha x$ in $W$ are joined by a path. Transport along this path gives an equivalence
\[
e_{\alpha,x}:A_i(x)\simeq A_j(D_\alpha x).
\]
A section therefore restricts to local sections $s_i$ whose values agree under transport, so $e_{\alpha,x}(s_i(x))=s_j(D_\alpha x)$. The higher simplices record the coherences among these equivalences and paths.

Conversely, descent lets us construct a family over $W$ from local families $A_i$ with coherent transition equivalences. We use these equivalences to glue the local total spaces $E_i=\Sigma_{x:D_i}A_i(x)$. Their projections $E_i\to D_i$ induce a map
\[
q:\hcolim_i E_i\longrightarrow W.
\]
The glued family assigns to each $w:W$ the fiber of $q$ over $w$.

In this infinitary dependent language, judgmental equations specify structural boundaries, and higher terms witness substitution comparisons and computation rules. Section~\ref{sec:language} gives the type expression for the glued family. Section~\ref{sec:formats} gives a single set-indexed scheme for the boundaries of terms and computations in every dimension.

\subsection{Main Results}
Fix Grothendieck universes $\U\in\V$ and write ``small'' for $\U$-small. A small theory $\T$ is given by a small simplicial presentation of generators, boundary equations, and axioms on specified diagrams. For a Grothendieck $\infty$-topos $\X$, the $\infty$-category $\Mod_\T(\X)$ consists of diagrams in $\X$ interpreting the generators subject to these equations and axioms, with all natural transformations between them. Write $\G_\T$ for the dependent syntax of $\T$. We also give a second presentation of $\T$, called the \emph{boundary syntax} and denoted by $\Sps_\T$, whose rules express the universal properties of finite-limit cones and small-colimit cocones. The classification theorem below identifies the two presentations.

Write $\Fun_\geo(\C,\D)$ for the $\infty$-category of finite-limit- and small-colimit-preserving functors from $\C$ to $\D$ and all natural transformations between them. For geometric morphisms, we use the inverse-image convention
\[
\Geom(\X,\B)=\Fun^{L,\lex}(\B,\X).
\]

\begin{theorem}[Classification]\label{thm:main}
For every small theory $\T$, there are equivalences
\[
\G_\T\simeq\Sps_\T\simeq\B_\T,
\]
where $\B_\T$ is the left-exact localization of a presheaf $\infty$-category at the small set $R_\T$ of axiom comparisons from Section~\ref{sec:classifier}. For every Grothendieck $\infty$-topos $\X$,
\[
\Fun_\geo(\G_\T,\X)\simeq\Mod_\T(\X)
\simeq\Geom(\X,\B_\T).
\]
Under these equivalences, model homomorphisms correspond to natural transformations of inverse-image functors, and the mapping spaces record their higher identifications.
\end{theorem}

\begin{theorem}[Duality]\label{thm:main-duality}
Let $\mathbf{Th}_{\geo}$ denote the $(\infty,2)$-category of small geometric type theories with interpretations, and $\mathbf{Top}_\infty$ the $(\infty,2)$-category of Grothendieck $\infty$-toposes relative to $\U$ with geometric morphisms. The classifier induces an equivalence
\[
  \mathbf{Th}_{\geo}\simeq\mathbf{Top}_\infty^{\op},
  \qquad \T\longmapsto\B_\T .
\]
An interpretation of $\T$ in $\T'$ is realized by the inverse image of a geometric morphism $\B_{\T'}\to\B_\T$. Transformations between interpretations correspond to natural transformations of inverse images. The opposite reverses $1$-morphisms and preserves the direction of $2$-morphisms.
\end{theorem}

In Theorem~\ref{thm:reconstruction}, descent extends the interpretation of the generators to a geometric functor from $\B_\T$ back to the syntax. This inverse identifies the category of contexts generated by the rules as a presentable $\infty$-category. Every Grothendieck $\infty$-topos relative to $\U$ admits a small presentation by Theorem~\ref{thm:alltopoi}. Corollary~\ref{cor:duality} combines these presentations with the correspondence on interpretations to prove Theorem~\ref{thm:main-duality}.
We construct two universal families of classifiers. The first is parametrized by \emph{indexed presentations}, in which the generators, axioms, and shapes of the specified colimits can vary over the base (Theorem~\ref{thm:uf-pres}). The second is parametrized by the \emph{Morita space} of a chosen family of theories. A point of this space represents a theory with its classifier, and a path between two points represents a geometric equivalence between their classifiers (Theorem~\ref{thm:uf-morita}).

Reconstruction also identifies dependent proofs in the universal model. Write $E:\G_\T\xrightarrow{\sim}\B_\T$ for the equivalence interpreting the syntax in the classifier. For a derived type $\Gamma\vdash A\ \type$, its \emph{display} is the projection $p_A:\Gamma.A\to\Gamma$ that forgets the variable $x:A$ in the extended context $\Gamma.A$. Let $\Tm_\T(\Gamma,A)$ be the space of terms of $A$ in $\Gamma$ and their higher identifications. By Theorem~\ref{thm:proofspaces}, taking sections of $E p_A$ over $E\Gamma$ gives a natural equivalence
\[
\Tm_\T(\Gamma,A)\simeq
\Map_{(\B_\T)_{/E\Gamma}}(\id_{E\Gamma},E p_A).
\]

For an ordinary geometric theory $T$, its \emph{set-sorted translation} $T^\Delta$ into geometric homotopy type theory (HoTT) interprets primitive sorts as $0$-truncated types and predicates as subobjects of their contexts. Let $(C_T,J_T)$ be a small geometric syntactic site presenting $\mathbf{Set}[T]$. With $\Sps_\U$ the $\infty$-category of small spaces and the subscript $\leq0$ denoting the full subcategory of $0$-truncated objects, Theorem~\ref{thm:ordinary} gives
\[
\B_{T^\Delta}\simeq\operatorname{Sh}(C_T,J_T;\Sps_\U),
\qquad (\B_{T^\Delta})_{\leq0}\simeq\mathbf{Set}[T].
\]
The translation is conservative on ordinary geometric sequents.

Two constructions connect the dependent rules to the classifier. For every substitution $f:X\to Y$, the type
\[
y:Y\vdash \Fib_f(y):=\Sigma_{x:X}\Id_Y(fx,y)
\]
has total context equivalent to $X$ over $Y$. Thus reindexing a type supplies pullbacks of arbitrary substitutions. For a diagram with a colimit, taking the colimit of local total spaces is left adjoint to restricting a family along the constructor maps. Descent computation makes the unit an equivalence by identifying the prescribed local family with the restriction of the glued family. Substitution stability makes the counit an equivalence by identifying the colimit of the restricted total spaces with the original total space. These comparisons are natural in family maps, so the two constructions give inverse equivalences of $\infty$-categories.

\subsection{Related Work}
A classifying $\infty$-topos represents the models of a theory as the ambient $\infty$-topos varies. Lurie gives such a classifier for structures associated to a geometry~\cite{LurieDAGV}. Di Liberti and Meadows construct classifiers by weighted limits and characterize the classifiable prestacks of models as those presented by small geometric sketches~\cite{DiLibertiMeadows}. Our reconstruction identifies the classifier of a small geometric type theory with the $\infty$-category of contexts and substitutions generated by its rules. Under this equivalence, a type in context gives a family, and its section space recovers the terms of that type and all their higher identifications. Adjoining a term of a closed type $A$ passes from the classifier to its slice over the object interpreting $A$.

For ordinary geometric theories, syntactic sites exhibit Morita equivalences~\cite{Caramello}, arithmetic-universe sketches describe change of base~\cite{Vickers}, and rounded sketches admit classifying logoi~\cite{DiLibertiLobbia}. Buchholtz and Schipp von Branitz give types for the classifying spaces of propositional geometric theories~\cite{BuchholtzSchipp}. Lynch and Myers take geometric theories as contexts in a category with families~\cite{LynchMyers}, and Riley gives two treatments of set-indexed syntax~\cite{Riley}. Our calculus recovers ordinary geometric logic by interpreting sorts as sets and predicates as subobjects. The translation preserves and reflects derivability of geometric sequents, and the $0$-truncated objects of its one-localic classifier form the ordinary classifying topos.

Kapulkin and Lumsdaine formulated the internal-language problem for finite limits~\cite{KapulkinLumsdaine}. Kapulkin and Szumi{\l}o treat it using tribes~\cite{KapulkinSzumilo}, and Nguyen and Uemura show that dependent type theory with intensional identity types gives an internal language for finitely complete $\infty$-categories~\cite{NguyenUemura}. Two precedents for our construction are the fiber representation of substitutions in extensional type theory~\cite{ClairambaultDybjer} and their factorization using intensional identity types~\cite{GambinoGarner}. In our calculus, representing a substitution by its family of fibers lets us construct its pullbacks by reindexing. Together with the terminal context, these pullbacks give all finite limits.

Type-theoretic constructions include graph-indexed homotopy limits~\cite{AvigadKapulkinLumsdaine} and colimits~\cite{Boulier}, and Hart and Hou construct colimits in coslices of a type universe~\cite{HartHou}. Sojakova characterizes induction for W-suspensions by homotopy initiality~\cite{Sojakova}, Lumsdaine and Shulman give semantics for higher inductive types including pushouts~\cite{LumsdaineShulman}, and Uemura constructs an initial finitely complete $\infty$-category with finitary higher inductive types~\cite{UemuraHIT}. Here, the same small simplicial shape indexes a colimit's constructors and the data for its dependent eliminator. To build a section over the colimit, one gives a local section at each vertex, a comparison along each edge, and compatible higher comparisons on the higher simplices.

Descent explains how to construct the family carrying those sections. Puppe and Rezk relate homotopy fibers to homotopy colimits~\cite{Puppe,RezkSheaves}. Heindel and Soboci\'nski characterize ordinary Van Kampen colimits using spans~\cite{HeindelSobocinski}, and Rezk and Lurie give descent criteria for higher toposes~\cite{HTT,RezkToposes}. Shulman identifies the unit and counit of the adjunction between total-space colimit and restriction with the comparisons for descent and stability~\cite{ShulmanDescent}. In type theory, the flattening lemma describes the total space of a family~\cite{HoTTBook}, and Rijke treats pushout descent via univalence and, with Spitters, descent for reflexive-graph colimits~\cite{Rijke,RijkeIntro}. Our calculus realizes the two adjunction comparisons as computations for every small simplicial diagram. The unit identifies the prescribed local families with the restrictions of the glued family, coherently with their transport. The counit reconstructs a global family from its restrictions.

The double cover of the circle makes the role of the descent rule concrete. The HoTT Book describes this cover~\cite{HoTTBook}, Licata and Shulman compute the fundamental group by transport in the universal cover~\cite{LicataShulman}, and Rezk compares pushouts of the span $1\leftarrow 2\to 1$ in sets and in spaces~\cite{RezkLeeds}. We use the identity and the transposition as the two gluing maps on a two-point fiber. The descent rule computes transport around the generating loop as the transposition. This transport distinguishes the loop from reflexivity. The other primitive rules admit a set-valued model in which the loop is reflexivity, so the same calculation proves the independence of descent computation.

Coherence also matters when a homomorphism carries a construction from one model to another. Bauer, Haselwarter, and Lumsdaine describe dependent theories by rules and derivations~\cite{BauerHaselwarterLumsdaine}, and Uemura constructs initial models and internal languages~\cite{UemuraFramework}. Algebraic fillers~\cite{Nikolaus}, marked diagrams for universal cones~\cite{Isaev}, and local universes for substitution~\cite{LumsdaineWarren} give related treatments of coherence. Our rules extend diagrams of inputs relative to outputs prescribed on a subdiagram. Applied to an arrow of models, this extension determines how the homomorphism acts on derived terms. Applied over higher simplices, it supplies the coherences that make restriction to the generators an equivalence of model $\infty$-categories.

Shulman constructs models with strict univalent universes in each Grothendieck $\infty$-topos~\cite{ShulmanUniverses}. Higher tripos theory produces elementary $\infty$-toposes with univalent universes from indexed logical data~\cite{MaruyamaHigherTripos}, and synthetic cohomology measures univalence for a given family~\cite{MaruyamaNasu}. Uemura uses modalities to express diagrams of $\infty$-logoses in dependent syntax~\cite{UemuraDiagrams}. Our classification associates a universal model to each small geometric theory. An interpretation of $\T$ in $\T'$ determines a geometric morphism $\B_{\T'}\to\B_\T$. Its inverse image sends the universal $\T$-model to the $\T$-model specified by the interpretation inside $\B_{\T'}$.

The inverse functor from the classifier to the syntax is where the descent calculation enters reconstruction. To\"en and Vezzosi characterize model topoi by Giraud axioms~\cite{ToenVezzosi}, Stenzel describes higher sheaves by covering diagrams and higher sites~\cite{StenzelSheaves,StenzelSites}, and homotopy-coherent sketches present accessible $\infty$-categories~\cite{CasacubertaEtAl}. Macpherson studies colimits subject to relations~\cite{Macpherson}, and Anel and Lejay present every $\infty$-logos as a left-exact localization of a free one~\cite{AnelLejay}. We apply the localization theorem of Anel, Biedermann, Finster, and Joyal~\cite{ABFJ} to the comparison maps expressing the theory's axioms. Their sequel relates localizations to extended Grothendieck topologies~\cite{ABFJII}. Raptis and Sch\"appi test left exactness of presheaf extensions into an $\infty$-topos on finite limits of representables~\cite{RaptisSchappi}. We prove left exactness with the generated syntax as target by checking pullbacks over representables and then applying descent over an arbitrary base. The resulting functor is inverse to the interpretation of contexts in the universal model, so the generated category of contexts inherits presentability from the localization. Conversely, writing the localization maps of a given Grothendieck $\infty$-topos as axioms produces a small dependent theory presenting it. The correspondence extends to interpretations and their transformations, giving the $(\infty,2)$-duality.

\subsection{Universes and Notation}
We work in ZFC with Grothendieck universes $\U\in\V$. The cardinal of $\U$ is denoted by $\kappa$. An auxiliary ambient universe accommodates functor categories of $\V$-small objects. A Grothendieck $\infty$-topos relative to $\U$ is an accessible left-exact localization of $\Psh(C)$ for a $\U$-small $\infty$-category $C$. For accessibility and presentability relative to $\U$, the witnessing regular cardinals lie below $\kappa$, the generating subcategories are $\U$-small, and the filtered colimits have $\U$-small indexing categories. In particular, $\Psh(C)=\Fun(C^{\op},\Sps_{\U})$. We use quasi-categories as concrete representatives. For a small simplicial set $K$ and a quasi-category $\D$, $\Fun(K,\D)$ denotes the quasi-category of $K$-shaped diagrams in $\D$. A fibrant replacement $K\hookrightarrow K'$ in the Joyal model structure represents the indexing $\infty$-category of $K$. Restriction gives an equivalence $\Fun(K',\D)\simeq\Fun(K,\D)$, so limits and colimits indexed by $K$ can be computed over $K'$.

A finite simplicial set has finitely many nondegenerate simplices. We fix representatives of the small shapes, and countably many representatives of the finite shapes. Every rule on another copy is transported along its specified isomorphism. We write $\hcolim$ for the type constructor and $\colim$ for its interpretation in an $\infty$-category, where colimits are homotopy colimits. The cellular construction in Section~\ref{sec:construction} uses ordinary colimits of presentations.

\subsection{Organization of the Paper}
Section~\ref{sec:syntax} introduces the dependent language, gives a simplicial grammar for its rules, and constructs the syntax generated by a theory. Section~\ref{sec:recognition} uses fiber representation to construct finite limits and to glue families over homotopy colimits. Section~\ref{sec:initiality} proves initiality, and Section~\ref{sec:boundarycomparison} compares the dependent calculus with its presentation by universal cones.

Section~\ref{sec:classification} constructs the classifying $\infty$-topos by presheaf localization and uses descent to prove its equivalence with the syntactic category. Section~\ref{sec:proofspaces} describes dependent proof spaces and context extension in terms of the classifier. Section~\ref{sec:alltopoi} gives small presentations of all Grothendieck $\infty$-toposes and identifies interpretations with inverse images, yielding the Morita correspondence and the duality. It then constructs two universal families of classifiers. In the first, the generators and axioms of a theory vary over the base. The second is parametrized by a Morita space of theories.

Section~\ref{sec:ordinary} recovers ordinary geometric logic. Section~\ref{sec:circle} computes the monodromy of a double cover of the circle and proves the independence of the descent rule. Appendix~\ref{app:typing} gives explicit derivations of the dependent operations, and Appendix~\ref{app:equivalent-parameters} describes coherence for equivalent parameters. Appendix~\ref{app:universal-families} gives the construction and base-change proofs for both families.

\section{The Syntax of Geometric HoTT}\label{sec:syntax}
The syntax of a small theory $\T$ is generated from its declarations by rules for dependent types and their higher terms. The simplicial form of the rules records the boundary of each new expression, so the same construction generates derivations in every dimension.

\subsection{Dependent Types and Homotopy Colimits}\label{sec:language}
The display of a type lets us describe substitution by pullback and dependent terms by sections.

The basic judgments are
\[
\Gamma\ \ctx,\qquad \sigma:\Delta\longrightarrow\Gamma,
\qquad \Gamma\vdash A\ \type,\qquad \Gamma\vdash a:A.
\]
There are also judgments for coherent diagrams of substitutions, higher terms with specified boundaries, and equivalence data. A declared type $A$ binds a total context $\Gamma.A$ and a display
\[
p_A:\Gamma.A\longrightarrow\Gamma.
\]
A typing derivation gives the projection $p_A$ its display certificate $\Disp(p_A)$. The display and its boundary determine the family and its terms, so derivations with the same display and boundary present the same family. The fiber-representation theorem constructs such a display for every substitution.

A term records a section and its section computation:
\[
s_a:\Gamma\longrightarrow\Gamma.A,
\qquad p_A s_a\simeq\id_\Gamma.
\]
This pair is one typed triangular judgment. A higher term records a coherent boundary of such triangles and an extension of that boundary. Section~\ref{sec:formats} specifies the boundary of each such judgment. A type name annotates a derived display, so different names can refer to the same display. Family maps are arbitrary maps of total contexts over the base, with their comparison triangles.

\subsubsection{Context Comprehension and Reindexing.}
Contexts can be named and can be extended by a variable. There is an empty context $1$. The context-type rule gives a closed type $\pack(\Gamma)$ with total context $\Gamma$. A \emph{telescope} is a finite sequence of variable declarations $\Gamma=(x_1:A_1,\ldots,x_n:A_n)$ in which each $A_i$ is a type in the context of the preceding variables $x_1,\ldots,x_{i-1}$. Its iterated dependent sum also has total context $\Gamma$. Since $\Map(\Gamma,1)$ is contractible, $\id_\Gamma$ induces a canonical equivalence between the two displays over $1$. This convention is often called democracy: every context can be represented by one closed type. It permits the same syntax to express both a list of dependent variables and a named structured context.

Substitution of generator symbols acts strictly functorially on derivation trees. Reindexing acts on types in context, with coherent comparisons for identities and composites. We write $\equiv$ for judgmental equality and $a=_A b$ for the identity type $\Id_A(a,b)$. In a type-theoretic computation, we abbreviate $a=_A b$ to $a=b$ when the type $A$ is clear. Between objects, $\simeq$ denotes an equivalence, with coherent inverse data. For families over the same context, it denotes an equivalence of their displays in the slice over that context. Between parallel maps, $\simeq$ denotes a homotopy in their mapping space. In particular, $p_A s_a\simeq\id_\Gamma$ records the specified triangle exhibiting $\id_\Gamma$ as a composite of $s_a$ and $p_A$.

Given a display over $\Gamma$, substitution along $\sigma:\Delta\to\Gamma$ reindexes it by the formation rule
\[
\inferrule{\Gamma\vdash A\ \type\\\sigma:\Delta\to\Gamma}
{\Delta\vdash A[\sigma]\ \type}.
\]
It simultaneously introduces a substitution $\sigma_A:\Delta.A[\sigma]\to\Gamma.A$ and a commuting square
\begin{equation}\label{eq:reindex-square}
\begin{tikzcd}[column sep=large,row sep=large]
\Delta.A[\sigma]\ar[r,"\sigma_A"]\ar[d,"p_{A[\sigma]}"']&\Gamma.A\ar[d,"p_A"]\\
\Delta\ar[r,"\sigma"']&\Gamma.
\end{tikzcd}
\end{equation}
Its introduction rule takes $u:Z\to\Gamma.A$, $v:Z\to\Delta$, and a specified comparison $p_Au\simeq\sigma v$ to a substitution into the new total context. Its computations recover $u,v$ and that comparison. Higher comparison rules extend every compatible boundary of introductions. In each comparison simplex, the last vertex carries the certified pullback cone to which the other cones are compared.

Reindexing a section gives $a[\sigma]:A[\sigma]$. When $p_A$ is reindexed along itself, taking $u=v=\id_{\Gamma.A}$ with the identity comparison on $p_A$ gives the diagonal section representing the variable in $\Gamma.A$. The comparisons
\[
A[\id]\simeq A,\qquad A[\sigma][\tau]\simeq A[\sigma\tau]
\]
are produced by the same universal introduction rules, together with their higher comparisons. The relative constructor scheme~\eqref{eq:relative} applies a rule to a family of inputs while retaining the outputs already prescribed on a parameter subshape.

\subsubsection{Unit, Dependent Sums, and Identity Types.}\label{subsec:identity}
The unit family $\Gamma\vdash 1_\Gamma\ \type$ has display $\id_\Gamma$. For $\Gamma\vdash A\ \type$ and $\Gamma,x:A\vdash B(x)\ \type$, form
\[
\Gamma\vdash\Sigma_{x:A}B(x)\ \type.
\]
Its total context is the iterated total context $\Gamma.A.B$; its display is the coherently specified composite $p_Ap_B$. Thus the operation generates a display for a composite of displays.

For a term $a:A$ with section $s_a:\Gamma\to\Gamma.A$, write $B[a]$ for $B[s_a]$. The pairing rule is
\[
\inferrule{\Gamma\vdash a:A\\\Gamma\vdash b:B[a]}
{\Gamma\vdash(a,b):\Sigma_{x:A}B(x)}.
\]
Pairing is constructed by the reindexing introduction and the composition triangle. Its projections and computations are the corresponding reindexing comparisons. Dependent elimination is expressed by the common total context of $\Gamma,z:\Sigma_{x:A}B(x)$ and $\Gamma,x:A,y:B(x)$. All its comparison boundaries are part of the same universal scheme.

To form the identity type of $p_A:E\to\Gamma$, first reindex the display along itself to obtain $Q=E\times_\Gamma E$. Universal introduction supplies its relative diagonal $\delta_A:E\to Q$. Identity formation introduces a display $p_{\Id}:I_A\to Q$ and a reflexivity equivalence
\begin{equation}\label{eq:identity-factor}
E\xrightarrow{r_A\;\simeq}I_A\xrightarrow{p_{\Id}}Q,
\qquad p_{\Id}r_A\simeq\delta_A.
\end{equation}
This is a formation rule for the family
\[
\Gamma,x:A,y:A\vdash\Id_A(x,y)\ \type.
\]
Reindexing produces $\Id_A(a,b)$ for specified terms. The equivalence in~\eqref{eq:identity-factor} identifies the total identity context with the reflexivity context; the fibers of $p_{\Id}$ retain the paths between arbitrary endpoints.

For a family $c:C\to I_A$, a term $d$ over reflexivity gives a section $s_d:E\to E\times_{I_A}C$. Put $\bar d=\operatorname{pr}_C s_d:E\to C$. The reindexing square and the section triangle give $c\bar d\simeq r_A$. Using the supplied inverse of $r_A$, set $j=\bar d\,r_A^{-1}:I_A\to C$. The composite $cj\simeq r_A r_A^{-1}\simeq\id_{I_A}$ is its section triangle. This gives the rule
\[
\inferrule{\Gamma,x:A,y:A,p:\Id_A(x,y)\vdash C\ \type\\
\Gamma,x:A\vdash d:C[x,x,\refl_x]}
{\Gamma,x:A,y:A,p:\Id_A(x,y)\vdash J_C(d,x,y,p):C}.
\]
The inverse computation for $r_A$ gives the computation of $J_C$ at reflexivity. Section~\ref{sec:formats} encodes the coherent inverse data for $r_A$ by binary strings. The strings $010$ and $101$ represent the homotopies $r_A^{-1}r_A\simeq\id$ and $r_A r_A^{-1}\simeq\id$, respectively. Longer strings record the higher coherences. The identity-formation rule supplies the reflexivity equivalence $r_A$, and its coherent inverse gives path induction.

\subsubsection{Homotopy-Colimit Types.}\label{subsec:hcolim}
Let $K$ be a small external simplicial shape. It specifies the shape of a family of premises and the equations on its boundary. A judgment
\[
\Gamma\vdash D:\mathsf{Diag}_K
\]
consists of types $D_k$ over $\Gamma$, substitutions between their total contexts over $\Gamma$, and their specified higher coherence cells. The complete input is an externally indexed array of type judgments, maps, and higher comparisons over $\Gamma$. The notation $\mathsf{Diag}_K$ abbreviates that array, with the cell equations specified in Section~\ref{sec:formats}.

There is a formation rule
\[
\inferrule{\Gamma\vdash D:\mathsf{Diag}_K}
{\Gamma\vdash W:=\hcolim_{k\in K}D_k\ \type}.
\]
It introduces constructors $\iota_k:D_k\to W$, one compatibility cell for each cell of $K$, and a formal certificate $\HC_K(\iota)$ of their elimination rules. Fix a family $\Gamma,w:W\vdash B(w)\ \type$. This family is the \emph{motive} of dependent elimination and specifies the type $B(w)$ of the eliminator's value at each $w$. Let $b^\sharp$ be a coherent array of local terms
\[
\Gamma,x:D_k\vdash b_k(x):B(\iota_kx)
\]
together with their section triangles and higher coherences. For the constructor cocone $\iota$, the derived term $\mathsf{ind}_{\iota,B}(b^\sharp)$ extends these data over $W$. Along each constructor $\iota_k$, the eliminator restricts to the given local term $b_k$, up to a specified identification. These identifications respect the section triangles and higher coherences in $b^\sharp$. Comparison rules extend every compatible boundary between such extensions. Their full boundary format distinguishes the initial constructor cocone.

Homotopy-colimit formation is stable under substitution:
\begin{equation}\label{eq:stable}
(\hcolim_K D)[\sigma]\simeq\hcolim_K(D[\sigma]).
\end{equation}
The stability rule states that the reindexed constructor diagram itself receives the homotopy-colimit elimination certificate. Universal comparison gives~\eqref{eq:stable}. The rule includes $K=\varnothing$. Discrete shapes give coproducts, and spans give pushout types.

\subsubsection{Gluing Families over Homotopy Colimits.}\label{subsec:glue}
Suppose $W=\hcolim_K D$. A compatible local family consists of
\[
\Gamma,x:D_k\vdash P_k(x)\ \type,
\]
with a specified equivalence $P_k\simeq D_e^*P_l$ for every edge $e:k\to l$ and specified compatibilities on all higher cells. Equivalently, the projections from the total spaces $E_k=\Sigma_{x:D_k}P_k(x)$ form a \emph{cartesian family} over $D$, a map of diagrams $E\to D$ whose naturality square at each edge of $K$ is a pullback. These squares are the reindexing squares of the local families. This is an explicit array of type judgments, substitutions, and coherent computations.

Apply dependent sum and homotopy-colimit formation to obtain
\begin{equation}\label{eq:glue-total}
 V:=\hcolim_{k\in K}E_k,\qquad
 g:V\longrightarrow W,
\end{equation}
where the map $g$ and its constructor comparisons are induced by the local projections. Define
\begin{equation}\label{eq:glue-rule}
 \Gamma,w:W\ \vdash\
 \glue_D(P)(w):=\Sigma_{v:V}\Id_W(gv,w)\ \type.
\end{equation}
The right-hand side is formed from $\Sigma$, $\Id$, and $\hcolim$. For each $k$, its constructors give a comparison
\[
 c_k:P_k\longrightarrow\glue_D(P)[\iota_k].
\]
For $u:P_k(x)$, let $v$ be the image of $(x,u)$ under the constructor $E_k\to V$. Then $c_k(u)$ is the pair consisting of $v$ and the specified path from $g(v)$ to $\iota_k(x)$.

The \emph{descent computation rule} certifies the squares
\[
\begin{tikzcd}[column sep=large]
 E_k\ar[r]\ar[d]&V\ar[d,"g"]\\
 D_k\ar[r,"\iota_k"']&W
\end{tikzcd}
\]
as reindexing squares whenever both rows are the specified homotopy-colimit cocones and every transition square of the local family is a reindexing square. Fiber representation and reindexing comparison then give the local computation equivalences
\begin{equation}\label{eq:glue-beta}
 \beta_k:\glue_D(P)[\iota_k]\simeq P_k,
 \qquad \beta_k\simeq c_k^{-1}.
\end{equation}
The rule retains the complete \emph{cocone prism}, the coherent map from the total-space cocone $E_k\to V$ to the base cocone $D_k\to W$. The square above is its restriction to the constructor edge at $k$. Consequently these comparisons carry the specified transition equivalences and their higher coherences. The relative rules let us fix the base cocone first and then construct the total-space cocone and its map to that base, retaining the outputs prescribed on a parameter subshape (Section~\ref{subsec:syntactic-gluing}). The same parameter scheme gives coherent reindexing, identity formation, and both colimit computations.

For a family $B$ over $W$, let $R_D B$ denote its coherent restriction along the constructor cocone. Its components are $B[\iota_k]$, with transition comparisons induced by the cocone. Global reconstruction gives the derived equivalence
\begin{equation}\label{eq:glue-eta}
 \glue_D(R_D B)\simeq B.
\end{equation}
It follows from substitution stability: the pullback of the constructor cocone along $B\to W$ is a colimit of the restricted total spaces with apex $B$. Theorem~\ref{thm:local-descent} shows that taking the colimit of the local total spaces is left adjoint to restricting a family along the constructors. Under fiber representation, the component of the unit at $k$ is the map $c_k:P_k\to\glue_D(P)[\iota_k]$. Its inverse is the local computation $\beta_k$ in~\eqref{eq:glue-beta}. The counit at $B$ is the reconstruction map~\eqref{eq:glue-eta} from the glued restrictions to $B$. Naturality of the unit and counit and their triangle identities give the computation laws on all family maps and higher terms.

Context representation and reindexing allow the constructors $1,\Sigma,\Id,\hcolim$ to act in every context. The expression~\eqref{eq:glue-rule} therefore assembles families in context as well. Geometric inverse images transport this construction together with its local and global computations.

To see how descent computes transport, consider a span $A\xleftarrow{f}C\xrightarrow{g}B$ in context $\Gamma$. Its pushout $W$ has constructor maps $i_A,i_B$ and paths
\[
\alpha_c:i_A(fc)=_W i_B(gc)\qquad(c:C).
\]
Given families $P_A$ over $A$ and $P_B$ over $B$, an equivalence
\[
c:C\ \vdash\ e_c:P_A(fc)\simeq P_B(gc)
\]
with its coherent dependence on $c$ is precisely the local input for gluing. For the resulting family $P=\glue(P_A,P_B,e)$, write
\[
 \beta_A:P[i_A]\simeq P_A,\qquad \beta_B:P[i_B]\simeq P_B
\]
for the local equivalences of~\eqref{eq:glue-beta}. At each $c:C$, these equivalences identify transport $\operatorname{tr}_P$ along $\alpha_c$ with the transition $e_c$ by a homotopy of maps from $P(i_A(fc))$ to $P_B(gc)$,
\[
 \beta_B(gc)\circ\operatorname{tr}_P(\alpha_c,-)
 \simeq e_c\circ\beta_A(fc).
\]
Once $P$ is formed, sections $a$ of $P_A$ and $b$ of $P_B$ glue to a section of $P$ by pushout elimination whenever they are equipped with coherent paths $e_c(a(fc))=b(gc)$. Thus the same transition equivalence specifies transport in the family and the compatibility needed to glue its sections. In Section~\ref{sec:circle}, choosing a transposition as one transition gives the double cover of the circle.

\subsection{A Simplicial Grammar for Higher Terms and Computation Rules}\label{sec:formats}
To specify the higher computations of these operations, we give each rule a boundary of input expressions and describe the cells that extend it. The shape of a cell records its dimension, and its faces record its typing and computation data.

A simplicial shape is a set of symbols in each dimension with face and degeneracy operations. Let $\Delta^n$ be the standard simplex, $\partial\Delta^n$ its boundary, and $\Lambda_i^n$ its $i$th horn. Use the ordinary join $K\star L$, with
\[
(K\star L)_n=\coprod_{p+q+1=n}K_p\times L_q,
\qquad K_{-1}=L_{-1}=\{*\}.
\]
Write $K^\triangleleft=\Delta^0\star K$ and $K^\triangleright=K\star\Delta^0$. The inverse-data interval is the explicit simplicial set
\[
J_n=\{0,1\}^{n+1},
\]
whose faces delete entries and degeneracies repeat them. The image of the inclusion $j:\Delta^1\hookrightarrow J$ consists of the nondecreasing binary strings.

The raw structural expressions have a dimension: dimension zero contains context symbols, dimension one contains substitution symbols with specified endpoints, and dimension $n\geq2$ contains coherence symbols with their specified faces. For every monotone map $\theta:[m]\to[n]$, the grammar includes a dimension-changing operation $\theta^*$. Generator cells and constructor outputs are the remaining raw expressions. Their simplicial and boundary equations are listed below. The type and term judgments refine these structural expressions by the generated certificates and boundary requirements.

\subsubsection{Context Diagrams and Higher Terms.}
A context diagram on $K$ is an array of formal expressions $a_\sigma$ for every $\sigma\in K_n$, with equations
\begin{equation}\label{eq:array}
\theta^*a_\sigma\equiv a_{\theta^*\sigma}.
\end{equation}
In dimensions zero and one, these expressions are contexts and substitutions. In higher dimensions, they are substitution coherence judgments with their prescribed faces. A type is a substitution with a derivation of $\Disp$. A dependent term of $p:E\to\Gamma$ is a triangular expression $a$ with
\begin{equation}\label{eq:term-triangle}
d_0a\equiv p,\qquad d_1a\equiv s_0\Gamma.
\end{equation}
Its remaining edge is the section. An $n$-dimensional term is an $(n+2)$-dimensional expression $a$ with
\begin{equation}\label{eq:termspace}
a|_{\{0,\ldots,n,n+2\}}\equiv\operatorname{const}_\Gamma,
\qquad a|_{\{n+1,n+2\}}\equiv p.
\end{equation}
With faces and degeneracies acting on the vertices $0,\ldots,n$ and fixing the final two vertices, the higher terms form the simplicial set $\Tm_\T(\Gamma,p)$.

\subsubsection{Formal Certificates and Relative Rules.}\label{subsec:relative}
A presentation is a simplicial set of such formal symbols, together with sets of judgments of the following shapes:
\[
\begin{array}{c|c|l}
\text{judgment}&\text{shape}&\text{syntactic role}\\\hline
\Term&\Delta^0&\text{empty-context comparison rules}\\
\Disp&\Delta^1&\text{well-formed family display}\\
\PB&\Delta^1\times\Delta^1&\text{reindexing comparison rules}\\
\HC_K&K^\triangleright&\text{homotopy-colimit comparison rules}\\
\Eq&\Delta^1&\text{equivalence-edge certificate}.
\end{array}
\]
The certificate $\HC_K$ is indexed by an arbitrary small shape $K$. Write $|P|$ for the underlying simplicial set of a presentation $P$. Each row of the table specifies a certificate $t$ of shape $H_t$, so $t(m)$ marks a diagram $m:H_t\to |P|$. We write $\cmark{K}{t_1(m_1),\ldots,t_s(m_s)}$ for the presentation with underlying simplicial set $K$ and the listed marks. Write $K^\flat$ for the unmarked shape $K$. For example, $\Disp(01)$ marks the edge from vertex $0$ to vertex $1$ as a display, and $\PB(\id)$ marks the whole square via the identity map of $\Delta^1\times\Delta^1$. Section~\ref{sec:recognition} identifies the categorical structure expressed by each mark.

A map of presentations is a simplicial map preserving the specified certificates. A rule is specified by a \emph{proof packet} consisting of an input presentation $A_r$, an output presentation $B_r$, and a boundary inclusion $r:A_r\hookrightarrow B_r$. The rule can introduce symbols or certify an existing array. For an input array $a:A_r\to P$, every output cell $\sigma\in(B_r)_m$ has a constructor name $\mathsf{out}_{r,\sigma}(a)$. The equations are precisely
\begin{align}
\mathsf{out}_{r,\sigma}(a)&\equiv a_\sigma &&(\sigma\in A_r),\label{eq:output1}\\
\theta^*\mathsf{out}_{r,\sigma}(a)&\equiv
\mathsf{out}_{r,\theta^*\sigma}(a),\label{eq:output2}\\
\id^*u&\equiv u,
&\phi^*(\theta^*u)&\equiv(\theta\phi)^*u.\label{eq:simplicial-eq}
\end{align}
Judgmental equality is the least congruence generated by these equations and the declared equations of the theory. Certificate derivations license rule applications; constructor names depend on the rule and its input array. Each attachment retains exactly the judgmental equalities between its previously present cells. A certificate records a formation or universal-comparison judgment. A dependent term and its higher identities are cells with the boundary specified in~\eqref{eq:termspace}.

A rule can also act on an array of inputs while keeping specified outputs fixed. For a presentation $P$ and a small shape $I$, let $P\otimes_v I$ have underlying shape $|P|\times I$. Its certificates are exactly the pairs $(m,i)$ with $m$ certified in $P$ and $i\in I_0$ constant. For each primitive rule $r:A_r\hookrightarrow B_r$ among the \hyperref[subsec:dependent-formats]{structural and dependent-type rules} or the \hyperref[subsec:colim-formats]{homotopy-colimit rules} and each small monomorphism $j:J_0\hookrightarrow I$, include
\begin{equation}\label{eq:relative}
r\squ j:
(A_r\otimes_v I)\amalg_{A_r\otimes_v J_0}(B_r\otimes_v J_0)
\hookrightarrow B_r\otimes_v I.
\end{equation}
This one scheme supplies the parameterized and relative forms of each constructor. The source contains the premises over $I$ and the outputs already given over $J_0$. All source certificates are inherited from the displayed pieces. The case $\partial\Delta^n\hookrightarrow\Delta^n$ specifies a computation with a prescribed higher boundary.

\subsubsection{Structural Rules and Rules for Dependent Types.}\label{subsec:dependent-formats}
The structural rules give substitutions their composites and higher coherences. The dependent sum of two types in successive contexts composes their displays, and reindexing pulls a display back along a substitution. For a display $p:E\to\Gamma$, identity formation factors the relative diagonal $E\to E\times_\Gamma E$ as a reflexivity equivalence followed by the display of the identity type.

\paragraph{Homotopy-Coherent Composition of Substitutions.}
For $n\geq2$ and $0<i<n$, include
\[
(\Lambda_i^n)^\flat\hookrightarrow(\Delta^n)^\flat.\tag{H}
\]
Thus composite substitutions and their higher coherences are explicit constructor outputs.

\paragraph{Empty Context and Its Comparisons.}
Include $\varnothing\to\cmark{\Delta^0}{\Term(*)}$ and, for $n\geq1$,
\[
\cmark{\partial\Delta^n}{\Term(n)}\hookrightarrow
\cmark{\Delta^n}{\Term(n)}.\tag{T}
\]
The unit-family rule is the identity-on-cells inclusion
\[
(\Delta^0)^\flat\longrightarrow\cmark{\Delta^0}{\Disp(s_0*)}.\tag{Unit}
\]
Thus every context has its identity display. Context packing is
\[
\cmark{\partial\Delta^1}{\Term(1)}\hookrightarrow
\cmark{\Delta^1}{\Term(1),\Disp(01)}.\tag{Pack}
\]
Its two input vertices are the context to be packed and the empty context.

\paragraph{Dependent Sum.}
For the composable displays $q:0\to1$ and $p:1\to2$, include
\[
\cmark{\Lambda^2_1}{\Disp(01),\Disp(12)}\hookrightarrow
\cmark{\Delta^2}{\Disp(01),\Disp(12),\Disp(02)}.\tag{Sum}
\]
The newly formed composite is the sum display.

\paragraph{Reindexing.}
Put $V=\Lambda^2_2$, with arrows $0\to2\leftarrow1$. The fixed identification
\[
\Delta^0\star V\cong\Delta^1\times\Delta^1
\]
identifies a cone with a square.
Label the vertices of $V$ by $0,1,2$, corresponding to $E,\Delta,\Gamma$. Reindexing formation is the inclusion
\[
\cmark{V}{\Disp(02)}\hookrightarrow
\cmark{\Delta^0\star V}{\Disp(02),\Disp(*1),\PB(\id)},
\]
where $*$ is the new cone vertex. Thus the total context, its projection to $\Delta$, and the reindexing square are produced together. The comparison rules, for $n\geq1$, are
\[
\cmark{\partial\Delta^n\star V}{\PB(\{n\}\star V)}
\hookrightarrow
\cmark{\Delta^n\star V}{\PB(\{n\}\star V)}.\tag{Reind}
\]
The last-vertex cone is the one to which an introduction is compared. These rules give the comparisons for every already certified reindexing square.

\paragraph{Identity Formation.}
The input is a display $p:E\to\Gamma$, its reindexing along itself, and a comparison $\delta:E\to Q$ from the identity cone to that square. The following finite quotient specifies this packet exactly. Let $u_0,u_1$ be the first two vertices of $\Delta^1\star V$, and write $a,b,c$ for the vertices of $V$. Let
\[
c_0:\{u_0\}\star V\longrightarrow\Delta^1
\]
send $u_0,a,b$ to $0$ and $c$ to $1$. Form
\begin{equation}\label{eq:identity-packet}
A_{\Id}=(\Delta^1\star V)
\amalg_{\{u_0\}\star V}\Delta^1.
\end{equation}
Certify the surviving edge $p:0\to1$ by $\Disp$ and the second cone $\{u_1\}\star V$ by $\PB$. Its vertex is $Q$, and the edge $u_0u_1$ is $\delta$. The quotient makes the first cone the identity cone on $p$. Attach $\Delta^2$ along its edge $02=\delta$. The new vertex is $I$, and the output edges are
\[
E\xrightarrow{r}I\xrightarrow{q}Q.\tag{Id}
\]
For a set $S$ of vertices of a simplex, write $\Delta^S$ for the face spanned by $S$. The target is $A_{\Id}\amalg_{\Delta^{\{0,2\}}}\Delta^2$, with the inherited certificates and $\Eq(01),\Disp(12)$. The attachment introduces exactly $I,r,q$ and the triangle $qr\simeq\delta$. Since the attached edge is a simplicial subobject of $\Delta^2$, the rule is monic on its underlying input cells.

\paragraph{Equivalences and Change of Presentation.}
Include
\[
\cmark{\Delta^1}{\Eq(\id)}\longrightarrow\cmark{J}{\Eq(j)},
\qquad J^\flat\longrightarrow\cmark{J}{\Eq(j)}.\tag{E}
\]
The first rule extends a marked edge to a $J$-diagram giving its coherent inverse data. The second certifies the distinguished edge of any such diagram.
For a certificate $t$ of shape $H_t$ and $\epsilon\in\{0,1\}$, let $\iota_\epsilon:H_t\to J\times H_t$ be the endpoint inclusion $h\mapsto(\epsilon,h)$. Include the rule
\[
\cmark{J\times H_t}{t(\iota_0)}\longrightarrow
\cmark{J\times H_t}{t(\iota_0),t(\iota_1)}.\tag{Tr}
\]
For $\Disp$, this transports a typing derivation along an explicit equivalence of family presentations. The $J$-array records the equivalence and its higher witnesses, with judgmental boundaries given by~\eqref{eq:output1}--\eqref{eq:simplicial-eq}.

\subsubsection{Formation and Computation Rules for Homotopy Colimits.}\label{subsec:colim-formats}
The formation rule constructs a colimit in a context. The cocone comparisons in (HC-comp) supply the section and its computation for dependent elimination. Reindexing preserves the colimit by (Stab), and (Desc) identifies the restrictions of a glued family with the prescribed local families.

\paragraph{Dependent Homotopy-Colimit Formation.}
For a diagram $D$ of types over $\Gamma$ indexed by a small shape $K$, the input has shape $K\star\{1\}$. Its vertex at $k$ is $\Gamma.D_k$, and its edge from $k$ to the final vertex is the display $p_{D_k}:\Gamma.D_k\to\Gamma$. Include
\begin{equation}\label{eq:HCformation}
\cmark{K\star\{1\}}{\Disp(k1)\ (k\in K_0)}
\hookrightarrow
\cmark{K\star\Delta^1}{\Disp(k1),\HC_K(K\star\{0\}),\Disp(01)}.
\tag{HC}
\end{equation}
For $W=\hcolim_K D$, the new vertex $0$ is the total context $\Gamma.W$. The edge $01$ is its display $p_W:\Gamma.W\to\Gamma$, and the triangles on $k01$ give the constructor comparisons over $\Gamma$. For $n\geq1$, include
\begin{equation}\label{eq:HCcomparison}
\cmark{K\star\partial\Delta^n}{\HC_K(K\star\{0\})}
\hookrightarrow
\cmark{K\star\Delta^n}{\HC_K(K\star\{0\})}.
\tag{HC-comp}
\end{equation}
The first-vertex cocone is distinguished. Proposition~\ref{prop:hc-derived} constructs the section map in dimension~$1$ and its section triangle in dimension~$2$. In a fixed outer context $\Gamma$, a further dimension-$2$ comparison expresses that the section lies over $\Gamma$, and a dimension-$3$ comparison makes this compatible with the section triangle. The relative forms~\eqref{eq:relative} retain the entire prescribed local computation array and any already specified output boundary.

\paragraph{Substitution Stability.}
On the shape $K^\triangleright\times\Delta^1$, denote the upper row by $P$, the lower row by $D$, and its cone vertex by $*$. Write $q_{k*}$ for the square on $(k\to*)\times\Delta^1$. Include the identity-on-cells rule
\begin{equation}\label{eq:U}
\frac{\HC_K(D)\qquad\PB(q_{k*})\ (k\in K_0)}{\HC_K(P)}.
\tag{Stab}
\end{equation}
All small shapes, including the empty shape, occur.

\paragraph{Descent Computation.}
On the same shape $K^\triangleright\times\Delta^1$, write $q_e$ for the square on $e\times\Delta^1$, for every edge $e$ of $K$. Include the identity-on-cells rule
\begin{equation}\label{eq:Dformat}
 \frac{\HC_K(P)\qquad\HC_K(D)\qquad
       \PB(q_e)\ (e\in K_1)}
      {\PB(q_{k*})\ (k\in K_0)}.
 \tag{Desc}
\end{equation}
The source records both colimit cocones, their mixed comparison cells, and the pullback certificates on transition squares. The rule certifies each cone square $q_{k*}$ in this prism, giving the local computation of the fiber expression~\eqref{eq:glue-rule}. For $K=\varnothing$, the family of cone squares is empty. Given an arrow $X\to 0$ to a context $0$ certified as an empty colimit, (Stab) certifies $X$ as an empty colimit and hence as an initial context.

\subsection{Transfinite Construction of the Syntax and Substitution on Derivations}\label{sec:construction}
A presentation supplies the generators and axioms to which these rules apply.

\begin{definition}[Geometric dependent syntax]\label{def:syntax}
The calculus $\mathsf{GHoTT}$ consists of the typed judgments of Section~\ref{sec:language}, with the primitive boundary formats (H)--(Desc), their relative forms~\eqref{eq:relative}, and equations~\eqref{eq:output1}--\eqref{eq:simplicial-eq}. A derivation is a well-founded tree of applications to explicit input arrays, with derivations of their required certificates. The names $A[\sigma]$, pairing, $J_C$, and $\hcolim$ denote the output packets and their derived elimination operations described above. The expression $\glue$ is the composite~\eqref{eq:glue-rule}; its local computations are derived using (Desc). A small theory $\T$ is specified by a small generating presentation $Q_\T$ of constants, boundary equations, and certificate axioms. Its syntax $\G_\T$ is the simplicial set of derived context diagrams modulo judgmental equality, with its derived certificates.
\end{definition}
To form the generating presentation from dependent declarations, take their total contexts, displays, terms, and boundary cells. The underlying simplicial set of $Q_\T$ is generated by the declared cells subject to their boundary equations, and its marks are the declared certificate axioms. The constructors subsequently obey~\eqref{eq:output1}--\eqref{eq:simplicial-eq}.

Let $\kappa$ be the inaccessible cardinal of $\U$ and choose a regular $\lambda>\kappa$ in $\V$. The rule formats form a set in $\V$, and each source, target, and collection of certificates is $\U$-small. Start with $P_0=Q_\T$. For $\alpha<\lambda$, let $I_\alpha$ be the set of pairs $(r,a)$ with $a:A_r\to P_\alpha$ whose input array and premise certificates are first available together at stage $\alpha$. Attach one copy of $B_r$ along each such $a$ by forming the pushout
\begin{equation}\label{eq:cellular}
\begin{tikzcd}[column sep=large]
 \coprod_{(r,a)\in I_\alpha} A_r
   \ar[r,"{[a]}"] \ar[d,hook,"{\coprod r}"']
 & P_\alpha \ar[d,hook] \\
 \coprod_{(r,a)\in I_\alpha} B_r \ar[r]
 & P_{\alpha+1}.
\end{tikzcd}
\end{equation}
On the $(r,a)$-summand, the top map is $a$ and the left map is the rule inclusion $r:A_r\hookrightarrow B_r$. Record each processed input and its constructor outputs. At each nonzero limit ordinal $\beta\leq\lambda$, set $P_\beta=\colim_{\alpha<\beta}P_\alpha$. These colimits of presentations are computed on the underlying simplicial sets, with precisely the certificates inherited from the pieces.

\begin{lemma}[Formation and closure]\label{lem:formation}
The presentation $P_\lambda$ is $\V$-small, is closed under every primitive and relative rule, and presents the derivation-tree syntax of Definition~\ref{def:syntax}. All transitions are injective on cells.
\end{lemma}
\begin{proof}
Every rule inclusion is monic on underlying simplicial sets. The identity-on-cells rules add certificates; the other rules adjoin the displayed cells with their faces. Pushouts and transfinite unions preserve these monomorphisms. In particular, an equality between two previously constructed cells holds at a later stage exactly when it already held at their common earlier stage.

The rule set is an object of $\V$. If $P_\alpha$ is $\V$-small, the maps from all its rule sources form a set in $\V$, and their coproduct of target packets is again $\V$-small. The pushout is therefore $\V$-small. At each limit $\beta\leq\lambda$, the union is indexed by an ordinal in $\V$, so it remains in $\V$. This proves the size assertion by transfinite induction.

Let $a:A_r\to P_\lambda$ be a rule input. Its source has fewer than $\kappa$ cells and certificates. Each image cell occurs at some stage, and each image certificate has been added at some stage together with a derivation. There are fewer than $\lambda$ such stages, so regularity gives one common bound $\alpha<\lambda$. The boundary equations already hold in $P_\alpha$, by injectivity of the transitions. Thus $a$ is a map of presentations to $P_\alpha$. If its instance was processed earlier, its output is in the record; otherwise, it is processed at stage $\alpha+1$. This proves closure, also for relative sources and for sources containing specified old output cells.

A new cell of a pushout is exactly a constructor output $\mathsf{out}_{r,\sigma}(a)$. Its identifications with the old cells and its face and degeneracy equations are precisely~\eqref{eq:output1}--\eqref{eq:output2}. The record ensures that the same rule on the same input array has one constructor name. Conversely, well-founded induction on a derivation puts all its premises at a common bounded stage and its conclusion at a subsequent stage. This identifies the derivation-tree presentation with $P_\lambda$ and its stated judgmental congruence.\qed
\end{proof}

The \emph{support} of a derivation is the set of generators and rule instances occurring in its derivation tree, including the subtrees that derive its premise certificates.

\begin{lemma}[Symbolic substitution and small support]\label{lem:substitution}
An assignment of generator symbols that respects their equations and derives their certificate axioms extends to all derivations. This substitution is judgmentally unital and associative. Each derivation has $\U$-small support; a small family of derivations has a small union of supports.
\end{lemma}
\begin{proof}
Define substitution recursively by
\[
\mathsf{out}_{r,\sigma}(a)[\rho]
=\mathsf{out}_{r,\sigma}(a[\rho]),\qquad
(\theta^*t)[\rho]=\theta^*(t[\rho]).
\]
The premises remain derivable by induction, so the output constructor exists by Lemma~\ref{lem:formation}. The same induction preserves the generating equations and proves the two functoriality identities. Every node of a derivation has a small family of immediate premises and a small label. Well-founded induction and closure of $\U$ under small indexed unions show that its dependency tree is small. For a small family of derivations, the union of their supports is again small.\qed
\end{proof}

The boundary equations and the input types determine the output types of every named constructor. For example, the output of (Sum) is a display with the stated total context, the output of (HC) is a display to the prescribed context, and (Desc) certifies the reindexing computations of the derived fiber family. These typing judgments provide the displays and comparison maps from which we now construct finite limits.

\section{Fiber Representation, Finite Limits, and Family Descent}\label{sec:recognition}
Representing a substitution by its fibers serves two purposes in $\G_\T$. For a cospan, it expresses the pullback as a dependent type. Together with the terminal context, these pullbacks give finite limits. For a cartesian family on a diagram, we take the fibers of the map from the colimit of the local total spaces to the colimit of the bases. The descent rule identifies the restrictions of this family with the prescribed local families.

\subsection{Fiber Representation of Substitutions and Finite Limits}
We first identify the terminal contexts and the pullback squares that reindex a display. Dependent sums and identity types then let us represent any substitution by a display and apply reindexing to an arbitrary cospan.

\begin{lemma}[Structural recognition]\label{lem:structural}
The underlying context diagrams form a quasi-category. The certificates $\Term$ and $\Eq$ designate exactly terminal objects and equivalence edges. Every certified $\PB$ square is a pullback; for a cospan with one certified reindexing cone, all of its pullback cones are certified. Every display can be reindexed along every substitution. Any two contexts admit a product.
\end{lemma}
\begin{proof}
Rule (H) is the inner horn condition. For (T), the prescribed boundary extension with final vertex $t$ is the terminal-object criterion. Thus a certified $t$ is terminal. Every terminal object is equivalent to the chosen one and is certified by (Tr). An edge is an equivalence exactly when it extends to the inverse-data interval $J$; the two rules (E) give both implications.

For a cospan $a:V\to\G_\T$, its join-defined cone category has $n$-simplices $\Delta^n\star V\to\G_\T$ extending $a$. The comparison part of (Reind) says that a certified cone is terminal there, hence a pullback. Any other pullback cone is equivalent to it in the cone quasi-category. The associated $J$-diagram of cones is a map $J\star V\to\G_\T$. It determines a map of squares by precomposition with
\begin{equation}\label{eq:cone-transport}
J\times(\Delta^0\star V)\longrightarrow J\star V.
\end{equation}
On a simplex, retain the entries of its $J$-coordinate over the cone-vertex block, and retain its $V$-coordinate over the remaining block. Deletion and repetition of entries commute with this formula. The endpoint restrictions are the two cones, so (Tr) certifies the second. The formation part of (Reind) supplies the required pullback whenever one cospan leg is a display. Finally, (Pack) makes $X\to1$ a display for any context $X$. Reindexing $X\to1$ along $Y\to1$ gives $X\times Y$.\qed
\end{proof}

\begin{proposition}[Coherent path induction]\label{prop:coherent-J}
Let $r:E\to I$ be the reflexivity substitution of an identity
formation packet. The coherent dependent elimination property
\[
 r^*:\Map_{\G_\T/I}(\id_I,p)
       \xrightarrow{\simeq}\Map_{\G_\T/I}(r,p)
\]
for every derived display $p:Z\to I$ is equivalent to $r$ being an
equivalence. The converse uses the terminal context, context packing, and
reindexing to construct the product displays that test path induction.
\end{proposition}
\begin{proof}
If $r$ is an equivalence, it is an equivalence from the object
$r$ to $\id_I$ in the slice, and precomposition with it induces
the displayed equivalence for every $p$. Its inverse gives the
eliminator; its inverse homotopies give computation at reflexivity
and the higher comparison data.

Conversely, let $T$ be any context. Packing both contexts over the terminal context and reindexing
form the constant display $I\times T\to I$. For this display, the
displayed map is naturally the precomposition map
\[
 \Map_{\G_\T}(I,T)\longrightarrow\Map_{\G_\T}(E,T).
\]
Path induction makes this map an equivalence for every $T$. The covariant form of Yoneda then shows that $r$ is an equivalence.\qed
\end{proof}

The equivalence certificate in (Id) therefore records precisely the full coherent path-induction property. The elimination and reflexivity computations are the inverse of this restriction on section spaces. Appendix~\ref{app:typing} realizes them by explicit section and computation cells with their prescribed boundaries.

To express an arbitrary substitution as a display, we take its fibers. The total context of this family recovers the domain of the substitution.

\begin{theorem}[Fiber representation]\label{thm:fiber}
For every substitution $f:X\to Y$, the calculus forms the type
\begin{equation}\label{eq:fiber}
y:Y\ \vdash\ \Fib_f(y):=\Sigma_{x:X}\Id_Y(fx,y)\ \type.
\end{equation}
Its display $p_f:F_f\to Y$ admits an equivalence
\[
\alpha_f:X\simeq F_f\quad\text{over }Y,
\qquad x\longmapsto (fx,x,\refl_{fx}).
\]
Consequently every substitution acquires a derivable display certificate. This statement holds coherently in parameters and relative to already specified boundaries.
\end{theorem}
\begin{proof}
Pack $X$ and $Y$ as closed types. Products exist by Lemma~\ref{lem:structural}. Apply identity formation to the closed type $Y$. It gives
\[
Y\xrightarrow{r_Y}I_Y\xrightarrow{q_Y}Y\times Y,
\qquad \Eq(r_Y),\quad\Disp(q_Y),\quad q_Yr_Y\simeq\delta_Y.
\]
The display $q_Y$ can be pulled back along the map
$h:X\times Y\to Y\times Y$, $(x,y)\mapsto(fx,y)$.
The result, followed by the display $X\times Y\to Y$, is a display $p_f$ by (Sum), and has exactly the type expression~\eqref{eq:fiber}. The identity cone and the reindexing comparison give $\alpha_f$ and its triangle over $Y$.

To show that $\alpha_f$ is an equivalence, we compute its effect on maps out of each test context $Z$. Put $A=\Map(Z,X)$ and $B=\Map(Z,Y)$. The universal properties of the product and of the pullback of the display $q_Y$ give
\begin{align*}
\Map(Z,F_f)
&\simeq (A\times B)\times_{B\times B}\Map(Z,I_Y)\\
&\simeq (A\times B)\times_{B\times B} B
\simeq A.
\end{align*}
In the second line, the map $B\to B\times B$ is the diagonal, because $r_Y$ is an equivalence with the stated triangle. The final equivalence projects $(x,y,p:fx=y)$ to $x$. Its fiber at $x$ is the contractible space of pairs $(y,p:fx=y)$. Here, $A\times B$ orders the coordinates as $(x,y)$, while the total context of $\Fib_f$ orders them as $(y,x)$. Thus the inverse $x\mapsto(x,fx,\refl_{fx})$ is $\Map(Z,\alpha_f)$ in pullback coordinates, naturally in $Z$. Yoneda proves that $\alpha_f$ is an equivalence.

In dependent notation, the inverse is the projection to $x:X$, and path induction on $p:fx=y$ identifies the composite on $F_f$ with its identity. Path induction is available because the reflexivity map in (Id) is an equivalence: precompose a section with its coherent inverse and use its inverse triangles for the computation. This also identifies the dependent and mapping-space descriptions.

The square $(\alpha_f,\id_Y)$ is an equivalence from the arrow $f$ to the display $p_f$ in the arrow quasi-category $\mathcal A=\Fun(\Delta^1,\G_\T)$. Extend it to a $J$-diagram and precompose with the automorphism of $J$ interchanging $0$ and $1$. The resulting diagram has $p_f$ at endpoint $0$ and $f$ at endpoint $1$, so (Tr) carries $\Disp(p_f)$ to $\Disp(f)$. The cells and certificate witnesses of this small diagram occur together at a bounded stage, giving a derivation of $\Disp(f)$.

For a small parameter inclusion $H_0\hookrightarrow H$, the relative constructors form the comparison squares over $H$ with their prescribed restriction to $H_0$. Each square is an equivalence by the calculation above. Let $\Fun(\Delta^1,\mathcal A)_{\rm eq}$ be the full subcategory on equivalence edges. Restriction along $j:\Delta^1\hookrightarrow J$ gives a trivial categorical fibration
\[
 j^*:\Fun(J,\mathcal A)\longrightarrow\Fun(\Delta^1,\mathcal A)_{\rm eq}.
\]
Since $\Delta^1\hookrightarrow J$ is a monomorphism and $\mathcal A$ is a quasi-category, restriction is a categorical fibration. Every $J$-diagram restricts to an equivalence edge, so pulling this fibration back to the full subcategory of equivalence edges gives $j^*$.

To see that $j^*$ is an equivalence, evaluate at the source. For $J$-diagrams, source evaluation is an equivalence because $J$ is equivalent to a point. For equivalence edges, its inverse sends $x$ to $\id_x$. Each equivalence $u:x\to y$ gives a natural equivalence of arrows $\id_x\to u$ with components $\id_x$ and $u$. Since $j^*$ commutes with source evaluation, it is a categorical equivalence and hence a trivial categorical fibration. Lifting the $H$-diagram of comparison squares along $j^*$ therefore extends the prescribed $J$-diagrams on $H_0$ exactly. The same endpoint interchange and the relative form of (Tr) now give the display certificates over $H$.\qed
\end{proof}

\begin{corollary}[Finite limits from the dependent operations]\label{cor:finite-limits}
The context quasi-category has finite limits. A pullback of arbitrary substitutions $f:X\to Z$ and $g:Y\to Z$ is represented by the dependent type
\[
\Sigma_{x:X}\Sigma_{y:Y}\Id_Z(fx,gy).
\]
Every pullback square has a derivable $\PB$ certificate. General finite-limit formation is consequently a derived rule.
\end{corollary}
\begin{proof}
By Theorem~\ref{thm:fiber}, either leg of an arbitrary cospan is a display. Reindexing therefore supplies its pullback, and the cone comparison and (Tr) certify any equivalent pullback cone. The displayed expression has the same universal mapping space: a point is a pair of maps with a path between their composites, with its full higher identifications. A quasi-category with a terminal object and pullbacks has all finite limits~\cite{Kerodon}. The next lemma gives the construction for finite diagrams.\qed
\end{proof}

\begin{lemma}[Finite homotopy records]\label{lem:records}
For every finite simplicial set $K$ and every coherent diagram $a:K\to\G_\T$, there is a derived context $\Rec_K(a)$ with a universal cone. It is built using the empty context, dependent sums, identity types, and reindexing. Finite-limit cone axioms can therefore be compiled into equivalence axioms between dependent expressions.
\end{lemma}
\begin{proof}
Build $K$ by finitely many nondegenerate simplex attachments. For $K=\varnothing$, take the terminal context. The limit of a diagram on $\Delta^n$ is its value at the initial vertex, with the coherent cone supplied by the diagram. At an attachment
\[
K=K'\amalg_{\partial\Delta^n}\Delta^n,
\]
let $\sigma:\Delta^n\to K$ denote the attached simplex. Restricting the cone with vertex $a(\sigma(0))$ to the boundary gives the right-hand map in the homotopy pullback
\begin{equation}\label{eq:record-recursion}
\Rec_K(a)=\Rec_{K'}(a|_{K'})
\times_{\Rec_{\partial\Delta^n}((a\circ\sigma)|_{\partial\Delta^n})}a(\sigma(0)).
\end{equation}
The induction is lexicographic: first on the dimension of $K$, then on the number of its top-dimensional nondegenerate simplices. Thus both $K'$ and $\partial\Delta^n$ are already covered. For a test context $Z$, coherent cones with vertex $Z$ are natural transformations $\operatorname{const}_Z\to a$. Mapping the simplex attachment into the context quasi-category gives the strict pullback
\[
 \Fun(K,\G_\T)\cong
 \Fun(K',\G_\T)
 \mathop{\times}_{\Fun(\partial\Delta^n,\G_\T)}
 \Fun(\Delta^n,\G_\T).
\]
Restriction from the last factor is a categorical fibration, since $\partial\Delta^n\hookrightarrow\Delta^n$ is monic. This strict pullback therefore computes the pullback of $\infty$-categories. Its mapping space from $\operatorname{const}_Z$ to the given $a$ is the homotopy pullback of the corresponding mapping spaces in the three factors. These are exactly the two cone spaces and their common boundary cone space. Mapping $Z$ into~\eqref{eq:record-recursion} gives this same homotopy pullback. This proves the universal cone property, including its higher simplices. Each pullback in the recursion has the dependent description of Corollary~\ref{cor:finite-limits}.

A specified cone $c$ induces a comparison from its vertex to $\Rec_K(a)$, including its cone coherence. The cone is limiting exactly when this comparison is an equivalence. The construction of the record, the comparison, and their boundary data involves finitely many universal operations; their relative versions retain prescribed faces. Thus an equivalence axiom for the comparison with the record expresses the finite-limit axiom in dependent syntax. The universal property identifies the records produced by different attachment orders. The comparisons are natural for all transformations of the input diagram: a transformation induces a map between their cone functors, and representability supplies the map between the records with a contractible space of coherent choices. The same reasoning applies after adjoining arbitrary small parameter shapes.\qed
\end{proof}

\subsection{Pullback Stability and Descent for Families}\label{sec:descent}
Every pullback square in $\G_\T$ now has a derivable certificate. This lets us apply the substitution-stability rule to the pullback of a colimit cocone along any map.

\begin{lemma}[Homotopy-colimit recognition]\label{lem:colimits}
The context quasi-category has all small colimits, and the $\HC$ certificates designate exactly its colimit cocones. These colimits are stable under arbitrary pullback.
\end{lemma}
\begin{proof}
For a small diagram $a:K\to\G_\T$, extend it to a cocone at a terminal context. This can be done coherently since maps into a terminal object form contractible spaces. Its edges are displays by Theorem~\ref{thm:fiber}. Apply (HC). Rule (HC-comp) is the initial-object boundary criterion in the cocone category, so the resulting cocone is a colimit. Every other colimit cocone is equivalent to it, and (Tr) derives its certificate. Conversely, the comparison rules show that every certified cocone is universal.

Given an arbitrary pullback of a colimit cocone, its apex squares are pullbacks and have derivable $\PB$ certificates by Corollary~\ref{cor:finite-limits}. Rule (Stab) certifies the upper cocone. This proves universality of all small colimits, including the empty one. Each diagram has small shape, so its certificate is derived at a bounded stage.\qed
\end{proof}

The two homotopy-colimit computation rules describe gluing in both directions. Descent computes the fibers of a glued family on each piece, and substitution stability reconstructs a family from those fibers and their transition maps.

\begin{definition}[Geometric universe]\label{def:geometric-universe}
A \emph{geometric universe} is a $\V$-small $\infty$-category $\D$ with finite limits and all $\U$-small colimits in which every $\U$-small colimit is Van Kampen. Explicitly, for every $\U$-small diagram $d:K\to\D$ with colimit $W$, restriction along the colimit cocone is an equivalence of $\infty$-categories
\[
R_d:\D_{/W}\xrightarrow{\ \simeq\ }\Cart(d).
\]
Here, $\Cart(d)$ is the full subcategory of $\Fun(K,\D)_{/d}$ on maps of diagrams $P\to d$ whose naturality squares are pullbacks.
\end{definition}

Pullback along the arrows of $d$ gives a diagram of slices with limit
\[
\Cart(d)\simeq\lim_{k\in K^{\op}}\D_{/d_k}.
\]
To see this, put $\mathcal E=\Fun(K,\D)$. By currying, sections of the codomain fibration over $d$ form the fat slice $\Fun(\Delta^1,\mathcal E)\times_{\mathrm{ev}_1,\mathcal E}\{d\}$. The dual of the ordinary-to-fat slice comparison~\cite{HTT} is an equivalence and is the identity on objects for this one-vertex base diagram. It restricts to the full subcategories whose transition squares are pullbacks. The cartesian-section formula~\cite{HTT} now gives the displayed limit with all maps between its families. For a simplicial indexing shape, the displayed categorical limit is taken over its represented indexing $\infty$-category. Grothendieck $\infty$-toposes are geometric universes by the descent form of Giraud's theorem for $\infty$-toposes~\cite{HTT}.

\begin{theorem}[Descent computation and full family descent]\label{thm:local-descent}
Let $\D$ have finite limits and small colimits, and assume that colimits are stable under pullback. For every small diagram $d:K\to\D$ with colimit cocone $\bar d:d\to W$, the following conditions are equivalent.
\begin{enumerate}[label=\textup{(\roman*)},leftmargin=*]
\item In every prism between colimit cocones $\bar P\to\bar d$, cartesian transition squares over the edges of $K$ imply cartesian squares over the cone edges $k\to *$.
\item Every cartesian family $P\to d$ extends to a family over $W$ whose cone squares are pullbacks.
\item Restriction $R_d:\D_{/W}\to\Cart(d)$ is an equivalence of $\infty$-categories.
\end{enumerate}
Its inverse is total-space colimit. Substitution stability gives the invertible counit, and condition \textup{(i)} gives the invertible unit with its prescribed cocone comparisons.
\end{theorem}
\begin{proof}
For $p:P\to d$, define
$L_d(P)=(\colim_KP\to W)$ by the colimit of the natural transformation $p$ and the specified identification $\colim_Kd\simeq W$. Colimits in the arrow category are computed pointwise, so this construction includes every map of cartesian families and its higher coherence. For $y:Y\to W$, the colimit universal property identifies maps from $L_d(P)$ to $y$ with the homotopy fiber, over the prescribed composite $p_W:P\to\operatorname{const}_W$, of
\[
\Map_{\Fun(K,\D)}(P,\operatorname{const}_Y)
\longrightarrow
\Map_{\Fun(K,\D)}(P,\operatorname{const}_W).
\]
The pointwise pullback
$d\times_{\operatorname{const}_W}\operatorname{const}_Y$
is $R_dY$. Its pullback universal property identifies this same fiber with the maps over $d$ from $P$ to $R_dY$. Both identifications are natural and give
\begin{equation}\label{eq:descent-adjunction}
\Map_{\D_{/W}}(L_dP,Y)
\simeq\Map_{\Fun(K,\D)_{/d}}(P,R_dY).
\end{equation}
Since $\Cart(d)$ is full on its objects, this is an adjunction $L_d\dashv R_d$ on the stated categories. Finite limits define the restriction functor, and the colimit functor defines its left adjoint. In particular, if $R_d$ is an equivalence, then its counit $L_dR_d(Y)\to Y$ is an equivalence for every $Y\to W$. This counit exhibits $Y$ as the colimit of the pulled-back cocone, so Van Kampen colimits are stable under pullback.

For a family $Y\to W$, pull back the constructor cocone along $Y$. Substitution stability makes its top row a colimit with apex $Y$. The counit
\[
 \epsilon_Y:L_dR_dY\longrightarrow Y
\]
is the comparison from the colimit of that top row to its prescribed apex and is therefore an equivalence. Thus $R_d$ is fully faithful.

Assume \textup{(i)}. The unit
\[
 \eta_P:P\longrightarrow R_dL_dP
\]
has component $P_k\to d_k\times_W\colim_KP$. Its comparison square is precisely the cone square of the colimit prism used to define $L_dP$. Condition \textup{(i)} makes every such square a pullback. Every component of $\eta_P$ is consequently an equivalence. Equivalences of diagrams are detected pointwise, so $\eta_P$ is an equivalence in $\Cart(d)$. Both unit and counit are equivalences, proving \textup{(iii)}.

Assume \textup{(iii)}. The left adjoint in~\eqref{eq:descent-adjunction} is an inverse, so its unit is an equivalence. Given any prism as in \textup{(i)}, its upper colimit cocone identifies its apex with $\colim_KP$, compatibly over the complete lower cocone. Under that identification, the unit components are exactly its cone-square comparisons. They are equivalences, making those squares pullbacks. This proves \textup{(i)}, and taking the apex $L_dP$ also proves \textup{(ii)}.

Finally, assume \textup{(ii)}. An extension with apex $G\to W$ gives comparisons $P_k\simeq d_k\times_WG$. The complete prism makes them a natural equivalence $P\simeq R_dG$. Hence $R_d$ is essentially surjective, and its already proved full faithfulness gives \textup{(iii)}.

The natural unit and counit satisfy
\[
 \epsilon_{L_dP}\,L_d(\eta_P)\simeq\id_{L_dP},
 \qquad
 R_d(\epsilon_Y)\,\eta_{R_dY}\simeq\id_{R_dY}
\]
with their adjunction coherences. Thus the local and global computations are compatible on all morphisms and higher simplices. For $K=\varnothing$, substitution stability identifies the slice over the initial object with the terminal category; the same adjunction and proof apply.\qed
\end{proof}

\begin{corollary}[Recognition of the typed syntax]\label{cor:geometric}
The syntax $\G_\T$ is a geometric universe. Its derivable $\Disp$ judgments include every substitution; its terminal, pullback, equivalence, and homotopy-colimit certificates are exactly their corresponding universal structures.
\end{corollary}
\begin{proof}
The structural assertions follow from Theorem~\ref{thm:fiber} and Lemmas~\ref{lem:structural} and~\ref{lem:colimits}. For a transformation between two colimit cocones whose restriction to the indexing shape $K$ is cartesian, structural recognition gives the two $\HC$ certificates and the transition $\PB$ certificates. Rule (Desc) then supplies all cone-square $\PB$ certificates. Thus condition \textup{(i)} of Theorem~\ref{thm:local-descent} holds. That theorem proves the Van Kampen property directly from the generated rules.\qed
\end{proof}

\subsection{Gluing Families by Homotopy Colimits of Total Spaces}\label{subsec:syntactic-gluing}
To realize the inverse to restriction in the dependent language, we take the colimit of the local total spaces and form the fibers of its map to the base colimit. The following lemma lets us construct this map while retaining the prescribed base cocone.

\begin{lemma}[Relative closed colimit formation]
\label{lem:relative-closed-colimit}
The closed colimit packet
\[
 K^\flat\hookrightarrow
 \cmark{K^\triangleright}{\HC_K(\id_{K^\triangleright})}
\]
and all its vertexwise relative forms are derived from
\textup{(T)}, \textup{(HC)}, and the dependent operations.
More precisely, for a small monomorphism $J_0\hookrightarrow I$,
an array on
\[
 P=(K\times I)\cup_{K\times J_0}
                   (K^\triangleright\times J_0)
\]
whose prescribed cocones over the vertices of $J_0$ are certified
colimits extends to $K^\triangleright\times I$, with certified
colimits at all parameter vertices and with its entire original
array fixed.
\end{lemma}
\begin{proof}
Choose a terminal context $t$. Extend the given array to
$P\star\{t\}$. This extension is an explicit sequence of
\textup{(T)} comparisons: for each nondegenerate simplex
$\sigma:\Delta^n\to P$, in order of dimension, attach
$\sigma\star\{t\}$ along its boundary. Its last vertex is $t$,
its opposite face is the prescribed simplex $\sigma$, and its
remaining faces were attached in lower dimensions. Thus each
step is the dimension-$(n+1)$ terminal comparison, and all
existing cells remain fixed. The small set of these steps
provides the complete terminal cocone. Every arrow to $t$ has
a derivable display certificate by Theorem~\ref{thm:fiber}.

The input for contextual colimit formation over the constant
context $t$, with parameter $I$, is
$(K\star\{1\})\times I$. Map it to $P\star\{t\}$ by retaining
the $K\times I$ part and collapsing the final-vertex block to
$t$. Over $J_0$, the already specified output is an array on
$(K\star\Delta^1)\times J_0$: retain
$(K\star\{0\})\times J_0=K^\triangleright\times J_0$ inside
$P$ and collapse the final-vertex-$1$ block to $t$. These are
simplicial maps. On each simplex, its first join-coordinate
separates an initial block in $K\star\{0\}$ from a final block
at $1$; the first block retains its parameter coordinate and
the second is sent to $t$. Deleting or repeating entries
commutes with this construction. The two arrays agree on
their common subshape $(K\star\{1\})\times J_0$.

At a parameter vertex in $J_0$, the cocone on $K\star\{0\}$ is
exactly the prescribed certified colimit. All edges to the
fixed context $t$ are certified displays. We have therefore
specified the complete source of the relative \textup{(HC)}
rule for $J_0\hookrightarrow I$. Its output is an array on
$(K\star\Delta^1)\times I$ equal to both supplied pieces.
Restrict it to $(K\star\{0\})\times I$. This is the required
closed-colimit array, retaining every original mixed simplex
as well as every previously given cocone cell.\qed
\end{proof}

Write $\Fam_\T(\Gamma)$ for the full subcategory of $(\G_\T)_{/\Gamma}$ on derived displays over $\Gamma$. For $d:K\to\G_\T$, write $\Cart_\T(d)$ for $\Cart(d)$ in $\G_\T$, the full subcategory of $\Fun(K,\G_\T)_{/d}$ on transformations $E\to d$ whose naturality squares are pullbacks.

\begin{theorem}[Syntactic gluing and reconstruction]\label{thm:family-descent}
For a derived colimit cocone $d\to W$ and a cartesian family $E_k=\Sigma_{x:d_k}P_k(x)\to d_k$, the dependent expression
\begin{equation}\label{eq:derived-glue-expression}
 \glue_d(P)(w)=\Sigma_{v:V}\Id_W(gv,w),
 \qquad V=\hcolim_k E_k,\quad g:V\to W,
\end{equation}
is a family with the prescribed local computations. Type restriction and this gluing construction induce inverse equivalences
\begin{equation}\label{eq:family-descent}
\Fam_\T(W)\simeq\Cart_\T(d)
\simeq\lim_{k\in K^{\op}}\Fam_\T(d_k).
\end{equation}
The computations are
\[
 \glue_d(P)[\iota_k]\simeq P_k,
 \qquad \glue_d(R_d B)\simeq B,
\]
natural for all family maps and higher terms. The construction retains any prescribed base cocone and extends relative to specified output parameter subshapes. For a fixed family $B$,
\begin{equation}\label{eq:section-descent}
\Tm_\T(W,B)\simeq
\lim_{k\in K^{\op}}\Tm_\T(d_k,B[\iota_k]).
\end{equation}
\end{theorem}
\begin{proof}
The given local family and the entire specified base cocone form an array on
\[
 A_K=(K\times\Delta^1)\cup_{K\times\{1\}}
                    (K^\triangleright\times\{1\}).
\]
Apply Lemma~\ref{lem:relative-closed-colimit} with $I=\Delta^1$ and $J_0=\{1\}$. It extends this exact array to $K^\triangleright\times\Delta^1$. Its upper row is a certified colimit $E\to V$, and its apex edge is $g:V\to W$. Both rows carry $\HC$ certificates; the transition squares retain the input $\PB$ certificates. Rule (Desc) certifies every apex square. This constructs the complete colimit prism by derived rule applications, with every original mixed simplex fixed.

Theorem~\ref{thm:fiber} presents $g$ by the family~\eqref{eq:derived-glue-expression}, with an equivalence of its total context to $V$ over $W$. Reindexing this equivalence along each $\iota_k$ and comparing the two certified pullback cones gives the local equivalence with $P_k$. The colimit prism supplies the full coherent transition array. Applying the parameterized reindexing comparisons gives an equivalence of local-family diagrams, including the specified edge comparisons and all higher faces. Appendix~\ref{app:glue-computation} spells out both directions and their inverse computations.

By Theorem~\ref{thm:fiber}, every arrow is a display, so $\Fam_\T(\Gamma)$ is the whole slice. The family in~\eqref{eq:derived-glue-expression} represents the total-space-colimit functor $L_d$ of Theorem~\ref{thm:local-descent}. The forward comparisons $c_k$ are the components of the unit after fiber presentation. Their inverses $\beta_k$ give the first displayed computation. For $B\to W$, rule (Stab) certifies the pulled-back constructor cocone with apex $B$. It therefore identifies the total colimit of its restrictions with $B$ over $W$. This is the counit and gives the second displayed computation. Naturality and the triangle identities in Theorem~\ref{thm:local-descent} prove these assertions for entire categories of families.

For an additional parameter pair $H_0\hookrightarrow H$, apply the closed-colimit lemma with
\[
 I=\Delta^1\times H,\qquad
 J_0=(\{1\}\times H)\cup(\Delta^1\times H_0).
\]
Its input is exactly the original prism over $H$, the entire base cocone over $H$, and the already specified upper cocones and mixed cells over $H_0$. The relative forms of (Desc), fiber representation, and the reindexing comparisons retain all these cells and certificates. This proves the stated relative assertion.

The limit description of $\Cart_\T(d)$ is the cartesian-section description for the codomain fibration $\Ar(\G_\T)\to\G_\T$: its cartesian edges are pullback squares, and its fiber at $d_k$ is the slice. Cartesian sections compute the indicated categorical limit~\cite{HTT}. All morphisms between the specified cartesian sections are retained.

Finally, the boundary formula~\eqref{eq:termspace} is the right mapping-space formula in the slice:
\begin{equation}\label{eq:terms-as-sections}
\Tm_\T(\Gamma,p)=\Map^R_{(\G_\T)_{/\Gamma}}(\id_\Gamma,p).
\end{equation}
It is therefore a Kan complex. Restriction sends the identity family to the compatible local identity families. Taking mapping spaces from these objects in~\eqref{eq:family-descent} proves~\eqref{eq:section-descent}.\qed
\end{proof}

\begin{proposition}[Sums, identities, and dependent-term spaces]\label{prop:typed-term-spaces}
Let $\Gamma\vdash A\ \type$, let $\Gamma,x:A\vdash B(x)\ \type$, and let $a,b$ be terms of $A$ in context $\Gamma$. The projection
\[
\Tm_\T(\Gamma,\Sigma_{x:A}B(x))\longrightarrow\Tm_\T(\Gamma,A)
\]
has homotopy fiber $\Tm_\T(\Gamma,B[a])$ over $a$. There is a natural equivalence
\begin{equation}\label{eq:identity-term-space}
\Tm_\T(\Gamma,\Id_A(a,b))
\simeq \operatorname{Path}_{\Tm_\T(\Gamma,A)}(a,b).
\end{equation}
Both descriptions respect substitution and higher terms.
\end{proposition}
\begin{proof}
Write $p:E\to\Gamma$ and $q:F\to E$ for the displays. The sum display is the chosen composite $pq$. In the slice over $\Gamma$, taking maps from $\id_\Gamma$ turns the pullback of $q$ along the section $a$ into the homotopy fiber of the displayed map. Formula~\eqref{eq:terms-as-sections} identifies this fiber with the stated dependent-term space.

For identities, put $S=\Tm_\T(\Gamma,A)$. Identity formation gives the equivalence $r:E\simeq I_A$ over $E\times_\Gamma E$, with comparison to the diagonal. Pulling the identity display back along the pair of sections $(a,b)$ and taking maps from $\id_\Gamma$ therefore gives
\[
\Tm_\T(\Gamma,\Id_A(a,b))
\simeq \{(a,b)\}\times_{S\times S}S,
\]
where $S\to S\times S$ is the diagonal. This homotopy fiber is the path space from $a$ to $b$, including its higher paths. The two calculations use mapping functors and pullbacks, so their naturality supplies the asserted substitution and higher-term comparisons.\qed
\end{proof}

Thus the dependent expression~\eqref{eq:derived-glue-expression} forms the family, (Desc) computes its restrictions, and (Stab) reconstructs an existing family from those restrictions. The \hyperref[thm:local-descent]{local descent theorem} suggests taking family formation as a primitive rule. In \S\ref{sec:initiality}, we establish the universal property of the generated syntax, which we use in \S\ref{sec:boundarycomparison} to compare the two presentations.

\section{Initiality of the Geometric HoTT Calculus}\label{sec:initiality}
A model assigns values to the generators of a theory. Initiality extends this assignment to every derived context, substitution, and term. Applying the same extension to diagrams of models determines how model homomorphisms act on derived expressions.
For a geometric universe $\D$ (Definition~\ref{def:geometric-universe}), give every arrow a display certificate and certify exactly the terminal objects, pullback squares, colimit cocones, and equivalences. Denote this presentation by $\D^g$. For a presentation $P$, let $\Mod_P(\D)$ be the full subcategory of $\Fun(|P|,\D)$ on certificate-preserving maps $P\to\D^g$. Taking $P=Q_\T$ gives the model $\infty$-category $\Mod_\T(\D)$ of the theory $\T$. Write $\mathcal C^{\simeq}$ for the maximal subgroupoid of an $\infty$-category $\mathcal C$ and define the space of $P$-models by
\[
M(P,\D)=\Mod_P(\D)^{\simeq}.
\]
It is a union of components of $\Fun(|P|,\D)^{\simeq}$, since the certificate conditions are invariant under natural equivalence.

\subsection{Marked Models and Strict Fibers}
\begin{lemma}[Pointwise geometric structure]\label{lem:functor-universe}
For every small simplicial shape $I$, $\Fun(I,\D)$ is a geometric universe. Terminal objects, finite limits, small colimits, and the descent comparisons are evaluated at the vertices of $I$.
\end{lemma}
\begin{proof}
Limits and colimits in the diagram $\infty$-category are computed pointwise. In particular, a square is a pullback precisely when every vertex evaluation is a pullback, and a cocone is a colimit precisely when every vertex evaluation is a colimit. Pulling back a colimit cocone therefore gives another colimit cocone, by the same property in $\D$.

Consider a transformation between colimit cocones in $\Fun(I,\D)$ whose restriction to the indexing shape $K$ is cartesian. Evaluation at each $i\in I_0$ gives two colimit cocones in $\D$ with cartesian transition squares. Descent in $\D$ makes every comparison square at the cone vertex a pullback. These squares are consequently pullbacks in $\Fun(I,\D)$. The total-colimit--restriction adjunction~\eqref{eq:descent-adjunction} now has invertible unit and counit: the preceding calculation gives its unit, and pullback stability gives its counit. It is an equivalence of whole family categories, which proves the Van Kampen condition. This argument also applies to the empty indexing diagram. Pointwise equivalence detection proves all the stated recognition assertions.\qed
\end{proof}

For a presentation $P$, let $\mathscr M_P(\D)$ denote the full subcategory $\Mod_P(\D)$ viewed as a simplicial set for the lifting arguments below. A certificate is a property of a diagram, invariant under natural equivalence. Thus the vertices of $\mathscr M_P(\D)$ specify the certificate-preserving diagrams, and all simplices between these vertices are retained.

\begin{lemma}[Restriction of models and their cores]\label{lem:mapping-cores}
Let $\D$ be a quasi-category and suppose that each certificate condition on diagrams in $\D$ is invariant under natural equivalence. If $A\hookrightarrow B$ is a monomorphism of presentations, restriction is a categorical fibration
\begin{equation}\label{eq:marked-restriction}
 p:\mathscr M_B(\D)\longrightarrow\mathscr M_A(\D).
\end{equation}
Its map on maximal subgroupoids is a Kan fibration. Consequently its strict fiber on cores over each prescribed input computes its homotopy fiber. For every diagram of presentations $(P_i)$,
\begin{equation}\label{eq:mapping-colimits}
 \mathscr M_{\colim_iP_i}(\D)\cong\lim_i\mathscr M_{P_i}(\D),
 \qquad M(\colim_iP_i,\D)\cong\lim_iM(P_i,\D).
\end{equation}
\end{lemma}
\begin{proof}
The ambient restriction
\[
 \Fun(|B|,\D)\longrightarrow\Fun(|A|,\D)
\]
is a categorical fibration. Indeed, the cartesian Joyal model structure has monomorphisms as cofibrations and every simplicial set is cofibrant. The pushout product of $|A|\hookrightarrow|B|$ with a trivial cofibration is again a trivial cofibration. Adjunction therefore gives the required lifting property for the displayed restriction into the fibrant object $\D$~\cite{HTT}.

We verify its restriction to marked models. A full subcategory of a quasi-category is a quasi-category. Given an inner-horn lifting problem for~\eqref{eq:marked-restriction}, solve it in the ambient functor categories. Every vertex of the solution is already a vertex of the horn, so every vertex is a $B$-model. Fullness implies that the whole solution lies in $\mathscr M_B(\D)$. Its restriction is the prescribed simplex in $\mathscr M_A(\D)$. Thus~\eqref{eq:marked-restriction} is an inner fibration.

Next, prescribe a source $B$-model $b$ and an equivalence with source $b|_A$ in $\mathscr M_A(\D)$. Extend this equivalence to a $J$-diagram and use the ambient categorical fibration to lift it with source $b$. Every resulting vertex is naturally equivalent to $b$. For each certificate $m:H_t\to B$, evaluation of this natural equivalence on $m$ identifies its two image diagrams. Certificate invariance therefore makes every resulting vertex a $B$-model. This proves equivalence lifting. The inner-fibration and equivalence-lifting criterion for quasi-categories proves that~\eqref{eq:marked-restriction} is a categorical fibration.

On maximal subgroupoids, inner horns have the lifts just constructed. Outer horns lift by the equivalence-horn criterion: an edge of the source core is an equivalence, so it can be inverted and the remaining composition problem filled by inner horns. For one-dimensional horns, the required lift is exactly equivalence lifting. Every edge of the resulting simplex is an equivalence, by two-out-of-three where an edge is newly formed. These are the Kan lifting properties. Hence the strict fibers on cores are Kan complexes and compute their homotopy fibers.

Finally, an $n$-simplex of a model category is a map $|P|\times\Delta^n\to\D$ whose vertex diagrams preserve the certificates. Cartesian product with $\Delta^n$ preserves colimits of simplicial sets. The underlying maps out of $\colim_iP_i$ are therefore exactly compatible tuples of the underlying maps from $P_i$. The certificates of a colimit presentation are exactly those inherited from its pieces. This proves the first isomorphism of~\eqref{eq:mapping-colimits} degree by degree. A natural transformation is an equivalence precisely when it is so at every object. Every object of the colimit has a representative in a piece. Taking the simplices all of whose parameter edges are such equivalences proves the second isomorphism.\qed
\end{proof}

\begin{proposition}[Boundary-compatible join comparison]\label{prop:join-fiber-comparison}
Fix a diagram $d:K\to\D$ in a quasi-category. Let $\mathcal C^\diamond_d$ be its fat cone category, characterized by
\[
 (\mathcal C^\diamond_d)_n
 =\{f:\Delta^n\diamond K\to\D\mid f|_K=d\}.
\]
For any simplicial set $A$, put
\[
 \mathcal E_A(d)=
 \Fun(A\star K,\D)\times_{\Fun(K,\D)}\{d\}.
\]
There is a natural categorical equivalence
\begin{equation}\label{eq:join-fixed-diagram-equivalence}
 \rho_A:\mathcal E_A(d)\longrightarrow
 \Fun(A,\mathcal C^\diamond_d).
\end{equation}
It commutes with restriction in $A$ and with the functor to $\Fun(A,\D)$ that retains the cone vertices. For a vertex $v\in A_0$, it restricts to an equivalence of the full subcategories on arrays whose $v$-cone is limiting and diagrams whose value at $v$ is terminal. Consequently, for $A'\hookrightarrow A$ containing $v$, these equivalences identify the homotopy fibers of the corresponding marked restriction maps at every specified boundary array. The dual statement holds for cocones, colimits, and initial objects.
\end{proposition}
\begin{proof}
The fat join is the pushout
\[
 A\diamond K=(A\amalg K)
 \amalg_{A\times K\times\partial\Delta^1}
 (A\times K\times\Delta^1),
\]
with the two endpoints attached by the projections to $A$ and $K$.
Let $c_A:A\diamond K\to A\star K$ be the natural map. It is the identity on $A\amalg K$ and a categorical equivalence~\cite{HTT}. Precomposition therefore gives a categorical equivalence
\[
 \Fun(A\star K,\D)\longrightarrow\Fun(A\diamond K,\D)
\]
over $\Fun(A,\D)\times\Fun(K,\D)$. The restriction maps to $\Fun(K,\D)$ are categorical fibrations, since the inclusions of $K$ are monic. Taking their strict fibers at $d$ consequently preserves this equivalence.

The resulting strict fiber for the fat join is $\Fun(A,\mathcal C^\diamond_d)$. To verify this in degree $m$, collapse the constant $K$-row by forming the pushout
\begin{equation}\label{eq:fat-join-parameter-quotient}
 \bigl((A\diamond K)\times\Delta^m\bigr)
 \amalg_{K\times\Delta^m}K
 \ \cong\ (A\times\Delta^m)\diamond K.
\end{equation}
Both sides consist of $A\times\Delta^m$, the fixed copy of $K$, and the cylinder $A\times\Delta^m\times K\times\Delta^1$, attached by the same endpoint projections. Maps from either side that restrict to $d$ on $K$ are therefore identical. By the defining adjunction for $\mathcal C^\diamond_d$, they are precisely maps $A\times\Delta^m\to\mathcal C^\diamond_d$. This verification commutes with every face, degeneracy, and restriction map, proving~\eqref{eq:join-fixed-diagram-equivalence} as a natural functor between the indicated quasi-categories.

The ordinary cone category $\D_{/d}$ maps equivalently to $\mathcal C^\diamond_d$~\cite{HTT}. Its terminal objects are the limiting cones on $d$. Restriction of $c_A$ to $\{v\}\diamond K$ is $c_{\{v\}}$, so $\rho_A$ carries precisely the specified limiting $v$-cones to terminal objects. Restricting an equivalence to corresponding full subcategories again gives an equivalence.

For $A'\hookrightarrow A$, the square
\begin{equation}\label{eq:join-restriction-square}
\begin{tikzcd}[column sep=large]
 \mathcal E_A(d)\ar[r,"\rho_A","\simeq"']\ar[d]
 &\Fun(A,\mathcal C^\diamond_d)\ar[d]\\
 \mathcal E_{A'}(d)\ar[r,"\rho_{A'}","\simeq"']
 &\Fun(A',\mathcal C^\diamond_d)
\end{tikzcd}
\end{equation}
commutes strictly, including after imposing the universal-cone condition at $v$. Taking cores gives a square of spaces whose horizontal maps are equivalences. The induced map of homotopy fibers at a boundary $b$ is therefore an equivalence, with the boundary on the right exactly $\rho_{A'}(b)$. The left restriction is a categorical fibration by Lemma~\ref{lem:mapping-cores}, applied after fixing the $K$-row; its core is a Kan fibration. Its strict fiber thus represents this same homotopy fiber over the prescribed boundary, including the $K$-diagram and the distinguished universal cone. Taking opposites proves the cocone version.\qed
\end{proof}

\begin{lemma}[Cone comparison fibers]\label{lem:cone-fibers}
Let $d:K\to\D$ be a diagram. If $d$ admits a limit, its limiting cones form a contractible space. If a boundary comparison on $\partial\Delta^n\star K$ has a distinguished limiting last-vertex cone, its space of extensions to $\Delta^n\star K$ is contractible. Both spaces are strict fibers of restriction over the specified inputs. The dual assertions hold for colimit cocones and a distinguished first vertex. Taking $K=\varnothing$ gives terminal-object formation and comparison.
\end{lemma}
\begin{proof}
Proposition~\ref{prop:join-fiber-comparison} identifies the fixed-diagram marked arrays with the corresponding diagrams in the fat cone category $\mathcal C=\mathcal C^\diamond_d$. Limiting cones correspond to terminal objects. The full subgroupoid on terminal objects, when nonempty, is contractible: its mapping spaces are contractible, and every such map is an equivalence. This gives the formation fiber.

For a comparison, the same proposition identifies its homotopy fiber with that of
\[
 \Fun(\Delta^n,\mathcal C)_{n\text{-terminal}}^\simeq
 \longrightarrow
 \Fun(\partial\Delta^n,\mathcal C)_{n\text{-terminal}}^\simeq
\]
at the induced boundary $b$. Put $t=b(n)$. Consider the fat slice $\mathcal C^{/t}$ and its source projection $\pi:\mathcal C^{/t}\to\mathcal C$. This is a right fibration~\cite{HTT}, whose fiber at $x$ models $\Map_{\mathcal C}(x,t)$. Since $t$ is terminal, these fibers are contractible. The projection is consequently equivalent over $\mathcal C$ to the identity right fibration and is a trivial categorical fibration; in particular, it has the right lifting property against every simplicial monomorphism.

Write $A=\Delta^{n-1}$ and $A_0=\partial\Delta^{n-1}$. The ordinary boundary decomposes as
\[
 \partial\Delta^n=A\cup_{A_0}(A_0\star\{n\}),
 \qquad \Delta^n=A\star\{n\}.
\]
Pulling the cone part of $b$ back along the fat-join comparison gives a diagram $a:A\to\mathcal C$ and a lift $A_0\to\mathcal C^{/t}$ of its boundary. Its fat-join extension problem is exactly
\begin{equation}\label{eq:terminal-lift}
\begin{tikzcd}
\partial\Delta^{n-1}\ar[r]\ar[d,hook]&\mathcal C^{/t}\ar[d,"\pi"]\\
\Delta^{n-1}\ar[r,"a"']\ar[ur,dashed]&\mathcal C.
\end{tikzcd}
\end{equation}
For $n=1$, the upper-left corner is empty. Its full solution object is the fiber of
\[
 \Fun(A,\mathcal C^{/t})\longrightarrow
 \Fun(A,\mathcal C)
 \times_{\Fun(A_0,\mathcal C)}\Fun(A_0,\mathcal C^{/t}).
\]
This map is a trivial fibration. Indeed, a lifting problem against any monomorphism is adjoint to lifting $\pi$ against its pushout product with $A_0\hookrightarrow A$, which is again monic. The solution object is therefore a contractible Kan complex, including every higher comparison parameter.

To identify this object with the original ordinary-simplex extension fiber, use Proposition~\ref{prop:join-fiber-comparison} for the singleton diagram $t:\Delta^0\to\mathcal C$. Its equivalences commute both with $A_0\hookrightarrow A$ and with the functors retaining the $A$-diagrams. They therefore compare the restriction maps with $a$ fixed and identify their homotopy fibers at the exact pulled-back boundary. Equivalently, the fat boundary $A\cup_{A_0}(A_0\diamond\Delta^0)$ maps to the ordinary boundary by the natural join comparison, compatibly with the corresponding full cones. The resulting square of mapping categories is an equivalence on both rows. Lemma~\ref{lem:mapping-cores} turns the ordinary marked core restriction into a Kan fibration, so its strict fiber over the original array is the same contractible homotopy fiber.

This calculation applies to $\partial\Delta^n\star K\hookrightarrow\Delta^n\star K$ through~\eqref{eq:join-restriction-square}; the quotient~\eqref{eq:fat-join-parameter-quotient} records the fixed $K$-row in every parameter degree. It is also the marked cone calculation of~\cite{Isaev}. Taking opposites exchanges terminal with initial cones and reverses the distinguished vertex, proving the colimit assertions. Taking $K=\varnothing$ gives terminal-object formation and comparisons.\qed
\end{proof}

The join comparison and cone-fiber calculation apply uniformly to the dependent formation and comparison packets. The colimit computation rules certify diagrams on a fixed prism. Their extension fibers are determined by the colimit and pullback properties of those diagrams.

\subsection{Contractibility of Extension Spaces for Primitive Rules}
\begin{lemma}[Universal rule extension]\label{lem:rule-extension}
For every primitive typed rule $r:A\hookrightarrow B$ and every geometric universe $\D$, restriction
\begin{equation}\label{eq:universal-core-restriction}
 M(B,\D)\longrightarrow M(A,\D)
\end{equation}
is a trivial Kan fibration. Its fiber over a specified input includes all output cells, all computation witnesses, and their higher identifications.
\end{lemma}
\begin{proof}
By Lemma~\ref{lem:mapping-cores}, each restriction is a Kan fibration. Its strict fiber is the space of extensions of the given marked boundary, with its prescribed vertices and degeneracies. We compute this fiber for each rule.

\par\smallskip\noindent\textit{Structural rules, unit, packing, and sums.}\enspace
For (H), the inner horn inclusion is a categorical trivial cofibration. Mapping it into $\D$ gives a categorical equivalence and hence an equivalence on cores. Terminal formation and its comparisons are the case $K=\varnothing$ of Lemma~\ref{lem:cone-fibers}. For (Pack), the target terminal object $t$ and the source $X$ are already fixed, and the fiber is $\Map_{\D}(X,t)$, which is contractible. Every arrow to $t$ carries a display certificate in $\D^g$. The (Unit) rule retains all cells and adds a certificate that every identity arrow satisfies. For (Sum), the fiber is the space of a composite of two fixed edges with its composition triangle. It is contractible by the inner-horn calculation, and all three arrows are displays in $\D^g$.

\par\smallskip\noindent\textit{Reindexing and every comparison dimension.}\enspace
A formation input fixes the entire cospan and its display leg. An output is a limiting cone on it, with its projection and pullback certificate. Its fiber is the formation fiber of Lemma~\ref{lem:cone-fibers}. The comparison formats (Reind), for every $n\geq1$, are exactly the last-vertex extension problems of that lemma. Equation~\eqref{eq:terminal-lift} identifies the space of fillers of each prescribed comparison boundary as contractible.

\par\smallskip\noindent\textit{Identity formation.}\enspace
The packet~\eqref{eq:identity-packet} fixes $\delta:E\to Q$ and all its input cells. An output is
\[
 E\xrightarrow{r\;\simeq}I\xrightarrow{q}Q,
 \qquad qr\simeq\delta.
\]
In $\D_{/Q}$, this is an equivalence with source $\delta$. Let $S=(\D_{/Q})^{\simeq}$. Its space of outputs is the space of paths in $S$ with initial point $\delta$, with the endpoint free; it is contractible by contraction to the constant path. Equivalently, equivalence arrows with fixed source form a contractible $\infty$-groupoid. Proposition~\ref{prop:join-fiber-comparison}, with the edge $02$ fixed, identifies this ordinary-join/fat-slice calculation with the marked triangle attached along $02$. Every such $q$ carries its display mark in $\D^g$, so this fiber classifies extensions of the given identity packet.

\par\smallskip\noindent\textit{Contextual homotopy-colimit formation.}\enspace
Write the fixed input as $b:K\star\{1\}\to\D$, with underlying diagram $d$ and target $\Gamma$. Factor the underlying extension through
\[
 K\star\{1\}\ \hookrightarrow\ K\star\partial\Delta^1
 \ \hookrightarrow\ K\star\Delta^1.
\]
At the middle stage, mark the first cocone as a colimit. Restriction from this middle presentation to the input is a pullback of the colimit-formation restriction over $d$. It is a trivial Kan fibration by the dual of Lemma~\ref{lem:cone-fibers}. Restriction from the final presentation to the middle one is the $n=1$ colimit-comparison restriction, again a trivial Kan fibration. Their composite is the (HC) restriction. Concretely, its data are an initial object $c$ in the cocone category $\D_{d/}$ and an arrow $c\to b$. The space of choices for $c$ is contractible, and for each such $c$, the space of arrows $c\to b$ is contractible. The factorization proves this jointly, retaining all compatibility cells to $\Gamma$. When $K$ is empty, the same argument chooses an initial object and its arrow to the fixed $\Gamma$.

\par\smallskip\noindent\textit{Higher homotopy-colimit comparisons.}\enspace
For (HC-comp) in dimension $n$, Proposition~\ref{prop:join-fiber-comparison} identifies the marked ordinary-join extension fiber with its cocone-category fiber, and the cocone at vertex $0$ is initial. Let $\mathcal C=\D_{d/}$, and let $c$ be the chosen initial cocone. Write $\mathcal C^{c/}$ for the fat coslice of arrows with source $c$. The dual of~\eqref{eq:terminal-lift} is the relative lifting problem for the target projection $\mathcal C^{c/}\to\mathcal C$. This projection is a trivial fibration. The complete boundary includes all the previously specified comparisons; the relative lifting calculation gives a contractible fiber with this boundary fixed. Thus the higher comparison formats have the same extension property as formation.

\par\smallskip\noindent\textit{Inverse data and transport of certificates.}\enspace
Restriction from $J$ to $\Delta^1$ identifies the category of $J$-diagrams with the full subcategory of equivalence arrows, up to categorical equivalence. Indeed, evaluation at either chosen vertex is an equivalence for both categories: $J$ is the contractible groupoid on two objects, and every equivalence arrow is equivalent to the identity of its source. Restriction commutes with this evaluation. Lemma~\ref{lem:mapping-cores} therefore gives a contractible strict fiber over each fixed equivalence edge. This proves the first rule (E). For its recognition rule, every $J$-diagram already makes the indicated edge an equivalence, so the two model categories are identical. In (Tr), the $J$-parameter identifies the two marked diagrams by a natural equivalence. Certificate invariance again makes restriction an identity of the same model subcategories. Rule (Stab) has this same property, by universality of colimits, including the empty colimit.

\par\smallskip\noindent\textit{Descent computation.}\enspace
An input to (Desc) is a prism between colimit cocones with cartesian transition squares. In a geometric universe $\D$, Theorem~\ref{thm:local-descent}\textup{(i)} makes each cone square a pullback. The source and target thus define the same full subcategory of diagrams on the prism. Restriction is the identity on these diagrams and all their transformations. For $K=\varnothing$, the collection of cone squares is empty and restriction is again the identity.

Each map~\eqref{eq:universal-core-restriction} is therefore a Kan fibration with nonempty contractible fibers, hence a trivial Kan fibration.\qed
\end{proof}

\subsection{Relative Rule Extension and Initiality}
Lemma~\ref{lem:rule-extension} gives a contractible space of extensions for each input model of a primitive rule. To recover arbitrary homomorphisms as well, we apply the same calculation in the arrow category $\Fun(\Delta^1,\D)$. The homotopy fiber of endpoint evaluation over a pair of models gives their mapping space.

\begin{lemma}[Diagrams of models]\label{lem:pointwise-models}
For every small shape $I$, there are canonical simplicial isomorphisms
\begin{align}
 \mathscr M_P(\Fun(I,\D))&\cong\Fun(I,\mathscr M_P(\D)),\label{eq:pointwise-models}\\
 \mathscr M_{P\otimes_v I}(\D)&\cong\Fun(I,\mathscr M_P(\D)).\label{eq:vertexwise-currying}
\end{align}
The first uses the pointwise certificate structure. In both formulas, the parameter arrows are arbitrary natural transformations. Passing to maximal subgroupoids gives the corresponding identities of Kan complexes.
\end{lemma}
\begin{proof}
An $n$-simplex of either side of~\eqref{eq:pointwise-models} is, after interchanging factors, an array $|P|\times I\times\Delta^n\to\D$. For each vertex $v\in(\Delta^n)_0$, preservation of a certificate of $P$ in $\Fun(I,\D)$ is equivalent to its preservation at every $i\in I_0$. This follows from pointwise terminal, pullback, colimit, and equivalence recognition; the display condition is automatic. Hence the vertices required on both sides are identical. Fullness retains exactly the same higher simplices, and face and degeneracy maps agree. For~\eqref{eq:vertexwise-currying}, the certificates of $P\otimes_v I$ are precisely the copies of each certificate of $P$ at the vertices of $I$. This is the identical degreewise test. Equivalences in functor categories are pointwise, so the identifications also agree on cores.\qed
\end{proof}

\begin{lemma}[Arbitrary relative parameters]\label{lem:relative-extension}
For a primitive rule $r:A\hookrightarrow B$, the restriction
\[
 p:\mathscr M_B(\D)\longrightarrow\mathscr M_A(\D)
\]
is a trivial categorical fibration. For every small monomorphism $j:J_0\hookrightarrow I$, the restriction associated with~\eqref{eq:relative} is a trivial categorical fibration, and its map on maximal subgroupoids is a trivial Kan fibration. Each lift restricts exactly to the specified input and prescribed output cells and preserves all stipulated certificates.
\end{lemma}
\begin{proof}
By Lemma~\ref{lem:mapping-cores}, $p$ is a categorical fibration. The primitive fiber calculations use the boundary-compatible join comparison of Proposition~\ref{prop:join-fiber-comparison}; (Stab) and (Desc) preserve the full model subcategories identically. Apply Lemma~\ref{lem:rule-extension} to $\Fun(\Delta^m,\D)$, for every $m\geq0$, and use Lemma~\ref{lem:pointwise-models}. We obtain weak equivalences
\begin{equation}\label{eq:rezk-levels}
 \Fun(\Delta^m,\mathscr M_B(\D))^{\simeq}
 \longrightarrow
 \Fun(\Delta^m,\mathscr M_A(\D))^{\simeq}.
\end{equation}
They are the levels of the classification diagram of $p$.

The case $m=0$ gives essential surjectivity. To obtain full faithfulness, take $m=1$. For vertices $x,y$ in the source, consider the two endpoint maps
\[
 \Fun(\Delta^1,\mathscr M_B(\D))^{\simeq}
 \longrightarrow\mathscr M_B(\D)^{\simeq}\times\mathscr M_B(\D)^{\simeq}
\]
and the corresponding map for $A$. They are Kan fibrations, and their fibers over $(x,y)$ and $(px,py)$ model the full mapping spaces. Formula~\eqref{eq:rezk-levels} for $m=1$ gives a weak equivalence on their total spaces; its case $m=0$ gives one on their bases. Comparing homotopy fibers yields
\[
 \Map_{\mathscr M_B(\D)}(x,y)\simeq
 \Map_{\mathscr M_A(\D)}(px,py).
\]
The mapping-space equivalence shows that $p$ is fully faithful, and the case $m=0$ gives essential surjectivity. Thus $p$ is an equivalence of $\infty$-categories of models. Being also a categorical fibration, it is a trivial fibration in the Joyal model structure. Such a map has the right lifting property against every simplicial monomorphism.

Under~\eqref{eq:vertexwise-currying}, the $\infty$-category of models of the source presentation in~\eqref{eq:relative} is the strict pullback
\[
 \Fun(I,\mathscr M_A(\D))
 \times_{\Fun(J_0,\mathscr M_A(\D))}
 \Fun(J_0,\mathscr M_B(\D)).
\]
Under this identification, restriction is the map
\begin{equation}\label{eq:relative-model-map}
 \Fun(I,\mathscr M_B(\D))\longrightarrow
 \Fun(I,\mathscr M_A(\D))
 \times_{\Fun(J_0,\mathscr M_A(\D))}
 \Fun(J_0,\mathscr M_B(\D)).
\end{equation}
For a further monomorphism $L\hookrightarrow K$, a lifting problem for~\eqref{eq:relative-model-map} is adjoint to a lifting problem for $p$ with domain inclusion
\begin{equation}\label{eq:double-pushout-product}
 (L\times I)\amalg_{L\times J_0}(K\times J_0)
 \hookrightarrow K\times I.
\end{equation}
This inclusion is monic degree by degree. The right lifting property of $p$ solves it. Therefore~\eqref{eq:relative-model-map} is itself a trivial fibration. Its core map is a trivial Kan fibration.

A vertex lift is a map $B\times I\to\D$ whose restrictions are equal to the given maps on $A\times I$ and $B\times J_0$. The simplicial pushout gives these equalities on both pieces and their intersection. At each parameter vertex, the lift is a $B$-model, so it preserves the target certificates as well as those inherited from the two pieces. For an interpretation in syntax, equations~\eqref{eq:output1}--\eqref{eq:output2} express these same equalities as the judgmental boundary equations of the constructor. Taking $L=\partial\Delta^n$ verifies the identical assertion for every prescribed higher boundary.\qed
\end{proof}

\begin{theorem}[Initiality of the dependent calculus]\label{thm:initiality}
For every geometric universe $\D$, restriction induces a natural equivalence
\begin{equation}\label{eq:initiality}
 \Fun_\geo(\G_\T,\D)\simeq\Mod_\T(\D),
\end{equation}
carrying natural transformations of geometric functors to the corresponding model homomorphisms. In the displayed simplicial models, restriction is a trivial categorical fibration. In particular, interpretations extend relative to any prescribed small parameter subshape.
\end{theorem}
\begin{proof}
For each stage $P_\alpha$ of~\eqref{eq:cellular}, consider the entire model quasi-category $\mathscr M_{P_\alpha}(\D)$. By~\eqref{eq:mapping-colimits}, at a successor stage, its restriction to $P_\alpha$ is the pullback of the product, over the newly processed instances $(r,a)$, of
\[
 \mathscr M_{B_r}(\D)\longrightarrow\mathscr M_{A_r}(\D).
\]
These maps are trivial categorical fibrations for both primitive and relative formats, by Lemma~\ref{lem:relative-extension}. Products and pullbacks preserve their right lifting property against all monomorphisms. Thus every successor restriction is a trivial fibration of simplicial sets.

At a limit ordinal $\beta$,~\eqref{eq:mapping-colimits} identifies the model simplicial set with the strict inverse limit of the preceding ones. To lift a monomorphism $L\hookrightarrow K$ against the final restriction to $P_0$, start with its prescribed $K$-diagram of $P_0$-models and its prescribed $L$-diagram of full models. At a successor stage, solve the lifting problem for that stage while retaining the $L$-restriction. At a limit stage, take the compatible tuple of preceding solutions. Transfinite recursion gives a $K$-diagram of $P_\lambda$-models with exactly the specified boundary. Hence
\begin{equation}\label{eq:initiality-trivial-fibration}
 \mathscr M_{P_\lambda}(\D)\longrightarrow\mathscr M_{Q_\T}(\D)
\end{equation}
has the right lifting property against every monomorphism. This is the asserted trivial categorical fibration, and taking cores also gives the corresponding trivial Kan fibration.

By Corollary~\ref{cor:geometric}, the certificates on $P_\lambda$ are exactly its terminal objects, pullback squares, equivalences, small colimit cocones, and all arrows as displays. Therefore its certificate-preserving functors are exactly the finite-limit- and small-colimit-preserving functors. Thus~\eqref{eq:initiality-trivial-fibration} identifies restriction from $\Fun_\geo(\G_\T,\D)$ to $\Mod_\T(\D)$ as a trivial categorical fibration. Its lifting property supplies the relative extensions, and composition makes~\eqref{eq:initiality} natural in $\D$.\qed
\end{proof}

\subsection{Definitional Extensions and Contextual Elaboration}
\begin{definition}[Definitional extension]\label{def:definitional-extension}
An elementary definitional extension of a presentation is a pushout of a primitive or relative constructor rule along a specified derivable input. A small definitional extension is a small transfinite composite of these attachments with small resulting presentation. Naming a derived expression means naming its dependency diagram in this manner and taking the indicated output cell as its name.
\end{definition}

\begin{lemma}[Cellular realization of a small derived diagram]\label{lem:support-realization}
Let $P$ be a small presentation, and write $\G_P$ for the dependent syntax generated by $P$. A small array of cells of $\G_P$, together with a small collection of derivable certificates and specified cells from $P$, is contained in a small definitional extension $P\hookrightarrow P^{\rm def}$. There is a certificate-preserving map $P^{\rm def}\to\G_P$ that is the identity on $P$ and identifies the selected cells with the given ones. Thus the selected array factors through the extension with its given face and degeneracy equations.
\end{lemma}
\begin{proof}
Use the recorded cellular construction of $\G_P$. For each selected cell, choose its recorded constructor and its boundary expressions; for each required certificate, choose a derivation. Collect their dependency trees. Lemma~\ref{lem:substitution} says that their union is small. Whenever a processed rule instance occurs in these dependencies, retain its whole target packet. Retain its source cells, the derivations of its premise certificates, and their dependencies as well. These are already subtrees of the collected derivations; adjoining the complete output packets adds a small union of small sets. We obtain a small set $S$ of processed instances closed under prerequisites.

Order $S$ by its construction stage, choosing any ordering within a stage. The set of stages that occur is small, so its order type is smaller than the inaccessible cardinal of $\U$. Every instance has all of its input data in the portion already constructed: each prerequisite was available before that instance in the original construction. Starting from $P$, perform these same pushouts in this order. Distinct rule instances at a given stage adjoin disjoint sets of new cells. The resulting presentation embeds on cells into the corresponding part of the original construction, because all rule inclusions and all construction transitions are monic. The prescribed face and degeneracy identifications are exactly the same pushout identifications.

There are fewer than $\kappa$ instances, each with fewer than $\kappa$ cells and certificates. Regularity of $\kappa$ makes the resulting extension small. Each selected certificate has its retained derivation, and each selected cell has its original constructor name and boundaries. The extension thus names the selected coherent diagram together with its boundary equations.\qed
\end{proof}

For the construction of Section~\ref{sec:construction}, call a schedule \emph{fair} if every rule instance whose input cells and premise certificates have appeared is eventually processed. A schedule indexed by a limit ordinal $\delta$ is \emph{sufficiently long} if every rule input in the final union $P_\delta$ factors through some $P_\alpha$ with $\alpha<\delta$, including all its premise certificates. Since every rule source has fewer than $\kappa$ cells and certificates, cofinality at least $\kappa$ suffices, as in Lemma~\ref{lem:formation}. Fairness and sufficient length ensure that the final union is closed under the rules.

\begin{corollary}[Definitions and coherence]\label{cor:definitions}
A small definitional extension preserves the entire model $\infty$-category and the syntactic geometric universe. Every small collection of derived expressions can be named by such an extension. The same holds for adjoining chosen relative coherence operations. Different sufficiently long fair construction schedules give equivalent syntaxes under the signature, with a contractible space of compatible equivalences.
\end{corollary}
\begin{proof}
For a single defining attachment, apply the appropriate universal extension lemma. Pushouts give pullbacks on model cores, and small composites give inverse towers. Applying the argument to all functor targets recovers the full $\infty$-categories of models. Initiality, applied with the two syntaxes as targets, extends their generator interpretations in both directions. Fully faithful restriction identifies both composites with the identities. Lemma~\ref{lem:support-realization} supplies a literal cellular realization of any specified small collection of derived expressions and their required certificates. The same proof applies to alternative construction schedules and to relative coherence attachments, and identifies the space of choices with a contractible extension fiber.\qed
\end{proof}

\begin{proposition}[Slices of Geometric Universes and Reindexing]\label{prop:slices}
If $\D$ is a geometric universe, then every slice $\D_{/\Gamma}$ is a geometric universe. For $f:\Delta\to\Gamma$, the reindexing functor $f^*$ preserves finite limits, small colimits, and the gluing comparisons.
\end{proposition}
\begin{proof}
Pullbacks in a slice are computed on total objects; its terminal object is $\id_\Gamma$. Colimits are also computed on total objects. For a diagram in the slice with colimit $W\to\Gamma$, the identities
$(\D_{/\Gamma})_{/W}\simeq\D_{/W}$ and
$(\D_{/\Gamma})_{/d_k}\simeq\D_{/d_k}$ identify its descent comparison with that in $\D$. Reindexing preserves terminal objects and pullbacks, and preserves colimits by their universality. Gluing is total colimit with its cartesian comparisons, so these are preserved too.\qed
\end{proof}

\begin{proposition}[Contextual elaboration]\label{prop:contextual-elaboration}
Every named operation admits a contextual form over a prescribed $\Gamma$, with its prescribed type and term boundaries retained. In particular, a cartesian family over a colimit diagram in $(\G_\T)_{/\Gamma}$ has a syntactic glued family in the same context, and its local computations and reconstruction are compatible with reindexing along every $\Delta\to\Gamma$.
\end{proposition}
\begin{proof}
Proposition~\ref{prop:slices} makes $(\G_\T)_{/\Gamma}$ a geometric universe. By the definition of the slice, a diagram of shape $A$ is an array on $A\star\Delta^0$ in $\G_\T$ with final vertex $\Gamma$. Apply the universal rule extension and its relative form with this slice as target. This supplies an output array with that final vertex and all specified boundary data unchanged.

Its cells are existing syntactic cells. Pullback and colimit certificates in the slice are certified on their total diagrams by Corollary~\ref{cor:geometric}. A terminal-object certificate in the slice, on $q:X\to\Gamma$, is represented by $\Eq(q)$: the object $q$ is terminal exactly when it is equivalent to the unit family $\id_\Gamma$. Equivalences in the slice are detected on total objects. Every relevant arrow has a display certificate by Theorem~\ref{thm:fiber}. The output thus consists of syntactic cells and derivable typing judgments in the original calculus. Smallness puts the output cells and these derivations at a bounded construction stage.

Reindexing between the slices preserves finite limits and small colimits by Proposition~\ref{prop:slices}. It therefore preserves the universal interpretation of each named operation. The relative extension lemma supplies its comparisons on all parameter boundaries, proving the stated compatibility.\qed
\end{proof}

\section{Gluing Rules and Boundary Syntax as Equivalent Presentations}\label{sec:boundarycomparison}
We first show that gluing can be taken as primitive or derived from descent, with the same local computations. We then compare the dependent calculus with a syntax expressed in terms of universal cones. This second comparison preserves families, their term spaces, and the gluing computations established by the first.

\subsection{Equivalence of Primitive and Derived Gluing}
The derived gluing operation also gives a presentation in which the assembly of a family and its local computations are introduced together. For every small $K$, set
\[
 A_K=(K\times\Delta^1)\cup_{K\times\{1\}}
          (K^\triangleright\times\{1\}),
 \qquad B_K=K^\triangleright\times\Delta^1.
\]
On $A_K$, certify the bottom cocone by $\HC_K$, every vertical edge over $K_0$ by $\Disp$, and every transition square over $K_1$ by $\PB$. The family-formation packet is the inclusion $A_K\hookrightarrow B_K$ with target certificates
\begin{equation}\label{eq:Gformat}
 \Disp(*\times\Delta^1),\qquad
 \PB(q_{k*})\quad(k\in K_0),
 \tag{Glue}
\end{equation}
together with the inherited source certificates. Let $\G^{\rm gl}_\T$ be the calculus with this packet and its relative forms in place of (Desc). All the other primitive rules and judgmental equations are the same.

\begin{theorem}[Primitive and derived gluing]\label{thm:family-presentation}
There is a geometric equivalence
\[
 \G_\T\simeq\G^{\rm gl}_\T
\]
under the common generating theory. This equivalence preserves the local gluing computations, global reconstruction, and full dependent-term spaces. In the descent-computation presentation, the packet \textup{(Glue)} is realized by homotopy-colimit formation, \textup{(Desc)}, and fiber representation. In the family-formation presentation, \textup{(Desc)} is derived from \textup{(Glue)} and \textup{(Stab)}.
\end{theorem}
\begin{proof}
On any prescribed (Glue) output parameter subshape, (Stab) first certifies the upper cocone as a colimit. In $\G_\T$, Theorem~\ref{thm:family-descent} then constructs the entire prism relative to its given input, including these prescribed output cells, and certifies its cone squares. Its apex map has a display certificate by fiber representation. The constructed prism therefore realizes (Glue) relative to the prescribed parameter boundary.

In the family-formation presentation, the proofs of finite-limit and homotopy-colimit recognition use the shared rules. The output of (Glue) supplies condition \textup{(ii)} of Theorem~\ref{thm:local-descent}. That theorem therefore makes $\G^{\rm gl}_\T$ a geometric universe and proves condition \textup{(i)}. For a specified (Desc) input, the resulting cone squares are pullbacks and structural recognition derives their certificates. Concretely, apply (Glue) to its local prism. Rule (Stab) makes the constructed upper cocone a colimit. Its comparison with the originally specified upper colimit is an equivalence over the fixed lower cocone; transport of the cone-square certificates gives exactly the conclusions of (Desc). The comparison is made in the cocone category over the base cocone, so it retains its full mixed coherence data. Each small parameterized instance is a small array of syntactic cells and certificate derivations.

We verify the universal property of the family-formation syntax on the full $\infty$-categories of models. In a geometric universe $\D$, the closed-colimit formation restriction
\[
 \Fun(K^\triangleright,\D)_{\rm colim}
 \longrightarrow\Fun(K,\D)
\]
is a trivial categorical fibration. Its core fibers are the colimit-formation fibers of Lemma~\ref{lem:cone-fibers}; applying that calculation in every $\Fun(\Delta^m,\D)$ gives the full categorical assertion by Lemma~\ref{lem:relative-extension}. The relative form with $\{1\}\hookrightarrow\Delta^1$ extends the given diagram on $A_K$ to a prism on $B_K$ with a colimit as its top row. Adding the prescribed transition-square certificates restricts the base to a full subcategory and takes a pullback of this trivial fibration.

By (Stab), if the squares over the cone edges are pullbacks, then the upper cocone is a colimit. Descent in $\D$ gives the converse. Thus the two output conditions select the identical full subcategory of $\Fun(B_K,\D)$ over the given input category. The apex display condition is satisfied by every arrow of $\D$. The restriction for (Glue) is consequently this same trivial categorical fibration. The argument includes the empty shape, for which $\D_{/0}$ is contractible. Lemma~\ref{lem:relative-extension} supplies all relative parameter forms, and the transfinite proof of Theorem~\ref{thm:initiality} gives
\[
 \Fun_\geo(\G^{\rm gl}_\T,\D)\simeq\Mod_\T(\D).
\]
Both syntaxes therefore have the same full model-valued universal property. Interpret their generators in each other and extend by initiality. Fully faithful restriction identifies both composites with the identities. Their slice mapping spaces are equivalent, and the common total-colimit--restriction adjunction identifies their local and global computations.\qed
\end{proof}

\subsection{Equivalence of Dependent Syntax and Boundary Syntax}
We first take $\T$ to be a small boundary presentation and write $\Sps_\T$ for its syntax. As before, $\Sps_\U$ denotes the $\infty$-category of small spaces. Finite homotopy records express the limit of each finite diagram as a dependent type. They give the translation from axioms about finite cones to axioms in the dependent syntax. The boundary presentation has certificates $\Lim_K$ for finite $K$, $\Col_K$ for small $K$, and $\Eq$. The terminal-object certificate is $\Lim_\varnothing$. We write $\PB$ for the pullback certificate $\Lim_V$, using $\Delta^0\star V\cong\Delta^1\times\Delta^1$ to identify its cone with a square.

Its structural rules are (H), (E), and (Tr). Its formation and comparison rules are
\begin{align}
K^\flat&\hookrightarrow
\cmark{\Delta^0\star K}{\Lim_K(\id)},&& K\text{ finite},\label{eq:boundaryL0}\\
\cmark{\partial\Delta^n\star K}{\Lim_K(\{n\}\star K)}
&\hookrightarrow\cmark{\Delta^n\star K}{\Lim_K(\{n\}\star K)},&& n\geq1,\label{eq:boundaryLn}\\
K^\flat&\hookrightarrow\cmark{K\star\Delta^0}{\Col_K(\id)},&&K\text{ small},\label{eq:boundaryC0}\\
\cmark{K\star\partial\Delta^n}{\Col_K(K\star\{0\})}
&\hookrightarrow\cmark{K\star\Delta^n}{\Col_K(K\star\{0\})},&&n\geq1.\label{eq:boundaryCn}
\end{align}
It has (Stab) and (Desc), written with $\Col$. The latter is the identity-on-cells rule on $K^\triangleright\times\Delta^1$:
\begin{equation}\label{eq:boundaryD}
\frac{\Col_K(P)\quad\Col_K(D)\quad
\PB(q_e)\ (e\in K_1)}{\PB(q_{k*})},\qquad k\in K_0.
\end{equation}
All these are set-indexed inclusions of explicitly marked shapes. The same well-founded construction as~\eqref{eq:cellular}, now for these rules, defines the boundary syntax. Relative versions can either be included or adjoined conservatively.

\begin{proposition}[Boundary recognition and initiality]\label{prop:boundary}
The boundary syntax is a geometric universe and satisfies
\[
\Fun_\geo(\Sps_\T,\D)\simeq\Mod_\T(\D)
\]
for every geometric universe $\D$, with all natural transformations.
\end{proposition}
\begin{proof}
Inner horns give a quasi-category. The last-vertex comparisons~\eqref{eq:boundaryLn} characterize terminal objects in finite cone categories, and the first-vertex comparisons~\eqref{eq:boundaryCn} characterize initial objects in cocone categories. Formation gives the required limits and colimits. Transport certifies all equivalent universal cones, and (E) recognizes exactly equivalences.

For a colimit $d\to W$, total-space colimit is left adjoint to restriction by~\eqref{eq:descent-adjunction}. Rule (Stab) makes its counit an equivalence. Rule~\eqref{eq:boundaryD} makes its unit an equivalence by certifying all apex pullbacks. This gives descent, so the syntax is a geometric universe.

For a target geometric universe, every formation rule has a contractible space of interpretations, and every comparison rule has the terminal- or initial-cone lifting property used in Lemma~\ref{lem:rule-extension}. The two descent rules retain the underlying cells and add valid certificates. Thus every rule induces a trivial Kan fibration on marked mapping cores. The cellular-tower argument proves initiality on cores. Repeating it in $\Fun(\Delta^m,\D)$ proves the asserted equivalence of $\infty$-categories. The argument of Lemma~\ref{lem:relative-extension} proves the relative versions and their conservativity.\qed
\end{proof}

\begin{proposition}[Compilation of theories]\label{prop:compilation}
Small dependent presentations and small boundary presentations translate into one another with naturally equivalent $\infty$-categories of models. Definitions added by translation have contractible spaces of choices, also in parameter families.
\end{proposition}
\begin{proof}
For a dependent presentation, retain its underlying cell constants and equations. Translate $\Term$ as $\Lim_\varnothing$, $\PB$ as $\Lim_V$, $\HC$ as $\Col$, and $\Eq$ as itself. Display axioms are interpreted automatically in a geometric universe, where every arrow represents a family. A term declaration retains its section and section triangle as chosen data. This yields a boundary presentation with exactly the same underlying model functors and transformations.

For a boundary presentation, translate each finite-limit axiom on $a:K^\triangleleft\to Q$ by the comparison
\[
a_*\longrightarrow\Rec_K(a|_K)
\]
and declare it $\Eq$. Name the record, its cone, and this comparison by their universal constructions from Lemma~\ref{lem:records}. Translate colimit and equivalence axioms directly. Any arrow intended as a type can be named using its fiber presentation, or given its derivable display certificate. The added constants and all their required coherences are small by Lemma~\ref{lem:substitution}. Universal interpretations of the auxiliary definitions are contractible by Lemmas~\ref{lem:rule-extension} and~\ref{lem:relative-extension}. A cone is limiting precisely when its comparison with the record is an equivalence, so the translated axioms impose exactly the original model condition.

The equivalences on auxiliary choice spaces remain equivalences after replacing a target by $\Fun(\Delta^m,\D)$. Thus translation identifies the model $\infty$-categories on objects, homomorphisms, and their higher simplices, naturally in the target.\qed
\end{proof}

\begin{theorem}[Dependent--boundary comparison]\label{thm:comparison-syntax}
For a theory in either presentation, interpreted through Proposition~\ref{prop:compilation}, there is a geometric equivalence
\begin{equation}\label{eq:comparison-syntax}
\G_\T\simeq\Sps_\T
\end{equation}
under its generating data. It preserves type families, entire dependent-term spaces, substitution comparisons, and the local and global gluing computations.
\end{theorem}
\begin{proof}
Each syntax is a geometric universe and has the same model-valued universal property by Theorem~\ref{thm:initiality} and Proposition~\ref{prop:boundary}. Interpret the generators of each in the other. Initiality gives geometric extensions in both directions. Their composites restrict to the original generator interpretations. Fully faithful restriction therefore identifies both composites with the corresponding identity functors.

An equivalence of context $\infty$-categories induces an equivalence of their slices and of all slice mapping spaces. By~\eqref{eq:terms-as-sections}, the equivalence on slice mapping spaces identifies the dependent-term spaces. Gluing is characterized by total-space colimit with its cartesian local comparisons. Geometric functors preserve these constructions and the resulting adjunction, including its reconstruction map and triangle identities.\qed
\end{proof}

At the level of types, the comparison gives terms in both directions and higher terms witnessing their inverse laws. The boundary equations of the constructors determine the judgmental equalities within each presentation.

\section{The Classifying \texorpdfstring{$\infty$}{∞}-Topos and Its Equivalence with the Syntactic Category}\label{sec:classification}
Localizing a presheaf category at the comparison maps of the axioms constructs the classifier directly from the small theory. The universal model in the localization gives a geometric functor from the syntactic category to the classifier. We construct its inverse from the interpretation of the same generators in $\G_\T$.

\subsection{Presheaf Localization at the Axiom Comparisons}\label{sec:classifier}
\begin{lemma}[Small free finite-limit completion]\label{lem:freelex}
For every $\U$-small simplicial set $Q$, there is a $\U$-small finite-limit $\infty$-category $C_Q$ and $j:Q\to C_Q$ such that, for every finite-limit $\infty$-category $\D$,
\begin{equation}\label{eq:freelex-property}
 \Fun_\lex(C_Q,\D)\simeq\Fun(Q,\D).
\end{equation}
Restriction is a trivial categorical fibration in the diagram models. Consequently the universal property includes all natural transformations and every prescribed relative parameter boundary.
\end{lemma}
\begin{proof}
Retain the inner horns, the finite-limit formation and comparison formats~\eqref{eq:boundaryL0}--\eqref{eq:boundaryLn}, and transport for finite-limit certificates. Call this collection $R_{\rm lex}$. These unparameterized formats generate $C_Q$. The resulting restriction functor is a trivial categorical fibration, which also gives extension over arbitrary relative parameter shapes. There are countably many rule formats: finite simplicial shapes have countably many representatives, comparison dimensions are natural numbers, and the interval $J$ is countable. Each source and target has countably many simplices, counting degeneracies, and countably many certificates. All are objects of $\U$.

To construct the completion inside $\U$, take
\[
 \mu=\max\!\left(\aleph_0,\left|\coprod_{n\geq0}Q_n\right|\right),
 \qquad \theta=\omega_1.
\]
The cardinal $\theta$ is regular and uncountable and belongs to $\U$. The parameter $\mu$ records the signature size independently of the closure ordinal $\theta$. Starting with $Q$ with empty certificate sets, perform the recorded once-per-input cellular construction for $\theta$ stages using $R_{\rm lex}$. Write $P_\alpha$ for its stages, $\alpha\leq\theta$.

Let $m_\alpha$ be an infinite cardinal bounding the numbers of simplices and certificates at stage $\alpha$. For a countable source, there are at most $m_\alpha^{\aleph_0}$ underlying arrays; preservation of the required certificates is a condition on such arrays. Adjoining their countable outputs gives the bound
\[
 m_{\alpha+1}\leq (2+m_\alpha)^{\aleph_0}\cdot\aleph_0.
\]
At a limit $\beta\leq\theta$, one may use
\[
 m_\beta\leq\max\!\left(\aleph_0,|\beta|,
                   \sup_{\alpha<\beta}m_\alpha\right).
\]
Induction gives $m_\alpha<\kappa$: the successor bound follows from the strong-limit property of $\kappa$, and the limit bound from its regularity and $\theta<\kappa$. Thus $P_\theta$ is $\U$-small.

The cells and certified diagrams of each rule input at $P_\theta$ form a countable set (possibly finite). Choose their occurrence stages. Their supremum is smaller than $\theta$ by regularity. The input therefore appears in some $P_\alpha$, $\alpha<\theta$, and is either already processed or processed at the following stage. This proves closure under every rule in $R_{\rm lex}$. Inner horns make $P_\theta$ a quasi-category. Formation gives every finite limit, comparison recognizes its universal cones, and transport certifies all equivalent limiting cones. Put $C_Q=|P_\theta|$.

For a finite-limit target $\D$, every rule in $R_{\rm lex}$ has a trivial Kan restriction on model cores by the finite-limit cases of Lemmas~\ref{lem:cone-fibers} and~\ref{lem:rule-extension}. The same calculation applies in $\Fun(\Delta^m,\D)$ because its finite limits are pointwise. The $m=0,1$ mapping-space argument in Lemma~\ref{lem:relative-extension} therefore makes each rule restriction a trivial categorical fibration on the full $\infty$-categories of models. At each cellular stage, take its pullback and product, and at limits, take the inverse limit. The transfinite lifting argument of Theorem~\ref{thm:initiality} gives a trivial categorical fibration from the $P_\theta$-models to $Q$-diagrams. The former are exactly finite-limit-preserving functors. This proves~\eqref{eq:freelex-property}, including the stated transformations and relative boundaries.\qed
\end{proof}

For a typed theory, use its direct boundary translation in Proposition~\ref{prop:compilation}; for an original boundary theory, retain that theory. Put $C=C_{|Q_\T|}$ and let $y:C\to\Psh(C)$ be Yoneda. Each certificate axiom determines one comparison:
\begin{align}
\Lim_K(a)&:\quad yj(a_*)\longrightarrow\lim_{k\in K}yj(a_k),\label{eq:axiomlim}\\
\Col_K(b)&:\quad\colim_{k\in K}yj(b_k)\longrightarrow yj(b_*),\label{eq:axiomcol}\\
\Eq(e)&:\quad yj(e).\label{eq:axiomeq}
\end{align}
These comparisons form a small set $R_\T$. In the dependent syntax, the primitive instances of~\eqref{eq:axiomlim} are terminal objects and pullbacks. The same formula expresses the translated axioms for all finite limits.

\begin{definition}[Classifier by localization]\label{def:classifier}
A class of maps is \emph{strongly saturated} if it contains all equivalences, satisfies two-out-of-three, and is closed under small colimits in the arrow category. Let $W_\T$ be the smallest such class in $\Psh(C)$ containing $R_\T$ and stable under base change. Set
\begin{equation}\label{eq:localization}
L:\Psh(C)\longrightarrow
\B_\T=\Psh(C)[W_\T^{-1}].
\end{equation}
\end{definition}
The closure contains every iterated diagonal of each of its maps. Indeed, for $f:A\to B$ in $W_\T$, the projection $p_1:A\times_B A\to A$ belongs to $W_\T$ by base change. The identity $p_1\Delta f\simeq\id_A$ and two-out-of-three give $\Delta f\in W_\T$; iteration gives all higher diagonals. By the small-generated congruence theorem, $W_\T$ is the inverted class of an accessible left-exact localization, and $\B_\T$ is a Grothendieck $\infty$-topos~\cite{ABFJ}. The congruence theorem identifies the saturated, base-change-stable class with its finite-limit formulation and shows that it has a small generating set as a strongly saturated class. This establishes accessibility as well as left exactness. Choices of comparison cones have contractible spaces and yield equivalent localizations.

\subsection{Left-Exact Extension by Descent}\label{sec:lex-extension}
To construct the inverse functor, we extend the interpretation of the generators in $\G_\T$ from representables to presheaves by taking colimits. Pullback stability gives left exactness over representables. Over an arbitrary base, descent lets us test the pullback comparison along the maps from its representable pieces.

\begin{lemma}[Descent extension]\label{lem:lex-extension}
Let $C$ be a small finite-limit $\infty$-category and $\D$ a geometric universe (Definition~\ref{def:geometric-universe}). A finite-limit-preserving functor $u:C\to\D$ has a small-colimit-preserving extension
\[
F=\Lan_yu:\Psh(C)\to\D,
\]
and this extension preserves finite limits. Restriction induces
\begin{equation}\label{eq:lex-extension}
\Fun_\geo(\Psh(C),\D)\simeq\Fun_\lex(C,\D).
\end{equation}
\end{lemma}
\begin{proof}
The presheaf universal property supplies the extension and its uniqueness on the whole functor $\infty$-category. Explicitly,
\begin{equation}\label{eq:elements}
F(P)=\colim_{(c,x)\in C_{/P}}u(c).
\end{equation}
Here, $C_{/P}$ denotes the category of elements, the right fibration over $C$ classified by $P:C^{\op}\to\Sps_\U$. It is small because $C$ and all the values of $P$ are small. Thus every colimit in~\eqref{eq:elements} is among the specified $\U$-small colimits of $\D$. The presheaf universal property applies in the ambient universe, with these same small indexing diagrams. Because $y1_C$ is terminal, $F(1)\simeq u(1_C)\simeq1_\D$.

First, consider $X\to yc\leftarrow Y$. In the slice over $yc$, take element presentations $X\simeq\colim_i ya_i$ and $Y\simeq\colim_j yb_j$. The arrows to $c$ are part of their indexing data. Universality of presheaf colimits gives
\[
X\times_{yc}Y\simeq\colim_{i,j}y(a_i\times_c b_j).
\]
Using finite-limit preservation by $u$ and universality of colimits in $\D$, in the two variables successively, yields
\begin{align}
F(X\times_{yc}Y)
&\simeq\colim_{i,j}\bigl(u(a_i)\times_{u(c)}u(b_j)\bigr)\notag\\
&\simeq F(X)\times_{u(c)}F(Y).\label{eq:representable-base}
\end{align}

For a general cospan $X\to Z\leftarrow Y$, let $I=C_{/Z}$ be the category of elements of $Z$. For $i\in I$ with image $c_i\in C$, put $A_i=yc_i$. The element presentation gives $Z\simeq\colim_{i\in I} A_i$. Put $X_i=X\times_Z A_i$ and $Y_i=Y\times_Z A_i$. For every transition $i\to j$, pullback pasting in $\Psh(C)$ gives
\[
X_i = X\times_Z A_i
\simeq (X\times_Z A_j)\times_{A_j}A_i
= X_j\times_{A_j}A_i.
\]
Thus the defining square is
\[
\begin{tikzcd}
X_i\ar[r]\ar[d]&X_j\ar[d]\\
A_i\ar[r]&A_j.
\end{tikzcd}
\]
Its base $A_j$ is representable, so~\eqref{eq:representable-base} shows that its image under $F$ is a pullback. Consequently $F(X_i)\to u(c_i)$ is a cartesian family over the entire element diagram. Since $X\simeq\colim_i X_i$ and $F$ preserves these colimits, descent in $\D$ gives
\begin{equation}\label{eq:localize-F}
F(X_i)\simeq F(X)\times_{F(Z)}u(c_i).
\end{equation}
The same argument applies to $Y$ and $X\times_ZY$.

Pull back the canonical comparison
\[
F(X\times_ZY)\longrightarrow F(X)\times_{F(Z)}F(Y)
\]
along every $u(c_i)\to F(Z)$. Formula~\eqref{eq:localize-F} identifies the result with~\eqref{eq:representable-base} for $X_i\to A_i\leftarrow Y_i$. It is an equivalence. The descent equivalence
$\D_{/F(Z)}\simeq\lim_{i\in I^{\op}}\D_{/u(c_i)}$
shows that these pullbacks jointly detect equivalences. This includes the empty element diagram: descent then makes the slice over its colimit contractible. The comparison is therefore an equivalence in every case, proving left exactness. Finally, the presheaf extension theorem is fully faithful on all natural transformations; restricting its objects to the identified left-exact ones proves~\eqref{eq:lex-extension}.\qed
\end{proof}

For $\infty$-topos targets, the result recovers the left-exact extension theorem~\cite{HTT}. Substitution stability gives preservation of pullbacks over representable bases, and family descent extends the result to arbitrary bases. For the geometric universe $\G_\T$, this extension constructs the inverse functor in the next theorem.

\subsection{Syntactic Reconstruction and Classification of Models}
\begin{theorem}[Syntactic reconstruction]\label{thm:reconstruction}
There are geometric functors
\[
E:\G_\T\longrightarrow\B_\T,
\qquad Q:\B_\T\longrightarrow\G_\T
\]
with $EQ\simeq\id_{\B_\T}$ and $QE\simeq\id_{\G_\T}$, compatibly with the generator interpretations.
\end{theorem}
\begin{proof}
Let $u:C\to\G_\T$ be the finite-limit extension of the generator interpretation, using Lemma~\ref{lem:freelex}. The functor $Lyj$ interprets the axioms: $L$ preserves finite limits and colimits and inverts~\eqref{eq:axiomlim}--\eqref{eq:axiomeq}. Initiality gives a geometric functor $E$, and the free finite-limit property gives
\begin{equation}\label{eq:Eu}
Eu\simeq Ly.
\end{equation}

By Corollary~\ref{cor:geometric}, the target $\G_\T$ is a geometric universe. Lemma~\ref{lem:lex-extension} makes $F=\Lan_yu$ left exact as well as colimit preserving. The theory's axioms hold syntactically, so $F$ inverts $R_\T$. Equivalences satisfy two-out-of-three and are closed under retracts. A pointwise equivalence between two diagrams induces an equivalence between their colimits. Since $F$ preserves small colimits and pullbacks, the class of maps it inverts is strongly saturated and stable under base change. Thus $F$ inverts $W_\T$ and factors as
\begin{equation}\label{eq:QL}
QL\simeq F.
\end{equation}

Let $i:\B_\T\hookrightarrow\Psh(C)$ be the fully faithful right adjoint of $L$. Then $Q\simeq Fi$, so $Q$ preserves finite limits. For any small diagram $b$ in $\B_\T$,
\[
Q\bigl(\colim_{\B_\T}b\bigr)
\simeq QL\bigl(\colim_{\Psh(C)}ib\bigr)
\simeq\colim Fib\simeq\colim Qb.
\]
Hence $Q$ is geometric.

The colimit-preserving functors $EF$ and $L$ agree on representables by~\eqref{eq:Eu}. The presheaf universal property gives $EF\simeq L$, hence $EQL\simeq L$ by~\eqref{eq:QL}. Precomposition with a localization is fully faithful on functor $\infty$-categories, yielding $EQ\simeq\id$. For the other composite,
\[
QEu\simeq QLy\simeq Fy\simeq u.
\]
Both $QE$ and the identity are geometric. Fully faithful restriction in Theorem~\ref{thm:initiality}, with target $\G_\T$, identifies them. The two inverse comparisons identify the contexts and substitutions of $\G_\T$ with the objects and maps of the presheaf localization.\qed
\end{proof}

\begin{corollary}[Presentability and classification]\label{cor:classification}
The syntax $\G_\T$ is equivalent to a $\U$-presentable, essentially $\U$-locally small $\infty$-category, and is a Grothendieck $\infty$-topos relative to $\U$. For every such $\X$,
\begin{equation}\label{eq:classification}
\Mod_\T(\X)\simeq
\Fun^{L,\lex}(\B_\T,\X)=\Geom(\X,\B_\T).
\end{equation}
Together with Theorems~\ref{thm:comparison-syntax} and~\ref{thm:reconstruction}, this equivalence proves Theorem~\ref{thm:main}.
\end{corollary}
\begin{proof}
Presentability and local smallness hold for the localization in Definition~\ref{def:classifier} and transfer across Theorem~\ref{thm:reconstruction}. Combine that theorem with initiality. A small-colimit-preserving functor between these presentable $\infty$-categories is a left adjoint~\cite{HTT}. Naturality is induced by restriction and composition. We use the inverse-image convention for transformations throughout, so~\eqref{eq:classification} includes arbitrary natural transformations of inverse-image functors.\qed
\end{proof}

\begin{proposition}[Preservation of the Calculus by Inverse Images]\label{prop:transport}
Every geometric inverse-image functor preserves the interpretation of the calculus, including its type families, terms, higher terms, and all computation and gluing comparisons, up to their specified coherent equivalences.
\end{proposition}
\begin{proof}
It preserves finite limits, small colimits, and equivalences. The derived characterization of displays and the universal interpretation lemmas show that each primitive output is carried to an output for the image input. The space of such comparisons is contractible, including its relative parameter boundaries. Induction on the derivation tree gives the asserted interpretation and its coherences.\qed
\end{proof}

\section{Proof-Space Completeness and Slice Classifiers}\label{sec:proofspaces}
A type over a context determines an object of the corresponding slice of the classifier. Its sections represent dependent proofs. The mapping space of sections also records paths between proofs and their higher identifications.
\begin{theorem}[Dependent proof-space completeness]\label{thm:proofspaces}
For every derived type $\Gamma\vdash A\ \type$, its simplicial set of higher terms is a Kan complex, and reconstruction induces a natural equivalence
\begin{equation}\label{eq:proofspaces}
\Tm_\T(\Gamma,A)\simeq
\Map_{(\B_\T)_{/E\Gamma}}(\id_{E\Gamma},E p_A).
\end{equation}
The equivalence is compatible with reindexing, dependent sum, identity, homotopy-colimit elimination, and gluing.
\end{theorem}
\begin{proof}
Formula~\eqref{eq:termspace} is exactly the right mapping-space model in the slice, as observed in~\eqref{eq:terms-as-sections}. Right mapping spaces of a quasi-category are Kan. An equivalence of quasi-categories induces equivalences on slices and on their mapping spaces, proving~\eqref{eq:proofspaces}. Reindexing is a pullback, sum is composition of displays, and the identity display is the diagonal up to its specified reflexivity equivalence. Homotopy-colimit elimination is its universal section construction, and gluing is total colimit with cartesian comparisons. The geometric comparison preserves these universal constructions and hence their coherent computation witnesses.\qed
\end{proof}

\begin{corollary}[All-model completeness]\label{cor:completeness}
For a syntactic edge $e$, the following are equivalent:
\[
\T\vdash\Eq(e)\quad\Longleftrightarrow\quad
\begin{gathered}
\text{every model in every Grothendieck $\infty$-topos}\\
\text{interprets $e$ as an equivalence.}
\end{gathered}
\]
A derived type in context has a term if and only if its interpreted family has a section in every such model.
\end{corollary}
\begin{proof}
Soundness follows from the universal interpretation lemmas. For the converse, evaluate in the universal model in $\B_\T$. Reconstruction reflects equivalences, so the corresponding syntactic edge extends to a $J$-array. Its cells occur together at a bounded stage, where (E) derives $\Eq(e)$. For a type in context,~\eqref{eq:proofspaces} identifies sections in the universal model with the syntactic term space up to equivalence. Inhabitation of the section space therefore supplies a term with the prescribed section boundary.\qed
\end{proof}

For a closed type, a section in a model is a point of the object interpreting the type. Freely adjoining a term makes the choice of that point part of the model.

\begin{theorem}[Context extension as a slice]\label{thm:sliceclassifier}
Let $A$ be a closed derived type. Corollary~\ref{cor:definitions} supplies a small definitional extension in which to name it. Adjoining a term $a:A$ gives
\[
\B_{\T[a:A]}\simeq(\B_\T)_{/EA},
\qquad
\G_{\T[a:A]}\simeq(\G_\T)_{/A}.
\]
A finite dependent telescope is classified by the slice over its total context.
\end{theorem}
\begin{proof}
Write $\B=\B_\T$ and suppress $E$. For an $\infty$-topos $\X$, we prove an equivalence of $\infty$-categories
\begin{equation}\label{eq:slicemodels}
\Fun_\geo(\B_{/A},\X)
\simeq\{(F,s):F\in\Fun_\geo(\B,\X),\ s:1_\X\to F(A)\}.
\end{equation}
The right side is the $\infty$-category of pointed evaluations. A transformation $(F,s)\to(G,t)$ consists of a natural transformation $\eta:F\to G$ and a path $\eta_A s\simeq t$. Its higher morphisms record the corresponding coherences of the chosen points.

Let $\pi^*:\B\to\B_{/A}$ send $C$ to $C\times A\to A$. Given $H$, take $F=H\pi^*$ and let $s$ be the image of the diagonal section $\id_A\to\pi^*A$. Conversely, set
\[
H_{F,s}(C\xrightarrow{p}A)=1_\X\times_{F(A)}F(C).
\]
This preserves finite limits, and preserves colimits because slice colimits are computed on total objects and pullback in $\X$ preserves them. In $\B_{/A}$, there is a pullback square
\[
\begin{tikzcd}
(C\to A)\ar[r]\ar[d]&(C\times A\to A)\ar[d,"p\times\id_A"]\\
\id_A\ar[r,"\delta_A"]&(A\times A\to A).
\end{tikzcd}
\]
Applying $H$ identifies it with the displayed formula for $H_{F,s}$. In the other direction, pulling $F(C\times A)$ back along $s$ recovers $F(C)$, and the diagonal recovers $s$. For a transformation $(\eta,h):(F,s)\to(G,t)$, the maps $\eta_C,\eta_A$, together with $h:\eta_As\simeq t$, give a map from the pullback defining $H_{F,s}(C\to A)$ to the pullback defining $H_{G,t}(C\to A)$. Its functoriality and higher simplices follow from the pullback universal property. This explicitly constructs the inverse on transformations; the preceding pullback comparisons are its natural unit and counit. Hence~\eqref{eq:slicemodels} is an equivalence of $\infty$-categories.

A model of $\T[a:A]$ is exactly a model of $\T$ with such a section. The adjoined triangle records the section witness itself, as in~\eqref{eq:term-triangle}. Classification and~\eqref{eq:slicemodels} identify its classifier with the slice. Reconstruction gives the syntactic statement. For a telescope, iterate the equivalence
\[
\left(\B_{/\Gamma}\right)_{/(A\xrightarrow{p}\Gamma)}
\simeq \B_{/A}.
\]
Here, $A$ on the right denotes the total object of the family $p$ over $\Gamma$.\qed
\end{proof}

\section{Duality between Geometric Type Theories and Grothendieck~\texorpdfstring{\mbox{$\infty$-Toposes}}{∞-Toposes}}\label{sec:alltopoi}
The classification also runs from geometry to syntax. Starting with a Grothendieck $\infty$-topos, we choose a small family of generators and express the relations defining its presheaf localization as equivalence axioms. Together with the correspondence between interpretations and inverse-image functors, these small presentations give the $(\infty,2)$-duality.

\subsection{Small Presentations of Grothendieck \texorpdfstring{$\infty$}{∞}-Toposes}
The following lemma gives a small set of maps whose inversion presents the chosen localization.
\begin{lemma}[Generators within the universe]\label{lem:universe-generators}
Let $C$ be a $\U$-small $\infty$-category and let
\[
 L_0:\Psh(C)\rightleftarrows\X:N
\]
be an accessible reflective localization relative to $\U$, with unit
$\eta:\id\to NL_0$. There are a regular cardinal $\lambda<\kappa$ and a
$\U$-small category $\mathcal A$ with a fully faithful functor
to $\Psh(C)$ whose essential image consists of the $\lambda$-compact
presheaves. Writing $A$ also for the image of an object of $\mathcal A$, the family
\begin{equation}\label{eq:universe-generators}
 R=\{\eta_A:A\longrightarrow NL_0A\mid A\in\mathcal A_0\}
\end{equation}
has representatives indexed and encoded in $\U$, and its strongly saturated
closure, formed using $\U$-small colimits, is exactly the class of maps
inverted by $L_0$.
If $L_0$ is left exact, this class is also the congruence generated by $R$.
The category-of-elements diagrams of all domains and codomains in $R$, together
with their induced comparison functors, admit a joint $\U$-small presentation.
\end{lemma}
\begin{proof}
Write $\mathcal E=\Psh(C)$ and $T=NL_0$. By accessibility relative to $\U$, choose regular cardinals $\alpha,\beta<\kappa$
such that $\mathcal E$ is $\alpha$-accessible and $N$ preserves
$\beta$-filtered colimits. The existence of $\beta$ follows from accessibility
of right adjoints between presentable categories, applied in $\U$
\cite{HTT}. Choose an infinite cardinal
$\mu<\kappa$ with $\mu\geq\alpha,\beta$, and put
\[
 \lambda=(2^\mu)^+<\kappa.
\]
The last inequality follows from the strong inaccessibility of $\kappa$.
For every cardinal $\rho<\lambda$ and every $\sigma<\alpha$, one has
\[
 \rho^\sigma\leq(2^\mu)^\mu=2^\mu<\lambda.
\]
Thus $\lambda$ satisfies the cardinal enlargement condition for
$\alpha$-accessible categories, so $\mathcal E$ is $\lambda$-accessible
\cite{HTT}. Every $\lambda$-filtered category is
$\beta$-filtered; hence $N$, and therefore $T$, preserves
$\U$-small $\lambda$-filtered colimits. The full subcategory of $\lambda$-compact objects is essentially
$\U$-small by the compact-generation characterization of accessibility
\cite{HTT}. Choose a $\U$-small equivalent
quasi-category $\mathcal A$ and compose the equivalence with the full
inclusion into $\mathcal E$.

Let $W$ be the maps inverted by $L_0$, and let $\langle R\rangle$ denote the
strongly saturated closure of~\eqref{eq:universe-generators}. Since
$L_0\eta_A$ is an equivalence, one has $\langle R\rangle\subseteq W$.
For any $P\in\mathcal E$, choose a $\U$-small $\lambda$-filtered presentation
$P\simeq\colim_{i\in I}A_i$ with $A_i\in\mathcal A$. Naturality of the unit
and preservation of this colimit by $T$ identify a colimit in the arrow
category:
\begin{equation}\label{eq:unit-filtered-colimit}
 (\eta_P:P\to TP)\simeq
 \colim_{i\in I}(\eta_{A_i}:A_i\to TA_i).
\end{equation}
A strongly saturated class is closed under these colimits
\cite{HTT}; hence every $\eta_P$ lies in
$\langle R\rangle$. Suppose $f:P\to Q$ belongs to $W$. In the naturality square
\[
\begin{tikzcd}[column sep=large]
 P\ar[r,"f"]\ar[d,"\eta_P"']&Q\ar[d,"\eta_Q"]\\
 TP\ar[r,"Tf"']&TQ
\end{tikzcd},
\]
$Tf$ is an equivalence and both vertical arrows lie in $\langle R\rangle$.
Closure under composition and the two-out-of-three property give
$f\in\langle R\rangle$. Thus $W=\langle R\rangle$. If $L_0$ is left exact,
$W$ is stable under base change. Consequently it is also the smallest
strongly saturated, base-change-stable class containing $R$. This is the
universe-relative generating-set construction underlying the localization
criterion in \cite{HTT}.

For the size of the displayed generators, $\mathcal A_0$ is a set in $\U$,
and $A$ and $TA$ are presheaves with values in $\U$-small spaces. For each of
these presheaves, choose a right-fibration model
$E_A\to C$ or $E_{TA}\to C$ of its category of elements. These models can be
chosen in $\U$: the base $C$ is $\U$-small, the fibers are $\U$-small, and
straightening and unstraightening are carried out with $\U$-small simplicial
sets. Equivalently, the size assertion follows from the small-fiber criterion
for right fibrations \cite{HTT} after choosing a regular
cardinal in $\U$ that bounds the base and its fibers. A representative of
$\eta_A$ gives a map $E_A\to E_{TA}$ over $C$, with its complete simplicial
coherence data.

There are fewer than $\kappa$ indices $A$, and every one of the chosen
simplicial sets and comparison diagrams has fewer than $\kappa$ simplices.
Regularity of $\kappa$ supplies a single infinite cardinal $\nu<\kappa$
bounding their disjoint union, the cells of $C$, and the index set
$\mathcal A_0$. Finite products, joins with finite simplices, and countable
collections of simplicial dimensions preserve a bound of the form
\begin{equation}\label{eq:auxiliary-size-bound}
 (2+\nu)^{\aleph_0}<\kappa.
\end{equation}
Thus the element diagrams, their cones, and their comparison arrays form one $\U$-small collection. To include the definitions that name these diagrams, take the union of their derivation supports as in Lemmas~\ref{lem:substitution} and~\ref{lem:support-realization}. A $\U$-small family of $\U$-small supports has $\U$-small union, so we can choose $\nu<\kappa$ to bound this union together with the original diagrams. Estimate~\eqref{eq:auxiliary-size-bound} then covers the joins, prisms, and comparison arrays built from the definitions. The generating maps and the declarations naming them therefore belong to $\U$.\qed
\end{proof}

\begin{theorem}[Small presentation theorem]\label{thm:alltopoi}
For every Grothendieck $\infty$-topos $\X$ relative to $\U$, there is a small dependent theory $\T_\X$ with
\[
\G_{\T_\X}\simeq\B_{\T_\X}\simeq\X.
\]
\end{theorem}
\begin{proof}
Choose a regular cardinal $\chi\in\U$ such that $\X$ is $\chi$-accessible and its $\chi$-compact objects form an essentially $\U$-small subcategory generating $\X$ under $\chi$-filtered colimits. Such a cardinal exists by presentability~\cite{HTT}. Choose a small equivalent model $C_0$ for the full subcategory of $\chi$-compact objects, with its fully faithful functor to $\X$. We write $c$ also for the image of $c\in C_0$. To verify density explicitly, write $X\simeq\colim_i c_i$ as a small $\chi$-filtered colimit with $c_i\in C_0$. The restricted nerve $N_{C_0}(X)(c)=\Map_\X(c,X)$ satisfies
\[
 N_{C_0}(X)\simeq\colim_i y c_i,
\]
because every $c\in C_0$ is $\chi$-compact. Hence, for every $Y$,
\[
 \Map_{\Psh(C_0)}(N_{C_0}X,N_{C_0}Y)
 \simeq\lim_i\Map_\X(c_i,Y)
 \simeq\Map_\X(X,Y).
\]
Thus $C_0$ is dense. We construct a small model for its finite-limit closure. Start with the objects in the image of $C_0$. At stage $n$, let $D_n$ be the full subcategory of $\X$ on the objects chosen so far. This family of objects is indexed by a $\U$-small set. Since $\X$ is locally $\U$-small, $D_n$ is essentially $\U$-small by \cite{HTT}. For each finite simplicial shape $K$, the category $\Fun(K,D_n)$ is therefore essentially $\U$-small. Choose a limit object in $\X$ for one diagram from each equivalence class and add these objects at stage $n+1$. There are countably many finite shapes up to isomorphism, so the new family and the union over all stages are each indexed by a $\U$-small set. Every finite diagram in the full subcategory on this union factors through one stage, since its vertex set is finite. Its limit is thus represented in the union. This full subcategory is essentially $\U$-small by the same criterion. Choose a $\U$-small equivalent quasi-category $C$ and compose with the inclusion into $\X$. The resulting functor $C\to\X$ is fully faithful and preserves finite limits, and $C_0\to\X$ factors through it up to natural equivalence.

We next verify density of $C\to\X$. The colimit extension of $C_0\to\X$ is left adjoint to $N_{C_0}$, and the established full faithfulness of $N_{C_0}$ makes its counit an equivalence. Consequently the canonical category-of-elements cocone $(C_0)_{/X}\to\X$ has colimit $X$ for every $X\in\X$. To compute the enlarged nerve, write $N_C(X)(c)=\Map_\X(c,X)$ and let $i:C_0\to C$ be this fully faithful factor. The canonical expression of each $c\in C$ as a colimit of objects of $C_0$ gives
\[
\Map_\X(c,X)\simeq
\lim_{(a\to c)\in(C_0)_{/c}^{\op}}\Map_\X(a,X).
\]
These equivalences identify $N_C$ with $\operatorname{Ran}_{i^{\op}}N_{C_0}$. Right Kan extension along the fully faithful $i^{\op}$ is fully faithful on these diagrams. It follows that
\[
\Map_{\Psh(C)}(N_CX,N_CY)
\simeq\Map_{\Psh(C_0)}(N_{C_0}X,N_{C_0}Y)
\simeq\Map_\X(X,Y).
\]
Thus the enlarged finite-limit subcategory $C$ is dense in $\X$.

The fully faithful functor $C\to\X$ extends to a colimit-preserving functor $L_0:\Psh(C)\to\X$. Lemma~\ref{lem:lex-extension} makes it left exact, and density makes its right adjoint fully faithful. It is consequently an accessible left-exact localization. Lemma~\ref{lem:universe-generators}, applied to this localization and its right adjoint $N_C$, gives a regular $\lambda<\kappa$ and the explicit $\U$-small set $R=\{A\to N_CL_0A\mid A\in\mathcal A_0\}$. Its strongly saturated closure is precisely the inverted congruence. Thus the small-generation assertion of~\cite{ABFJ} is realized here inside $\U$.

Declare the cells of $C$ as constants and its finite-limit cones as axioms, compiling these into dependent records by Proposition~\ref{prop:compilation}. For every $r:P\to Q$ in $R$, use the element presentations
\[
P\simeq\colim_{(c,x)\in C_{/P}}yc,
\qquad Q\simeq\colim_{(c,y)\in C_{/Q}}yc.
\]
The map $r$ induces a small functor $C_{/P}\to C_{/Q}$ over $C$. Name these two colimit objects by (HC), and use its comparison rule to name the induced map $r_{\rm syn}$, including the compatibility with the element-diagram functor. Declare $\Eq(r_{\rm syn})$. Each step has small input, and there is a $\U$-small set of steps. More explicitly, Lemma~\ref{lem:universe-generators} bounds the cells of $C$, the set $R$, both element categories for every $r$, and their comparison functors by a single cardinal below $\kappa$. The finite-record compilation and the naming of the comparison arrays select a $\U$-small family of derivations. Their complete dependency supports have $\U$-small union by Lemmas~\ref{lem:substitution} and~\ref{lem:support-realization}. Choose an infinite $\nu<\kappa$ bounding this union together with the preceding data. Finite products, finite joins, and the countably many simplicial dimensions of the resulting packets have total size at most
\[
 (2+\nu)^{\aleph_0}<\kappa.
\]
There are consequently fewer than $\kappa$ cells, certificates, and boundary equations in the resulting theory $\T_\X$.

A finite-limit-preserving functor $C\to\D$ to a geometric universe extends to a model of $\T_\X$ precisely when its colimit extension inverts $R$. In that case, the defining constructions have a contractible space of choices. By Lemma~\ref{lem:lex-extension}, this extension is geometric and therefore inverts the generated congruence. Localization gives
\[
\Mod_{\T_\X}(\D)\simeq\Fun_\geo(\X,\D).
\]
The factor through the localization is geometric by the same right-adjoint and colimit calculation used in Theorem~\ref{thm:reconstruction}. The equivalence holds on all natural transformations. Compare it with initiality, using the two targets $\X$ and $\G_{\T_\X}$, to obtain mutually inverse geometric functors. Reconstruction then identifies $\B_{\T_\X}$ as well.\qed
\end{proof}

\subsection{Interpretations, Morita Equivalence, and Duality}
An interpretation of $\T$ in $\T'$ assigns its generating cells to syntactic cells of $\G_{\T'}$, respects the equations, and derives the specified certificate axioms. Let $\operatorname{Int}(\T,\T')$ be the full subcategory of $\Fun(|Q_\T|,\G_{\T'})$ on these interpretations; its higher simplices are parameterized substitution arrays.

\begin{corollary}[Interpretation and Morita correspondence]\label{cor:morita}
There are natural equivalences
\[
\operatorname{Int}(\T,\T')
\simeq\Fun_\geo(\G_\T,\G_{\T'})
\simeq\Geom(\B_{\T'},\B_\T).
\]
The classifiers are equivalent exactly when the theories have interpretations inverse up to coherent equivalence. Equivalently, their model-valued functors on $\infty$-toposes are pseudonaturally equivalent.
\end{corollary}
\begin{proof}
Apply initiality with target $\G_{\T'}$ and then reconstruction. Interpretations compose through their geometric extensions; their extension spaces are contractible, giving coherent composition. Inverse interpretations correspond to inverse geometric functors. Finally, a pseudonatural equivalence of the model functors, evaluated on each of their universal models, gives the same pair of inverse interpretations. Conversely, an equivalence of classifiers gives the pseudonatural equivalence by precomposition.\qed
\end{proof}

The equivalence $\G_{\T_\X}\simeq\X$ of Theorem~\ref{thm:alltopoi} carries gluing in $\T_\X$ to descent in $\X$. These small presentations give essential surjectivity of the classifier assignment, and Corollary~\ref{cor:morita} identifies its mapping $\infty$-categories.

Let $\mathbf{Th}_{\geo}$ denote the $(\infty,2)$-category of $\U$-small geometric HoTT theories with mapping $\infty$-categories $\operatorname{Int}(\T,\T')$ and composition by interpretation. Let $\mathbf{Top}_\infty$ denote the $(\infty,2)$-category of Grothendieck $\infty$-toposes relative to $\U$, with $\Geom(\X,\mathcal Y)=\Fun^{L,\lex}(\mathcal Y,\X)$ and all natural transformations of inverse images. We write $\op$ for the opposite that reverses $1$-morphisms and preserves the direction of $2$-morphisms.

\begin{corollary}[Duality between geometric HoTT theories and Grothendieck $\infty$-toposes]\label{cor:duality}
The classifier assignment induces an equivalence of $(\infty,2)$-categories
\[
  \mathbf{Th}_{\geo}\simeq\mathbf{Top}_\infty^{\op},
  \qquad \T\longmapsto\B_\T .
\]
An interpretation $I$ of $\T$ in $\T'$ corresponds to a geometric morphism $f_I:\B_{\T'}\to\B_\T$. Its inverse image carries the universal $\T$-model to the $\T$-model obtained by applying $I$ to the universal $\T'$-model. Two theories have equivalent classifiers precisely when they are Morita equivalent, that is, when they admit interpretations inverse up to coherent equivalence.
\end{corollary}
\begin{proof}
Write $j_\T:|Q_\T|\to\G_\T$ for the inclusion of generators. By Theorem~\ref{thm:initiality} and Corollary~\ref{cor:morita}, restriction $F\mapsto F j_\T$ gives an equivalence from $\Fun_{\geo}(\G_\T,\G_{\T'})$ to $\operatorname{Int}(\T,\T')$, including all transformations. For a further geometric functor $H$, the identity $(HF)j_\T=H(Fj_\T)$ holds in every simplicial degree. Thus we may realize $\mathbf{Th}_{\geo}$ with objects labelled by theories and with geometric functors between their syntactic categories composing by ordinary functor composition.

Put $\mathcal L=\mathbf{Top}_\infty^{\op}$ and let $J$ be the groupoid freely generated by an isomorphism $0\to1$. Use the $(\infty,2)$-category $\Fun(J,\mathcal L)$ of coherent diagrams and transformations with invertible naturality comparisons. Since each inclusion $\{i\}\hookrightarrow J$ is an equivalence, evaluation at either endpoint is an equivalence with inverse given by constant diagrams. By Theorem~\ref{thm:reconstruction}, each $E_\T:\G_\T\simeq\B_\T$ extends to a $J$-diagram. Choose one such diagram for each $\T$, and let $\D$ have these theory-labelled objects and the full mapping $\infty$-categories between them in $\Fun(J,\mathcal L)$. The endpoint evaluations restrict to
\[
  \mathbf{Th}_{\geo}\xleftarrow{\ \mathrm{ev}_0\ }\D
  \xrightarrow{\ \mathrm{ev}_1\ }\mathcal L .
\]
Both functors induce equivalences on mapping $\infty$-categories by fullness. The first reaches every theory label, and the second reaches every Grothendieck $\infty$-topos up to equivalence by Theorem~\ref{thm:alltopoi}. Hence both are equivalences of $(\infty,2)$-categories.

The composite $\mathrm{ev}_1\mathrm{ev}_0^{-1}$ sends $\T$ to $\B_\T$ up to equivalence. For an interpretation $I$ with extension $F:\G_\T\to\G_{\T'}$, lift $F$ along $\mathrm{ev}_0$ and write $f_I^*$ for its other endpoint. Naturality along $0\to1$ gives $f_I^*E_\T\simeq E_{\T'}F$. On generators, this identifies the pullback of the universal $\T$-model with the interpretation of $I$ in the universal $\T'$-model. The two evaluations are functors on coherent diagrams, so the resulting equivalences
\[
  \operatorname{Int}(\T,\T')\simeq
  \Fun^{L,\lex}(\B_\T,\B_{\T'})=\Geom(\B_{\T'},\B_\T)
\]
respect units, composition, and their higher coherences. The assertion about Morita equivalence follows from Corollary~\ref{cor:morita}.\qed
\end{proof}

\subsection{A Universal Family of Classifying \texorpdfstring{$\infty$}{∞}-Toposes}\label{sec:uf-pres}

Suppose a theory has one arrow $h:X_0\to X_1$. Over a topos $\X$ with an open $U\hookrightarrow 1$, we can require $h$ to be an equivalence on $U$. A model is then an arrow in $\X$ whose restriction to $\X_{/U}$ is an equivalence. Pulling back along a geometric morphism carries the axiom to the inverse image of $U$. We now allow the generators and the indexing diagrams of the axioms to vary in this way as well.

\begin{definition}[Indexed boundary presentation]\label{def:uf-presentation}
Write $s\X=\Fun(\Delta^{\op},\X)$ and let $\Delta_{\X}:\Sps_\U\to\X$ be the constant-object functor. An \emph{indexed boundary presentation} $P$ in $\X$ consists of the following data.
\begin{enumerate}
\item A simplicial object $Q$ in the category of set objects $\X_{\leq0}$, together with a set object $R$ and a map $R\to Q_1$ marking equivalence edges.
\item For each $J$ in a set of representatives of the finite simplicial sets, a set object $L_J$ and a map $L_J\times J^\triangleleft\to Q$ marking limit cones. In degree $n$, the domain is the coproduct of copies of $L_J$ indexed by $(J^\triangleleft)_n$.
\item A set object $A$ of colimit axioms, a simplicial object $K:\Delta^{\op}\to(\X_{/A})_{\leq0}$ in the category of set objects over $A$, and a map $b:K^{\triangleright_A}\to Q$ marking their cocones. Here, the right cone is formed in the slice and its total objects are used as the domain of $b$. In degree $n$, these are $A\amalg K_0\amalg\cdots\amalg K_n$.
\end{enumerate}
A morphism of indexed presentations gives maps on these objects commuting with the simplex operators and all incidence maps, including a map of the shapes $K$ over the map of axiom objects. Write $\operatorname{Pres}(\X)$ for this ordinary category and $N\operatorname{Pres}(\X)$ for its nerve.
\end{definition}

In spaces, a collection of cocones $K_a^\triangleright\to Q$ is encoded by taking $K_n=\coprod_{a\in A}(K_a)_n$. Giving each axiom of a boundary presentation its own index therefore defines a point of this parameter problem. The indices retain repeated occurrences of an axiom. A map of indexed presentations records these data and can change a colimit shape. An interpretation as in Corollary~\ref{cor:morita} assigns generators to derived expressions in the target syntax and derives the source axioms there.

To interpret the generators of $P$, choose a map $E\to Q$ of simplicial objects of $\X$ for which the initial vertex induces, in every degree, an equivalence
\begin{equation}\label{eq:uf-left}
 E_n\xrightarrow{\ \sim\ } Q_n\times_{Q_0}E_0 .
\end{equation}
We call such a map a \emph{left family}. Its fibers over the generators can be arbitrary objects of $\X$. On an external shape, this condition describes a covariant diagram, with all its natural transformations (Lemma~\ref{lem:uf-external}). Apply it in $\X_{/R}$ and $\X_{/L_J}$ to require that the marked edges be equivalences and the marked cones be limiting. For the colimit test, restrict $E$ first to the given family of cocones and then to their bases, obtaining $D^+$ over $K^{\triangleright_A}$ and $D^-$ over $K$. Require $|D^-|\to |D^+|$ to be an equivalence, with geometric realizations in $\X_{/A}$. Lemma~\ref{lem:uf-cone} identifies the target with the value at the apex. Write $\Mod_P(\X)$ for the full subcategory of $(s\X)_{/Q}$ on these models. In the $\infty$-category $\operatorname{Pair}(\X)$ of pairs $(P,E)$, a morphism $(P,E)\to(P',E')$ over $\varphi:P\to P'$ is a map $E\to\varphi_Q^*E'$ in $(s\X)_{/Q}$, where $\varphi_Q:Q\to Q'$ is the induced simplicial map.

\begin{theorem}[Classification of presentations and their models]\label{thm:uf-pres}
There are small theories $\Theta_{\rm pres}$ and $\Theta_{\rm pair}$ and an interpretation $i:\Theta_{\rm pres}\to\Theta_{\rm pair}$ whose geometric morphism is
\[
 p_{\rm pres}:\mathfrak M=\B_{\Theta_{\rm pair}}
       \longrightarrow \mathfrak P=\B_{\Theta_{\rm pres}} .
\]
There are natural equivalences of $\infty$-categories
\[
 \Geom(\X,\mathfrak P)\simeq N\operatorname{Pres}(\X),\qquad
 \Geom(\X,\mathfrak M)\simeq\operatorname{Pair}(\X).
\]
Composition with $p_{\rm pres}$ forgets the model. If $u:\X\to\mathfrak P$ classifies $P$, the pullback $\mathfrak M_P:=\X\times_{\mathfrak P}\mathfrak M$ is a relative classifier. For every geometric morphism $g:\mathcal Y\to\X$, the $\infty$-category of lifts of $g$ to $\mathfrak M_P$, with transformations over $\id_g$, is equivalent to $\Mod_{g^*P}(\mathcal Y)$. For an indexed boundary presentation $P$ of a small theory $\T$, let $t_P:\Sps_\U\to\mathfrak P$ be the point classifying $P$. The following square is a pullback. The generic pair restricts to the universal $\T$-model on its fiber.
\begin{equation}\label{eq:uf-pres-square}
\begin{tikzcd}[column sep=large]
\B_\T\ar[r]\ar[d]&\mathfrak M\ar[d,"p_{\rm pres}"]\\
\Sps_\U\ar[r,"t_P"']&\mathfrak P
\end{tikzcd}
\end{equation}
\end{theorem}
\begin{proof}
Proposition~\ref{prop:uf-pair} constructs the two theories and identifies restriction along $i$ with $(P,E)\mapsto P$. Classification and Corollary~\ref{cor:duality} give the displayed equivalences and the geometric morphism with the indicated direction. Appendix~\ref{app:universal-families} proves the relative assertion by computing the full $\infty$-category of lifts.

For the point $t_P$, pulling the presentation back to $\mathcal Y$ makes each set of marks constant. The slices over these sets split as products, one factor per mark. Lemma~\ref{lem:uf-external} identifies a family satisfying~\eqref{eq:uf-left} with a diagram on the generating simplicial set. On each factor, the three tests now impose exactly the marked equivalence, finite limit, or small colimit of the external presentation. Proposition~\ref{prop:compilation} therefore identifies this lift category with $\Mod_\T(\mathcal Y)$, naturally in $\mathcal Y$. By Corollary~\ref{cor:classification} and Yoneda, the fiber is $\B_\T$. Its identity corresponds to the restriction of the generic pair, proving the claim about the model.\qed
\end{proof}

Theorem~\ref{thm:alltopoi} now exhibits every $\U$-relative Grothendieck $\infty$-topos as a fiber of this single morphism. Two chosen presentation points have equivalent fibers precisely when the corresponding theories are Morita equivalent, by Corollary~\ref{cor:morita}.

\begin{corollary}[Axioms supported on opens]\label{cor:uf-conditional}
Let $\C$ be a small $\infty$-category and choose a small family $f_r:B_r\to B'_r$ in $\Psh(\C)$, indexed by $I$. Let $\mathfrak P_I$ be the $\infty$-topos of sheaves of spaces on the product of $I$ copies of the Sierpi\'nski locale, with universal opens $U_r$. Form the product $\mathcal A=\Psh(\C)\times^{\mathbf{Top}}\mathfrak P_I$ of topoi, with projections $q$ and $\pi$, and let
\[
 \mathfrak U_{\C,I}
   =L^{\lex}_{\{\,\pi^*U_r\times q^*f_r\,\mid\,r\in I\,\}}\mathcal A .
\]
Here, $L^{\lex}$ denotes the accessible left-exact localization generated by the indicated maps. After choosing a small presentation of $\Psh(\C)$ in which the $f_r$ are named, there is a map $\chi:\mathfrak P_I\to\mathfrak P$ and an equivalence over $\mathfrak P_I$
\[
 \mathfrak U_{\C,I}\simeq\mathfrak P_I\times_{\mathfrak P}\mathfrak M .
\]
At the point $s_J$ where precisely the opens indexed by $J\subseteq I$ are true, the fiber is $L^{\lex}_{\{f_r\mid r\in J\}}\Psh(\C)$.
\end{corollary}
\begin{proof}
By Theorem~\ref{thm:alltopoi}, choose a theory presenting $\Psh(\C)$. Use reconstruction and Lemma~\ref{lem:support-realization} to name representatives of the $f_r$ with their boundaries. These definitions preserve all models by Corollary~\ref{cor:definitions}. Over $\mathfrak P_I$, take the constant indexed presentation so obtained and add an equivalence mark for $f_r$ indexed by $U_r$. The new mark object is the coproduct of the old mark object with $\coprod_r U_r$. This presentation defines $\chi$.

A map $\mathcal Y\to\mathfrak P_I$ specifies opens $V_r\hookrightarrow 1_{\mathcal Y}$, by the free-frame property of the product Sierpi\'nski locale and the localic universal property~\cite{HTT}. Over this map, a model of the chosen presentation is a geometric inverse image $F:\Psh(\C)\to\mathcal Y$ for which each $F(f_r)$ restricts to an equivalence over $V_r$. In the product $\mathcal A$, the corresponding inverse image takes $\pi^*U_r\times q^*f_r$ to $V_r\times F(f_r)$. Thus the left-exact localization~\cite{ABFJ} and the pullback along $\chi$ classify the same data over every such $\mathcal Y$. Relative Yoneda gives the equivalence. At $s_J$, a mark indexed by $r\in J$ imposes invertibility of $F(f_r)$, while a mark indexed by $r\notin J$ has empty support. This gives the stated localization.\qed
\end{proof}

\begin{example}[A change of Morita type on the Sierpi\'nski locale]\label{ex:uf-switch}
Take $\C$ discrete on two objects, so that $\Psh(\C)=\Sps_\U\times\Sps_\U$, and take the single map $f:(0,0)\to(1,0)$. At the point where $U$ is empty, the comparison is $0\to0$ and the fiber is $\Sps_\U\times\Sps_\U$. At the point where $U$ is true, the fiber is the left-exact localization generated by $f$. An $f$-local object $(X,Y)$ has $X\simeq 1$, since mapping $f$ into $(X,Y)$ gives $X\to1$. Projection to the second factor is the corresponding reflector and is left exact, so this fiber is $\Sps_\U$. The two fibers have four and two subterminal isomorphism classes, respectively.
\end{example}

\subsection{Morita Transport and Higher Automorphisms}\label{sec:uf-morita}
Corollary~\ref{cor:duality} also describes transport between equivalent classifiers. Restrict the duality to equivalences of theories, invertible transformations, and their higher identifications. This retains the paths between Morita-equivalent theories together with the space of ways to transport their classifiers along those paths.

We specify the size of this space and of its fibers. Work in a universe $\mathbb W$ and choose a $\mathbb W$-small family of theory presentations. Write $\B_t^{\mathbb W}$ for the classifier of the presentation at $t$, formed with $\mathbb W$-small spaces and colimits. Let $K$ be the full $\infty$-groupoid on these theory-labelled classifiers, with geometric equivalences and their invertible higher transformations. Its path spaces are $\mathbb W$-small. Indeed, for a pair of classifiers, choose a common regular cardinal $\lambda$ in $\mathbb W$ at which both are accessible. An equivalence restricts to an equivalence of their essentially small subcategories of $\lambda$-compact objects. Restriction is fully faithful on equivalences and their transformations, since the classifiers are the $\lambda$-filtered-colimit completions of these subcategories. Thus $K$ is essentially $\mathbb W$-small. One can use $\mathbb W=\U$ for a small family in the original universe. To index all $\U$-small presentation codes at once, take $\mathbb W=\V$, choose representatives for the codes, and compute the classifiers, their equivalences, and their geometric fibers in $\V$. In the formulas below, $\Sps=\Sps_{\mathbb W}$.

Since the classifiers $\B_t^{\mathbb W}$ can be $\mathbb W$-large, we take $\mathrm{Cat}_{\infty}$ in the larger ambient universe. A path represented by a geometric equivalence $f$ acts by its direct image $f_*$, an inverse to $f^*$. This gives a coherent diagram $D:K\to\mathrm{Cat}_{\infty}$ with $D(t)=\B_t^{\mathbb W}$. Its limit is the category of sections, whose objects are families $X_t\in D(t)$ equipped with coherent transport along paths in $K$. Put
\[
  \mathfrak P_{\mathrm{Mor}}=\Sps_{/K}\simeq\Fun(K,\Sps),
  \qquad
  \mathfrak M_{\mathrm{Mor}}=\lim_{t\in K}D(t).
\]
For a topos $\mathcal E$, let $\Delta_{\mathcal E}:\Sps\to\mathcal E$ denote its constant-object inverse image. Geometric equivalences preserve constant objects, so these inverse images determine a functor
\[
 p_{\mathrm{Mor}}^*:\Fun(K,\Sps)\longrightarrow\mathfrak M_{\mathrm{Mor}},
 \qquad V\longmapsto\bigl(t\longmapsto\Delta_{D(t)}V(t)\bigr).
\]

\begin{theorem}[The family over Morita space]\label{thm:uf-morita}
The functor $p_{\mathrm{Mor}}^*$ is the inverse image of a geometric morphism
$p_{\mathrm{Mor}}:\mathfrak M_{\mathrm{Mor}}\to\mathfrak P_{\mathrm{Mor}}$.
For every $\mathbb W$-relative Grothendieck $\infty$-topos $\X$, the space of geometric morphisms from $\X$ to its base is
\[
 \Geom(\X,\mathfrak P_{\mathrm{Mor}})^{\simeq}
     \simeq \Map_{\X}(1,\Delta_{\X}K).
\]
If $a_t:\Sps\to\mathfrak P_{\mathrm{Mor}}$ is the point corresponding to $t\in K$, the geometric fiber satisfies
\[
 a_t^*\mathfrak M_{\mathrm{Mor}}\coloneqq \Sps\times_{\mathfrak P_{\mathrm{Mor}}}\mathfrak M_{\mathrm{Mor}}\simeq D(t)=\B_t^{\mathbb W}.
\]
Transport along paths induces an equivalence of spaces
\[
 \Map_K(t,t')\simeq
 \operatorname{Eq}_{\mathrm{geom}}\bigl(a_t^*\mathfrak M_{\mathrm{Mor}},
                                      a_{t'}^*\mathfrak M_{\mathrm{Mor}}\bigr),
\]
where the right-hand side consists of geometric equivalences and invertible transformations between them.
\end{theorem}
\begin{proof}
Lemma~\ref{lem:uf-eval} shows that the section category is a Grothendieck $\infty$-topos and that the indicated functor is a geometric inverse image. The universal property of the \'{e}tale slice over $K$ identifies a lift of the unique geometric morphism $\X\to\Sps$ with a section of $\Delta_{\X}K$~\cite{HTT}. In the description by functors on $K$, the object $(*\xrightarrow{t} K)$ is $h_t(s)=\Map_K(t,s)$. Its slice projection is $a_t$. Pullback of that projection gives the slice at $p_{\mathrm{Mor}}^*h_t$~\cite{HTT}. By Lemma~\ref{lem:uf-eval}, evaluation identifies this slice with $D(t)$ and sends transport along a path to its given equivalence of classifiers. The displayed map on path spaces is therefore the inclusion of the mapping space of the full groupoid $K$ into the space of geometric equivalences between the two classifiers, which is an equivalence by fullness.\qed
\end{proof}

\begin{proposition}[Change of Morita type on a connected base]\label{prop:uf-nonfactor}
The one-axiom family of Example~\ref{ex:uf-switch}, formed in $\mathbb W$, cannot be the pullback of $p_{\mathrm{Mor}}$ along a geometric morphism from its Sierpi\'nski base.
\end{proposition}
\begin{proof}
For the Sierpi\'nski locale $S$, a map $\operatorname{Sh}(S;\Sps)\to\mathfrak P_{\mathrm{Mor}}$ gives a section of the constant object on $K$. Since $S$ is connected, the image of this section in the constant sheaf on the set $\pi_0K$ takes the same value at both points. Theorem~\ref{thm:uf-morita} then gives equivalent fibers at these points. The family of Example~\ref{ex:uf-switch} has fibers $\Sps\times\Sps$ and $\Sps$. They have four and two subterminal isomorphism classes, respectively, so a pullback identification is impossible.\qed
\end{proof}

Within any one Morita component, the loop space at $t$ is the space of geometric self-equivalences of $D(t)$. We compute it for group actions. For a $\mathbb W$-small discrete group $G$, let $BG$ be the nerve of the one-object groupoid with automorphism group $G$. Write $\operatorname{Aut}(G)$ for the discrete automorphism group of $G$ and $\operatorname{Aut}(BG)$ for the space of self-equivalences of $BG$ under composition. Its delooping is $B\operatorname{Aut}(BG)$. Include in the chosen family a presentation of $\mathcal E_G=\Fun(BG,\Sps)\simeq\Sps_{/BG}$, which exists by Theorem~\ref{thm:alltopoi} in the working universe. Denote its point of $K$ by $t_G$.

\begin{proposition}[The Morita component of group actions]\label{prop:uf-group}
The component at $t_G$ has homotopy type $B\operatorname{Aut}(BG)$. Its positive-degree homotopy groups are
\[
 \pi_1(K,t_G)\cong\operatorname{Out}(G),\qquad
 \pi_2(K,t_G)\cong Z(G),\qquad \pi_n(K,t_G)=0\quad(n\geq3).
\]
\end{proposition}
\begin{proof}
The slice universal property identifies maps $\mathcal E_G\to\mathcal E_G$ in the geometric mapping space with sections of $BG\times BG\to BG$, hence with maps $BG\to BG$. Under this identification, a map $u$ gives pullback along $u$ between slices, so composition of geometric morphisms corresponds to composition of maps and self-equivalences correspond to $\operatorname{Aut}(BG)$. Delooping gives the component of $K$. A functor between the one-object groupoids $BG$ is a homomorphism of $G$, and a natural isomorphism between two such functors is a conjugation. Consequently $\operatorname{Aut}(BG)$ is the nerve of the action groupoid of $G$ on $\operatorname{Aut}(G)$ by postcomposition with inner automorphisms. Its components form $\operatorname{Out}(G)$, the stabilizer at the identity is $Z(G)$, and its higher homotopy groups vanish. Delooping gives the asserted groups.\qed
\end{proof}

\begin{corollary}\label{cor:uf-c2}
For the cyclic group $C_2$, the component at $t_{C_2}$ is the Eilenberg--Mac~Lane space $K(C_2,2)$.
\end{corollary}
\begin{proof}
The group $C_2$ has trivial outer automorphism group and center $C_2$. Hence its Morita component is a simply connected $2$-type with second homotopy group $C_2$.\qed
\end{proof}

\begin{corollary}[A Morita component with nonzero third homotopy group]\label{cor:uf-morita-pi3}
Let $X=K(C_2,2)$. Choose the family defining $K$ to contain a $\mathbb W$-small presentation of $\Sps_{/X}$, and let $t\in K$ be the corresponding point. The connected component of $K$ at $t$ has homotopy type $K(C_2,3)$. Its only nonzero homotopy group in positive degree is $\pi_3(K,t)\cong C_2$.
\end{corollary}
\begin{proof}
Theorem~\ref{thm:alltopoi} supplies the required small presentation. Since $\Sps$ is terminal among Grothendieck $\infty$-toposes, the slice universal property~\cite{HTT} identifies the space of geometric endomorphisms of $\Sps_{/X}$ with the space of sections of the first projection $X\times X\to X$. The graph of a map $u:X\to X$ corresponds to the geometric morphism with inverse image $u^*$. This correspondence respects composition, so it identifies geometric self-equivalences of $\Sps_{/X}$ with unpointed self-equivalences of $X$. By Theorem~\ref{thm:uf-morita}, the loop space $\Omega_t K$ is thus $\operatorname{Aut}(X)$ under composition, and its delooping $B\operatorname{Aut}(X)$ is the component at $t$.

For $X=K(C_2,2)$, the split fiber sequence of M{\o}ller and Scherer~\cite{MollerScherer} takes the form
\[
 K(C_2,3)\longrightarrow B\operatorname{Aut}(X)
           \longrightarrow B\operatorname{Aut}(C_2),
\]
where the discrete group $\operatorname{Aut}(C_2)$ is trivial. The base is therefore contractible, so the fiber inclusion gives $B\operatorname{Aut}(X)\simeq K(C_2,3)$ and, on looping, $\operatorname{Aut}(X)\simeq K(C_2,2)$.\qed
\end{proof}

Note that ordinary Grothendieck toposes, geometric equivalences, and natural isomorphisms form a weak $2$-groupoid. Its classifying space is a $2$-type, so its third homotopy group vanishes at every point.

\section{A Conservative Translation of Ordinary Geometric Logic}\label{sec:ordinary}
By interpreting primitive sorts as sets and predicates as subobjects, we can express ordinary geometric sequents in the dependent calculus. Their translation identifies the ordinary classifying topos with the $0$-truncated objects of the classifier of the translated theory.

\subsection{Sorts, Predicates, and Images}
For $f:A\to B$, write $\Delta f:A\to A\times_B A$ for its diagonal. Declare a predicate $m:R\to\Gamma$ to be monic by the axiom $\Eq(\Delta m)$. Declare a primitive sort $A$, with terminal map $p_A$, to be a set by
\begin{equation}\label{eq:setaxiom}
\Eq(\Delta^2p_A).
\end{equation}
Indeed, $f$ is $(-1)$-truncated precisely when $\Delta f$ is an equivalence; it is $0$-truncated precisely when $\Delta f$ is $(-1)$-truncated. Function symbols are arrows between finite products. Interpret variables by projections and compound terms by composition with the tuple of their arguments. For terms $t,u:\Gamma\to S$ and a tuple $\bar t:\Gamma\to D$ in the context of a relation $m_R:R\hookrightarrow D$, define the atomic predicates over $\Gamma$ by
\[
 \lbrack\!\lbrack t=u\rbrack\!\rbrack=(t,u)^*(\Delta_S),
 \qquad
 \lbrack\!\lbrack R(\bar t)\rbrack\!\rbrack=\bar t^*(m_R).
\]
Since $S$ is $0$-truncated, its diagonal is monic. Pullback preserves monomorphisms, so both formulas define subobjects. A section of the first subobject asserts equality throughout $\Gamma$. It is represented by a path witness between the two terms, whose space of choices is contractible when inhabited.

Images have the derived expression
\begin{equation}\label{eq:image}
\operatorname{im}(f)=
\colim_{[n]\in\Delta^{\op}} A^{\times_B(n+1)}\longrightarrow B.
\end{equation}

In an $\infty$-topos, this is the image monomorphism, by the effective-epimorphism--monomorphism factorization~\cite{HTT}. To express images in the language, we construct the augmented diagram coherently from its finite limits. Its universal property ensures that naming the diagram preserves the models and their homomorphisms.

\begin{lemma}[Definitional extension by pointwise right Kan extension]\label{lem:kan-definitions}
Let $i:I\hookrightarrow J$ be a full inclusion of small ordinary categories and assume that $N(j\downarrow i)$ is finite for every $j\in J$. Let $a:N(I)\to P$ be a declared diagram in a small presentation. There is a small definitional extension of $P$ containing a literal extension $\bar a:N(J)\to P^{\rm def}$, together with derivable finite-limit comparisons exhibiting it as the pointwise right Kan extension of $a$. The definition is conservative on full model $\infty$-categories and on their parameterized diagrams.
\end{lemma}
\begin{proof}
\noindent\textit{The universal extension category.}\enspace
For a finite-limit quasi-category $\D$, let $\mathscr R_i(\D)$ be the full subcategory of $\Fun(N(J),\D)$ on the right Kan extensions of diagrams on $N(I)$. The pointwise formula is
\begin{equation}\label{eq:finite-ran-formula}
 \bar a(j)\simeq\lim_{(j\to i(k))\in(j\downarrow i)}a(k).
\end{equation}
The limits exist by the finiteness assumption. Right Kan extension is right adjoint to restriction, and, since $i$ is fully faithful, its counit is an equivalence. Therefore restriction gives an equivalence
\[
 \mathscr R_i(\D)\longrightarrow\Fun(N(I),\D)
\]
of whole $\infty$-categories. It is also a categorical fibration. The ambient restriction is one because $N(I)\hookrightarrow N(J)$ is monic. Inner-horn lifts retain their prescribed model vertices, and equivalence lifts remain right Kan extensions, since~\eqref{eq:finite-ran-formula} is invariant under natural equivalence. The proof of Lemma~\ref{lem:mapping-cores} applies. Thus restriction is a trivial categorical fibration; in particular, it has a strict section over each specified input diagram and the full relative lifting property.

The limiting cones in~\eqref{eq:finite-ran-formula} are restrictions of the coherent diagram $\bar a$. Their shapes are the maps
\[
 N(j\downarrow i)^\triangleleft\longrightarrow N(J)
\]
with cone vertex $j$ and the specified arrows $j\to i(k)$. Every required cone certificate can be expressed in the dependent syntax by its finite record and equivalence comparison, as in Lemma~\ref{lem:records}. There is a small set of these comparisons.

\par\smallskip\noindent\textit{Realizing the diagram by constructors.}\enspace
Apply the preceding trivial fibration with $\D=\G_P$ and the given input $a$. The lift is a simplicial map $\bar a:N(J)\to\G_P$ with $\bar a|_{N(I)}=a$. Choose its finite-record comparisons. They are equivalences, so structural recognition and (E) derive their certificates. The diagram, its record data, and these certificate derivations form a small collection. Lemma~\ref{lem:support-realization} extracts their dependency-closed set of primitive and relative rule instances. Performing those instances produces a small definitional extension $P\hookrightarrow P^{\rm def}$ containing exactly this diagram and its derivable comparisons. The cellular realization inherits the face and degeneracy equations of the selected simplicial map.

Every model of $P^{\rm def}$ therefore interprets $\bar a$ as the indicated right Kan extension. Conversely, a model of $P$ extends to $P^{\rm def}$ with a contractible space of choices, including all transformations, by Corollary~\ref{cor:definitions}. The cellular definition thus preserves the $\infty$-categories of models, including all transformations between models. One may also describe the diagram by adjoining its cells and the cone comparisons as a single presentation: its $\infty$-category of models is the pullback of $\mathscr R_i(\D)\to\Fun(N(I),\D)$ along the given input. This presentation and the cellular realization consequently have equivalent $\infty$-categories of models over $\Mod_P(\D)$.

Finally, the same trivial fibration and its pushout-product lifting give the construction for arbitrary parameter shapes with outputs already fixed on a subshape. Pointwise finite limits identify those constructions with the corresponding diagrams of right Kan extensions. The cellular support argument retains the prescribed input and output boundaries and their certificates. This establishes the parameterized assertion.\qed
\end{proof}

\paragraph{The Augmented \v{C}ech Diagram.}
Take $J=\Delta_+^{\op}$ and let $I$ be its full subcategory on $[0]$ and $[-1]$. The input diagram sends the unique nonidentity arrow $[0]\to[-1]$ to the declared map $f:A\to B$. For $n\geq0$, the comma category $([n]\downarrow I)$ has objects
\[
 (0,0),\ldots,(0,n),\quad(-1,*),
\]
and exactly the nonidentity arrows $(0,j)\to(-1,*)$. Indeed, maps $[n]\to[0]$ in $\Delta_+^{\op}$ correspond to the $n+1$ maps $[0]\to[n]$ in $\Delta_+$, while the map $[n]\to[-1]$ is unique. At $n=-1$, the comma category is a point. Every one of these nerves is finite. Lemma~\ref{lem:kan-definitions} therefore constructs a single augmented diagram $C_\bullet(f)$ with
\begin{equation}\label{eq:cech-levels}
 C_{-1}(f)=B,\qquad C_0(f)=A,\qquad
 C_n(f)\simeq A^{\times_B(n+1)}\quad(n\geq1).
\end{equation}
The equalities at $-1$ and $0$ are inherited input equalities, and the remaining comparisons are the certified finite-limit record comparisons.

The record description also gives explicit structure maps. If $\theta:[m]\to[n]$ is monotone, the map
\begin{equation}\label{eq:cech-coordinate-map}
 C(\theta):C_n(f)\longrightarrow C_m(f)
\end{equation}
has $j$th projection equal, with its specified comparison, to the $\theta(j)$th projection of $C_n(f)$, and has the same augmentation to $B$. The finite-limit comparison introduces this map and its projection triangles. In particular, $d_i$ deletes coordinate $i$ and $s_i$ repeats coordinate $i$. For two composable ordinal maps, the two coordinate prescriptions agree. Universal comparison into the terminal record cone supplies the composition $2$-cell. For a chain of ordinal maps, its already specified faces form a boundary comparison into that same terminal cone, and the corresponding higher comparison supplies the next cell. The relative formats retain every previously given face. In the nerve of $\Delta_+^{\op}$, degeneracies insert identity arrows into composable strings. The coordinate description also determines the action of endomorphisms. For example, the constant map $[1]\to[1]$ of value $0$ repeats the zeroth projection of $C_1(f)$. The cells of this one coherent diagram realize the simplicial identities with their specified boundaries.

The positive levels are finite fiber products over $B$, and $\Delta_+^{\op}$ has countably many objects, ordinal maps, and finite composable chains. Thus the array and its finite-limit records are small, and the union of the supports of their chosen derivations is $\U$-small by Lemma~\ref{lem:substitution}. By Lemma~\ref{lem:support-realization}, a single small definitional extension constructs the augmented \v{C}ech nerve together with all its higher coherences.

Apply (HC) to the nonaugmented diagram over its given context $B$. Its new total context is $|C_\bullet(f)|$ and its display is the arrow in~\eqref{eq:image}; the output triangles are the augmentation compatibilities. The image expression is consequently a derived expression with a specified small syntax. In every $\infty$-topos, its interpretation is the image factorization of $f$.

\paragraph{Geometric Connectives and Sequents.}
Images interpret small disjunctions and existential quantification. For predicates over $\Gamma$, define
\begin{align}
\top_\Gamma&=\id_\Gamma,&
\bot_\Gamma&=(0\to\Gamma),\notag\\
\varphi\wedge\psi&=\varphi\times_\Gamma\psi,&
\bigvee_i\varphi_i&=\operatorname{im}\!\left(\coprod_i\varphi_i\to\Gamma\right).\label{eq:ordinary-ops}\\
\intertext{For a map $p:A\to\Gamma$ and a predicate $u:U\hookrightarrow A$, define the existential predicate over $\Gamma$ by}
\exists_p(U)&=\operatorname{im}(p\circ u)\hookrightarrow\Gamma.
&&\notag
\end{align}
An ordinary geometric sequent is translated by
\begin{equation}\label{eq:ordinary-sequent}
\varphi\vdash_\Gamma\psi
\quad\longmapsto\quad
\Eq(\varphi\times_\Gamma\psi\longrightarrow\varphi).
\end{equation}
For subobjects, this condition says that $\varphi$ is contained in $\psi$. Thus an existential conclusion asserts membership in the image of a projection. Adjoining a particular witness extends the theory by a term, as described by Theorem~\ref{thm:sliceclassifier}.

\begin{lemma}[Classical compilation]\label{lem:classical-compilation}
For every small ordinary geometric theory $T$, the constructions above give a small dependent theory $T^\Delta$ with natural equivalences
\begin{equation}\label{eq:classical-models}
\Mod_{T^\Delta}(\X)\simeq N\Mod_T(\X_{\leq0})
\end{equation}
for all $\infty$-toposes $\X$. Here, $\X_{\leq0}$ is the ordinary topos of $0$-truncated objects, and $N$ denotes the nerve functor.
\end{lemma}
\begin{proof}
Declare the primitive sorts, functions, relation arrows, and their finite context products, and impose~\eqref{eq:setaxiom} and the monomorphism conditions. For every formula appearing in an axiom, name the operations~\eqref{eq:ordinary-ops}, using the coherent image construction~\eqref{eq:image}. Impose~\eqref{eq:ordinary-sequent}. The union of the subformula trees of the small axiom set is small; each image definition has countable shape. Thus the compiled presentation is small. A further sequent can be considered by conservatively naming its subexpressions.

The primitive models are precisely many-sorted structures in $\X_{\leq0}$. The named constructions have contractible spaces of interpretations. Their values agree with ordinary conjunction, disjunction, image, and existential quantification in that topos. Induction on formulas proves the object part of~\eqref{eq:classical-models}.

For morphisms, consider a relation $m:R\to\Gamma$ and its target interpretation $m':R'\to\Gamma'$. A map on primitive sorts induces $\eta_\Gamma:\Gamma\to\Gamma'$. The possible relation components and their naturality triangles form the homotopy fiber of
\[
 \Map_\X(R,R')\longrightarrow\Map_\X(R,\Gamma')
\]
at $\eta_\Gamma m$. Since $m'$ is monic, this fiber is either empty or contractible, and it is inhabited precisely when the underlying map preserves the relation. Function-symbol coherence is likewise uniquely determined when it exists, because the relevant targets are $0$-truncated.

The auxiliary products, pullbacks, colimits, and images are small definitional extensions. Their forgetful functors are equivalences of entire model $\infty$-categories by Corollary~\ref{cor:definitions}, so they extend every primitive homomorphism and all its higher comparisons through contractible choice spaces. The translated sequent axioms select a full subcategory of model objects. Mapping spaces on the primitive data are $0$-truncated; the defining extensions preserve them. Restriction therefore identifies the full model $\infty$-category with the nerve of the ordinary category of $T$-models. Geometric inverse images preserve the defining constructions, proving naturality.\qed
\end{proof}

\subsection{The Ordinary Classifying Topos and Conservativity}
Let $\mathcal E_T=\mathbf{Set}[T]$ and choose a small finite-limit geometric syntactic site $(C_T,J_T)$ presenting it. Set
\[
\mathcal E_T^\infty=\operatorname{Sh}(C_T,J_T;\Sps_\U).
\]
This is the canonical one-localic enhancement of $\mathcal E_T$. Its $0$-truncated objects form $\mathcal E_T$, and that ordinary topos determines the enhancement up to equivalence. For every Grothendieck $\infty$-topos $\X$, restricting inverse images to $0$-truncated objects gives an equivalence of $\infty$-categories~\cite{HTT}
\[
\Geom(\X,\mathcal E_T^\infty)
\simeq N\Geom(\X_{\leq0},\mathcal E_T).
\]
The mate correspondence reverses the direction of transformations between right adjoints in the cited theorem, giving our convention for transformations of inverse images.

\begin{theorem}[Recovery of the ordinary correspondence]\label{thm:ordinary}
For every small ordinary geometric theory $T$,
\begin{equation}\label{eq:ordinary-classifier}
\G_{T^\Delta}\simeq\B_{T^\Delta}\simeq\mathcal E_T^\infty,
\qquad (\B_{T^\Delta})_{\leq0}\simeq\mathbf{Set}[T].
\end{equation}
Moreover,
\[
\Geom(\X,\B_{T^\Delta})\simeq N\Mod_T(\X_{\leq0}).
\]
The translation is conservative on ordinary geometric sequents.
\end{theorem}
\begin{proof}
Combining the one-localic universal property with the ordinary classifying-topos theorem~\cite{Johnstone,MakkaiReyes} and Lemma~\ref{lem:classical-compilation} gives $\Geom(\X,\mathcal E_T^\infty)\simeq\Mod_{T^\Delta}(\X)$ naturally in $\X$. Classification identifies this same functor with the one represented by $\B_{T^\Delta}$. Evaluating the two universal models in each other's classifiers gives inverse geometric morphisms. Reconstruction and taking $0$-truncated objects prove~\eqref{eq:ordinary-classifier}.

For conservativity, ordinary completeness identifies derivability of a geometric sequent with validity in every ordinary Grothendieck topos model relative to $\U$. Its translation is valid in every $\infty$-topos model exactly under that condition: one direction passes to $\X_{\leq0}$, and the other uses the canonical one-localic enhancement of each such Grothendieck topos. Apply Corollary~\ref{cor:completeness} to its translated equivalence assertion. This gives
\[
T\vdash_{\rm geom}(\varphi\vdash_\Gamma\psi)
\quad\Longleftrightarrow\quad
T^\Delta\vdash\Eq(\varphi\times_\Gamma\psi\to\varphi).
\tag*{\squareforqed}
\]
\end{proof}

Choose an indexed boundary presentation $P$ of $T^\Delta$. Theorem~\ref{thm:uf-pres} identifies the fiber at $t_P$ with $\B_{T^\Delta}$, so~\eqref{eq:ordinary-classifier} recovers $\mathbf{Set}[T]$ as the category of $0$-truncated objects in that fiber.

A theory freely generated by one closed type has $\X$ as its category of models in $\X$, since interpreting the type amounts to choosing an object. Its classifier is the presheaf $\infty$-category on the free finite-limit category generated by one object. Adding the set-sort axiom gives the ordinary set-sort instance of Theorem~\ref{thm:ordinary}.

\section{Independence of Descent Computation via Circle Monodromy}\label{sec:circle}
For the empty theory, the free finite-limit category is equivalent to the terminal category. Its presheaf classifier is the $\infty$-category of small spaces, so reconstruction gives
\[
\G_\varnothing\simeq\B_\varnothing\simeq\Sps_\U.
\]

The empty theory already expresses the circle and its double cover. Form $2=1+1$ with points $0,1$, and the pushout type
\[
W=1\amalg_2 1.
\]
It has points $N,S$ and paths $\alpha_0,\alpha_1:N=_W S$. Write $\gamma\cdot\delta$ for the path that traverses $\gamma$ first and then $\delta$, and put $\ell=\alpha_0\cdot\alpha_1^{-1}:N=_W N$. Path composition and inverse are derived from identity elimination.

\begin{theorem}[Monodromy and independence of descent]\label{thm:circle}
Gluing two-point fibers over $N$ and $S$ with transitions given by the identity and the transposition produces a double cover of $W$. Transport around $\ell$ exchanges its sheets. This distinguishes $\ell$ from reflexivity by the judgment
\begin{equation}\label{eq:nontrivial-loop}
q:\Id_{\Id_W(N,N)}(\ell,\refl_N)
\ \vdash\ \mathsf{absurd}(q):0.
\end{equation}
The structural and dependent rules, together with homotopy-colimit formation, comparison, and substitution stability, admit a set-valued model in which $\ell$ is reflexivity and $0$ is empty. Comparing the judgment with its interpretation in this model proves the independence of descent computation.
\end{theorem}
\begin{proof}
We construct the double cover by gluing a two-point fiber at $N$ to one at $S$. Its transition maps determine transport along the two paths from $N$ to $S$.

\par\smallskip\noindent\textit{The cartesian input.}\enspace
Coproduct elimination defines $\mathsf{sw}:2\to2$ by $\mathsf{sw}(0)=1$ and $\mathsf{sw}(1)=0$. Applying its comparison rule to the two summands gives $\mathsf{sw}^2\simeq\id_2$. This identifies $\mathsf{sw}$ as an equivalence; structural recognition and (E) supply its coherent inverse data. Put $t_0=\id_2$ and $t_1=\mathsf{sw}$. Over the defining span $1\leftarrow2\to1$, take the total-space span
\begin{equation}\label{eq:twisted-total-span}
 2\xleftarrow{\ h_N\ }2\times2\xrightarrow{\ h_S\ }2,
 \qquad h_N(i,u)=u,\quad h_S(i,u)=t_i(u).
\end{equation}
Choose a map $e:2\to1$ for both endpoint projections. The space $\Map(2,1)$ of such choices is contractible by terminality. At the middle vertex, take the first projection $2\times2\to2$. Rule (T) supplies the comparison triangles into $1$ for the two squares. The square for $h_N$ is the product pullback. The map $(i,u)\mapsto(i,t_i(u))$ is an equivalence over $2$, with the same map as its inverse up to the coproduct computation. It identifies the square for $h_S$ with the product pullback. Thus both squares have derivable $\PB$ certificates, by reindexing comparison and (Tr). This specifies the local input to the derived gluing construction in Theorem~\ref{thm:family-descent}.

\par\smallskip\noindent\textit{The base and total comparison paths.}\enspace
Write $c:2\to W$ for the middle leg of the chosen base cocone. Its two comparison triangles give paths
\[
 \rho_i^N:c(i)=_W N,\qquad \rho_i^S:c(i)=_W S,
 \qquad \alpha_i=(\rho_i^N)^{-1}\cdot\rho_i^S.
\]
These are the paths of the pushout presentation used in the statement. The colimit-prism construction of Theorem~\ref{thm:family-descent}, applied to~\eqref{eq:twisted-total-span}, gives $g:V\to W$. Rule (Desc) certifies its cone squares. Put
\[
 P(w)=\Sigma_{v:V}\Id_W(gv,w),
\]
and let $p:F_g\to W$ be its display. The equivalence $V\simeq F_g$ over $W$ transfers the certified cone squares to this display. The certified cone squares and the explicit two-way comparisons of Appendix~\ref{app:typing} give equivalences
\[
 b_N:2\simeq P(N),\qquad b_S:2\simeq P(S),\qquad
 b_C(i,-):2\simeq P(c(i)).
\]
These are the forward comparisons $c_k$ introduced after~\eqref{eq:glue-rule}, here in the three chosen fiber presentations. Each includes its section triangles, its inverse, and their higher computations. In particular, the transition maps of these presentations are the maps $h_N$ and $h_S$ specified in the input, with their specified comparisons.

Use the fiber-presentation equivalence to pass from $V$ to the dependent total context $\Sigma_{w:W}P(w)$. The two upper cocone triangles are paths in this total context from the image of $(i,u)$ to the images of $u$ and $t_i(u)$ at the endpoints. The mixed faces of the derived colimit prism identify their projections to $W$ with $\rho_i^N$ and $\rho_i^S$, respectively. Proposition~\ref{prop:total-paths}, applied with these fixed projected paths, therefore gives the dependent terms
\begin{align}
 \operatorname{tr}_P(\rho_i^N,b_C(i,u))
   &=b_N(u),\label{eq:circle-half-N}\\
 \operatorname{tr}_P(\rho_i^S,b_C(i,u))
   &=b_S(t_i(u)).\label{eq:circle-half-S}
\end{align}
The $2$-faces of the prism give the total paths, and its $3$-faces compare their projections with the chosen base paths. Identity elimination turns those comparisons into the two transport equations. The higher faces give coherence in $u$, equivalently the naturality of the local $\PB$ comparisons.

\par\smallskip\noindent\textit{The monodromy equation.}\enspace
The inverse and concatenation laws for transport, derived by path induction in Appendix~\ref{app:typing}, turn~\eqref{eq:circle-half-N}--\eqref{eq:circle-half-S} into
\begin{equation}\label{eq:circle-transport-conjugated}
 \operatorname{tr}_P(\alpha_i,b_N(u))
       =b_S(t_i(u)).
\end{equation}
Indeed, inverse transport along $\rho_i^N$ first takes $b_N(u)$ to $b_C(i,u)$ by~\eqref{eq:circle-half-N}, and transport along $\rho_i^S$ then gives~\eqref{eq:circle-half-S}. Thus the two transport maps, expressed in the fixed fiber presentations, are $t_0$ and $t_1$. The equivalences $b_N$ and $b_S$ identify these computations with transport in the glued family.

For $\ell=\alpha_0\cdot\alpha_1^{-1}$, apply~\eqref{eq:circle-transport-conjugated} first with $i=0$, then use its inverse form for $i=1$. The resulting computation is
\begin{equation}\label{eq:circle-monodromy-computation}
 \operatorname{tr}_P(\ell,b_N(0))
    =b_N(t_1^{-1}t_0(0))=b_N(1).
\end{equation}
A term $q:\Id_{\Id_W(N,N)}(\ell,\refl_N)$ yields, by identity elimination, an identification of the left side with transport along $\refl_N$, whose value is $b_N(0)$. Composing with~\eqref{eq:circle-monodromy-computation} and applying an inverse of $b_N$ gives a term of $\Id_2(0,1)$. The transport laws make this construction coherent under substitution and give its higher computations.

\par\smallskip\noindent\textit{Distinguishing the loop from reflexivity.}\enspace
Apply gluing to the discrete two-point diagram with local families $1$ and $0$. Its output is a family $H$ over $2$ with local equivalences $H(0)\simeq1$ and $H(1)\simeq0$. Choose the term of $H(0)$ corresponding to the unit term. Transport it along the derived path $0=_2 1$, and then apply $H(1)\simeq0$. The output is the term $\mathsf{absurd}(q)$ in~\eqref{eq:nontrivial-loop}. The total context of $H$ is equivalent over $1+1$ to $1+0$. (Stab) certifies the resulting upper coproduct cocone.

\par\smallskip\noindent\textit{The set-valued model and independence.}\enspace
Interpret the structural and dependent rules, (HC), (HC-comp), and (Stab) in $\D=N\mathbf{Set}_\U$, with all arrows as displays, ordinary finite limits and colimits, and identity formation given by the diagonal with identity reflexivity. A diagram $K\to N\mathbf{Set}_\U$ is equivalently a functor $hK\to\mathbf{Set}_\U$. The category $hK$ has the vertices of $K$ as objects and its edges as generating arrows, with degenerate edges acting as identities and each $2$-simplex imposing $e_{02}=e_{12}\circ e_{01}$ on its edges. The cone rules express the ordinary universal properties. Colimits of sets are stable under pullback, so (Stab), including its empty instance, holds.

For each retained primitive rule $r:A\hookrightarrow B$, the calculation for $r$ in the proof of Lemma~\ref{lem:rule-extension} gives contractible extension fibers in $\D$. It applies also in $\Fun(\Delta^1,\D)$, where limits and colimits are computed pointwise and isomorphisms are detected pointwise. By Lemma~\ref{lem:mapping-cores}, restriction $\mathscr M_B(\D)\to\mathscr M_A(\D)$ is a categorical fibration. The two fiber calculations give the $m=0,1$ equivalences on cores in~\eqref{eq:rezk-levels}. The mapping-space argument in the proof of Lemma~\ref{lem:relative-extension} therefore makes this restriction a trivial categorical fibration. Its lifting property against every small monomorphism $J_0\hookrightarrow I$ gives the relative rule~\eqref{eq:relative} with the prescribed outputs on $J_0$.

In this model, $W=1\amalg_2 1$ is a singleton and both paths are reflexivity. The hypothesis of~\eqref{eq:nontrivial-loop} is therefore inhabited, and the conclusion is the empty set. By soundness, this model witnesses independence. The pushouts describe the geometry of the calculation as well. In sets, the identifications in~\eqref{eq:twisted-total-span} join all four endpoint elements into one equivalence class. In spaces, the base pushout is a circle, and descent glues the two-point fibers into its double cover. Composing the two transport maps in~\eqref{eq:circle-transport-conjugated} gives the transposition around $\ell$.\qed
\end{proof}

\section{Conclusion}
The contexts and substitutions of a small geometric type theory present its classifying $\infty$-topos. In this presentation, the section space of a family represents the terms of the corresponding type, the paths between terms, and their higher identifications. When a finite telescope of variables is adjoined to the theory, the new classifier is the slice over the telescope's total context.

Descent gives this correspondence computational content. A cartesian family on a diagram glues by taking the colimit of its total spaces and then the fibers over the base colimit. The computation rules recover the local families with their prescribed transport. On the circle, transport along the first constructor path fixes the fiber labels, and transport back along the second exchanges them. The resulting transposition is the monodromy of the double cover. The same rules identify gluing and restriction as inverse equivalences of family categories for every small colimit. Consequently, introducing gluing as a primitive rule gives the same families and term spaces as deriving it from total spaces.

Finite homotopy records translate finite-limit axioms into dependent types, and every Grothendieck $\infty$-topos has a small presentation in the resulting language. Under the duality, an interpretation acts as the inverse image of a geometric morphism, and transformations between interpretations give the corresponding transformations of inverse images. For an ordinary geometric theory, this correspondence recovers both the classical classifying topos, as the $0$-truncated objects of the one-localic classifier, and the ordinary homomorphisms between models.

Applying the duality to indexed boundary presentations and to presentations equipped with a model gives a single geometric morphism with each small theory's classifier as a fiber. Over the Sierpi\'nski locale, activating a single equivalence axiom changes that fiber from $\Sps_\U\times\Sps_\U$ to $\Sps_\U$. Restricting the duality to Morita equivalences gives a family with coherent transport. The Morita components of the topos of $C_2$-spaces and of the slice $\Sps_{/K(C_2,2)}$ have homotopy types $K(C_2,2)$ and $K(C_2,3)$, respectively (Corollaries~\ref{cor:uf-c2} and~\ref{cor:uf-morita-pi3}).

\appendix
\section{Deriving Path Induction, Dependent Colimit Elimination, and Gluing}\label{app:typing}
In the generated syntax, the faces of a constructor encode its projections, section equations, and higher computations. We use these faces to derive substitution, path induction, dependent colimit elimination, and the local computations for gluing. Working in a slice gives the same operations over a specified context.

\subsection{Realizing Derived Proof Packets}
The designated diagrams of a proof packet state its typing and universal-comparison judgments. A relative packet also specifies a parameter shape and the subshape on which output cells have already been supplied.

\begin{lemma}[A criterion for realizing proof packets]\label{lem:admissible-packets}
Let $A\hookrightarrow B$ be a small presentation inclusion whose restriction on the $\infty$-categories of models of the marked presentations is a trivial categorical fibration for every geometric universe. Every derivable $A$-array in $\G_\T$ is the restriction of a $B$-array of syntactic cells and derivable certificates. For inputs varying over a small parameter shape, the extension agrees with the prescribed outputs on the given subshape. A small collection of these outputs can be named by a small definitional extension. The same statement applies in a prescribed contextual slice, using its slice certificate interpretations.
\end{lemma}
\begin{proof}
Take $\D=\G_\T$. A derivable input is a vertex of the target of the restriction fibration, so a vertex lift is a simplicial map $B\to\G_\T$ restricting to the input. Every value of that map is a syntactic cell. Every certificate it preserves is derivable, by recognition of the certificate sets in $\G_\T$. For a monomorphism $J_0\hookrightarrow I$, the adjoint lifting problem is~\eqref{eq:double-pushout-product}. This gives the relative assertion with exactly the prescribed cells on $B\times J_0$. The outputs and their certificate witnesses form a small set, so Lemma~\ref{lem:support-realization} names them together with the constructors that produce them.

For the slice over $\Gamma$, apply the same argument in the geometric universe $(\G_\T)_{/\Gamma}$. A slice cell is a cell on its shape joined with the final vertex $\Gamma$. The output therefore retains that final vertex. Pullback and colimit certificates are represented on their total diagrams, an equivalence certificate is represented by its total equivalence, and a terminal slice object $q:X\to\Gamma$ is represented by $\Eq(q)$. The unit display represents $\id_\Gamma$. These translations give derivable certificates in the original syntax, as in Proposition~\ref{prop:contextual-elaboration}. Their supports are small, so the same realization argument applies.\qed
\end{proof}

\begin{lemma}[Composition and cancellation packets]\label{lem:composition-cancellation}
An inner composition horn and the marked outer horn
\begin{equation}\label{eq:equivalence-outer-horn}
 \langle\Lambda^2_0\mid\Eq(01)\rangle
 \hookrightarrow\langle\Delta^2\mid\Eq(01)\rangle
\end{equation}
admit relative extensions by Lemma~\ref{lem:admissible-packets}. In particular, given $u:x\xrightarrow{\simeq}y$ and $d:x\to z$, the latter packet produces $j:y\to z$ and a triangle $ju\simeq d$, with all prescribed higher boundaries.
\end{lemma}
\begin{proof}
The inner composition packet is (H). For~\eqref{eq:equivalence-outer-horn}, restriction is a categorical fibration by the marked lifting argument. Fix $x,y,z,u,d$. Its homotopy fiber is the homotopy fiber, over $d$, of precomposition
\[
 \Map_\D(y,z)\xrightarrow{u^*}\Map_\D(x,z).
\]
Composition with an inverse of $u$ is an inverse to $u^*$, with inverse homotopies supplied by the two inverse triangles and their coherence. Thus $u^*$ is an equivalence, and its homotopy fibers are contractible. Equivalently, compose $d$ with $u^{-1}$ using an inner horn and use $u^{-1}u\simeq\id_x$ to obtain the displayed triangle. Applying the same calculation in every $\Fun(\Delta^m,\D)$ and using the $m=0,1$ argument of Lemma~\ref{lem:relative-extension} proves that the marked restriction is a trivial categorical fibration. This description applies in any quasi-category. Lemma~\ref{lem:admissible-packets} gives its syntactic and contextual realizations, and the pushout-product calculation gives all relative instances.\qed
\end{proof}

The defining pushout keeps the input diagram, the chosen outputs on its boundary, and their computation witnesses together. Equations~\eqref{eq:output1}--\eqref{eq:output2} identify the faces of each new output with the prescribed cells. We use this description first to compute reindexing.

\subsection{Reindexing, Contextual Substitution, and Dependent Sums}
Let $p:E\to\Gamma$ be the display of a type $A$ and let $f:\Delta\to\Gamma$. The (Reind) formation packet produces
\begin{equation}\label{eq:appendix-reindex}
\begin{tikzcd}
 E_f\ar[r,"\bar f"]\ar[d,"p_{A[f]}"']&E\ar[d,"p"]\\
 \Delta\ar[r,"f"']&\Gamma
\end{tikzcd}
\end{equation}
with $\Disp(p_{A[f]})$ and $\PB$. The display $p_{A[f]}:E_f\to\Delta$ represents the reindexed type $A[f]$, whose total context is $E_f$. Given another cone $c$ on this cospan, place $c$ and the certified cone $t$ at the two vertices of $\partial\Delta^1\star V$, with $t$ last. The $n=1$ comparison (Reind) produces the mediating edge and all three compatibility triangles. Two such comparisons with a compatible triangular boundary are compared by the $n=2$ rule, and the rule in dimension $n$ supplies each subsequent boundary comparison.

Suppose $a$ is a term of $p$, represented by $s:\Gamma\to E$ and its section triangle $ps\simeq\id_\Gamma$. An (H) instance forms $u=sf$. Whiskering the section triangle gives $pu\simeq f$. The cone with vertex $\Delta$, legs $u$ and $\id_\Delta$, and this comparison is an explicit input to the preceding $n=1$ rule. Its output contains
\[
 s_f:\Delta\to E_f,\qquad p_{A[f]} s_f\simeq\id_\Delta,
 \qquad \bar f s_f\simeq sf,
\]
with the comparison of these triangles to the original section computation. The first two entries are the term $a[f]$, and the third is its reindexing computation. The coherence cells are part of the cone comparison, so the section triangle has exactly the term boundary~\eqref{eq:term-triangle}.

For the variable, set $f=p$ and use the identity cone with vertex $E$, both legs $\id_E$, and common composite $p$. Its comparison into the reindexing cone gives
\[
 v:E\longrightarrow E\times_\Gamma E,
 \qquad \pr_2v\simeq\id_E.
\]
This is the variable term in the extended context. The remaining projection computation identifies its value in the original family with the identity substitution of $E$.

For $g:\Theta\to\Delta$, the pasted reindexing squares for $p$, $f$, and $g$ are a pullback cone on $E\to\Gamma\leftarrow\Theta$. The directly formed square for $fg$ is another such cone. Comparing them in both directions by (Reind) yields
\begin{equation}\label{eq:substitution-comparison}
 A[f][g]\simeq A[fg].
\end{equation}
The comparison respects both the total-context projections and their triangles. The identity case compares the pullback along $\id_\Gamma$ with the identity square and gives $A[\id_\Gamma]\simeq A$.

Let $E\to\Gamma_0$ be a display and let $h:\Gamma_r\to\Gamma_0$ be a chosen composite of a string $\Gamma_r\to\cdots\to\Gamma_0$ of three or more substitutions. By pullback pasting, every parenthesization of the iterated pullback presents a limiting cone on the cospan $E\to\Gamma_0\xleftarrow{h}\Gamma_r$. The comparison $2$-cells, including the associativity comparison, are the boundaries between maps into this limiting cone. Their next boundaries include the pentagon and the unit compatibility. The corresponding comparison formats fill those boundaries. More explicitly, apply the relative forms of (Reind) to $\partial\Delta^n\hookrightarrow\Delta^n$ in the category of the relevant cone diagrams. At each step, the input is the common base diagram with its previously given faces, and the output is its universal comparison. Lemmas~\ref{lem:cone-fibers} and~\ref{lem:admissible-packets} prove the extension with those faces fixed.

A composable string of contextual substitutions gives a parameter diagram of the input family and its reindexing problems. The functor category in Lemma~\ref{lem:pointwise-models} retains every map of that diagram. Applying (Reind) with these parameters constructs all the reindexed displays and their comparison triangles. On a subdiagram with prescribed displays and triangles, the pushout product~\eqref{eq:relative} makes those cells the boundary values of the construction. Thus $I=\Delta^n$ records composable strings and their higher composition data, while an arbitrary small $I$ permits a whole diagram of substitutions at once. This gives naturality in every contextual substitution. On the derivation trees for these comparisons, substitution of generator names acts by the judgmentally functorial recursion of Lemma~\ref{lem:substitution}.

Reindexing also gives the pairing and projections for a dependent sum. For $\Disp(p)$ and $\Disp(q)$, with $F\xrightarrow{q}E\xrightarrow{p}\Gamma$, (Sum) forms a triangle with composite $h:F\to\Gamma$ and $\Disp(h)$. The sum type has total context $F$. A section $a$ of $p$ and a section $b$ of $q[a]$ give, by~\eqref{eq:appendix-reindex}, a map $t:\Gamma\to F$ and a triangle $qt\simeq a$. Composing this triangle with the section triangle of $a$ and with the (Sum) triangle produces $ht\simeq\id_\Gamma$. An inner-horn filling supplies the common comparison simplex. Thus $(t,ht\simeq\id)$ is the pair $(a,b)$.

Conversely, a section $t$ of $h$ gives a section $qt$ of $p$. The cone consisting of $t$, $\id_\Gamma$, and the chosen composite comparison gives a section of the pullback $q[qt]$. These are the two projections. Pairing followed by projection and projection followed by pairing are compared in the appropriate reindexing cone categories. Their $n=1$ universal comparisons give the maps; their $n=2$ comparison cells give the inverse computations; and the higher rules supply the compatible further cells. The mapping-space calculation in Proposition~\ref{prop:typed-term-spaces} identifies precisely the fiberwise content of these operations.

The contexts $\Gamma,z:\Sigma_{x:A}B(x)$ and $\Gamma,x:A,y:B(x)$ have the same total object $F$ and their projections are related by the chosen (Sum) triangle. Consequently a motive on either is a family on that total context, and its sections give the dependent sum eliminator. Reindexing along pairing yields its computation term. This is a passage through the specified total contexts and their comparison, so relative (Sum) and (Reind) packets retain arbitrary motive and term boundaries.

\subsection{Path Induction and Its Computation Rule}
For $p:E\to\Gamma$, form $Q=E\times_\Gamma E$ by reindexing. The quotient~\eqref{eq:identity-packet} identifies its first cone with the identity cone on $p$ and keeps the second cone certified as a pullback. Its comparison edge is the relative diagonal $\delta:E\to Q$. The (Id) attachment yields
\[
 E\xrightarrow{r\;\simeq}I\xrightarrow{q}Q,
 \qquad qr\simeq\delta,
 \qquad\Disp(q).
\]
The maps from $I$ to the two copies of $E$ give its two endpoint variables; the map $r$ is reflexivity. Inverse data for $r$ are generated by (E).

Let $c:C\to I$ be the display of a motive and let $d:E\to C$, with a specified triangle $\vartheta:cd\simeq r$, be the term over reflexivity. Work in the slice $\mathcal E=(\G_\T)_{/I}$. Its objects
\[
 a=(E\xrightarrow{r}I),\qquad
 b=(I\xrightarrow{\id}I),\qquad
 e=(C\xrightarrow{c}I)
\]
have an equivalence $u:a\to b$ with underlying edge $r$, and the triangle $\vartheta$ is a morphism $d:a\to e$. They form the outer horn~\eqref{eq:equivalence-outer-horn} in $\mathcal E$. Its extension supplies
\begin{equation}\label{eq:J-output}
 j:b\longrightarrow e,\qquad j u\simeq d
 \quad\text{in }\mathcal E.
\end{equation}
This is the path eliminator $J_C$ and its computation over reflexivity.

We spell out the raw cell boundaries. A $2$-simplex in this join-defined slice is a $3$-cell with vertices
\[
 (E,I,C,I).
\]
The fixed edges are $01=r$, $02=d$, $03=r$, $13=\id_I$, and $23=c$. Its fixed face $013$ is the degenerate triangle on $r$, and its fixed face $023$ is $\vartheta$. The new edge $12=j$ and new faces are
\begin{equation}\label{eq:J-tetrahedron}
 123:\ cj\simeq\id_I,
 \qquad
 012:\ jr\simeq d.
\end{equation}
The full tetrahedron states the compatibility of the computation $012$ with the section witness $123$ and the given local witness $023$. The two faces therefore give a section term and its computation at reflexivity, and the tetrahedron witnesses their compatibility. Proposition~\ref{prop:typed-term-spaces} converts the latter section-space path to the identity-type computation in the displayed type-theoretic judgment.

There is also a concrete description using the inverse strings of (E). Write $g:I\to E$ for the inverse edge, $\eta:gr\simeq\id_E$ for the string $010$, and $\epsilon:rg\simeq\id_I$ for $101$. One can take $j$ to be the composite $dg$. Its section witness is the pasting
\[
 cdg\xRightarrow{\vartheta g}rg
     \xRightarrow{\epsilon}\id_I.
\]
Its computation is the pasting $dgr\xRightarrow{d\eta}d$, together with the associativity comparisons. The string $0101$ supplies $r\eta\simeq\epsilon r$, which is exactly the compatibility needed in~\eqref{eq:J-tetrahedron}. Longer strings and the relative composition packets give the higher compatibilities of these choices. This inverse description and the outer-horn description lie in the same contractible extension fiber.

For a parameterized motive and local term over a shape $H$, with eliminator outputs already prescribed over $H_0\hookrightarrow H$, apply the relative version of~\eqref{eq:equivalence-outer-horn} in the slice. It produces the entire family~\eqref{eq:J-output} retaining those outputs. Taking $H_0=\partial\Delta^n$ gives each higher computation with its given lower-dimensional boundary. Pulling the whole packet back along a contextual substitution gives the corresponding path-induction packet in that context. Reindexing preserves the reflexivity equivalence and its relative comparison, so the same lifting property gives the substitution computation and its higher compatibility. This establishes the full dependent elimination and computation scheme for the identity formation rule.

\subsection{Transport and Identities in a Total Type}\label{app:transport}
The path-induction packet defines transport as the judgment
\[
 x:A,\ u:P(x),\ y:A,\ \gamma:\Id_A(x,y)
 \ \vdash\ \operatorname{tr}_P(\gamma,u):P(y).
\]
Its reflexivity datum is the variable $u$. To retain the parameter $u$, pull the reflexivity equivalence back along the family on the first endpoint before applying~\eqref{eq:J-output}. Its computation at reflexivity is therefore a term $\operatorname{tr}_P(\refl_x,u)=u$. The same construction defines path inversion, concatenation, and the action of a map on paths. We use the convention that $\gamma\cdot\delta$ traverses $\gamma$ first and then $\delta$. Identity elimination on the path variables gives the computations
\begin{align}
 \operatorname{tr}_P(\gamma\cdot\delta,u)
  &=\operatorname{tr}_P(\delta,\operatorname{tr}_P(\gamma,u)),\label{eq:transport-composition}\\
 \operatorname{tr}_P(\gamma^{-1},\operatorname{tr}_P(\gamma,u))&=u,\label{eq:transport-inverse}\\
 h_y(\operatorname{tr}_P(\gamma,u))
  &=\operatorname{tr}_Q(\gamma,h_x(u))\label{eq:transport-naturality}
\end{align}
for a family map $h:P\to Q$ over $A$. Each right-hand side is a term in the same context as the left-hand side. At reflexivity, the required identities are the already derived unit computations. The path-induction packet extends those identities to the stated path contexts; its relative form gives the coherent comparisons between these computations.

\begin{proposition}[Paths in a total type]\label{prop:total-paths}
For a family $x:A\vdash P(x)\ \type$ and terms $u:P(x)$, $v:P(y)$, there is a derived equivalence
\begin{equation}\label{eq:total-paths}
\begin{split}
 &\Id_{\Sigma_{a:A}P(a)}((x,u),(y,v))\\
 &\hspace{8mm}\simeq
 \Sigma_{\gamma:\Id_A(x,y)}
 \Id_{P(y)}(\operatorname{tr}_P(\gamma,u),v).
\end{split}
\end{equation}
Its first projection is the action of the total-space projection on paths. The equivalence and its two inverse computations are natural in parameters and compatible with reindexing.
\end{proposition}
\begin{proof}
Write $\beta_u:\Id_{P(x)}(\operatorname{tr}_P(\refl_x,u),u)$ for the reflexivity computation of transport. The pair $(\refl_x,\beta_u)$ gives the datum for identity elimination on a total path $z$ at $z=\refl_{(x,u)}$. Extending this datum by path induction gives the forward map.

For the reverse map, first eliminate $\gamma$. At $\gamma=\refl_x$, the fiber datum is a path $h:\operatorname{tr}_P(\refl_x,u)=v$. Applying $w\mapsto(x,w)$ to the path $\beta_u^{-1}\cdot h:u=v$ gives the required total path. At $h=\beta_u$, inverse cancellation and the action of pairing on reflexivity identify this total path with $\refl_{(x,u)}$. Together with the computation of the forward map, this proves the inverse law on total paths by induction on $z$. For the inverse law on pairs, eliminate $\gamma$ and use the equivalence $q\mapsto\beta_u\cdot q$ from paths $q:u=v$ to paths $h:\operatorname{tr}_P(\refl_x,u)=v$. Path induction on $q$ reduces the claim to $q=\refl_u$, where the right unit law identifies $h$ with $\beta_u$ and the two computations just given recover $(\refl_x,\beta_u)$. Relative identity elimination gives the higher compatibility of the inverse laws, and the contextual path-induction packet gives the stated parameter and substitution naturality.\qed
\end{proof}

A commuting prism between a total-space cocone and a base cocone supplies a total path with its specified projected base path. Under~\eqref{eq:total-paths}, its remaining component is exactly a transport computation. The prism's next-dimensional cells supply the compatibility of that computation with the projected path. This is the derivation used in the monodromy calculation.

\subsection{Dependent Elimination for Homotopy Colimits with a Fixed Motive}
Let $d:K\to\G_\T$ have a certified constructor cocone
$c:K\star\Delta^0\to\G_\T$ with apex $W$, and let
$p:E\to W$ be the specified display.  A coherent collection
of local section triangles is an array
\[
 b^\sharp:K\star\Delta^{\{1,2\}}\longrightarrow\G_\T,
 \qquad b^\sharp(1)\equiv E,\quad b^\sharp(2)\equiv W,
 \qquad b^\sharp|_{12}\equiv p,
 \quad b^\sharp|_{K\star\{2\}}\equiv c.
\]
Its restriction to $K\star\{1\}$ is the local cocone $b$ with apex
$E$.  For a vertex $k$ of $K$, the triangle of $b^\sharp$ on $k12$ gives the supplied
computation $\theta_k:p b_k\simeq\iota_k$.  The higher simplices of $b^\sharp$
record the prescribed coherences of these triangles.  Local sections
of the pullback displays $\iota_k^*p$ have precisely this form after
composition with their pullback projections, with the pullback squares
providing the displayed triangles.

\begin{proposition}[Rule-level dependent colimit elimination]
\label{prop:hc-derived}
From $c$, $p$, and $b^\sharp$, the rules produce a term
$\mathsf{ind}_{c,p}(b^\sharp):p$ over $W$, together with its complete local
computation array.  The construction consists of an instance of
\textup{(HC-comp)} in dimension $1$, followed by an instance in
dimension $2$.  Its boundary restricts to the given $p$, $c$, and $b^\sharp$.
Every prescribed higher boundary of complete elimination packets
extends by two higher comparison instances.  For every small
monomorphism $J_0\hookrightarrow I$, the same construction applies
to $I$-parameterized inputs relative to elimination packets already
specified on $J_0$.  It also admits a direct rule-level realization
in a fixed outer context.
\end{proposition}
\begin{proof}
\noindent\textit{The input horn.}\enspace
Repeat the apex of $c$ by the simplicial collapse
$\Delta^{\{0,2\}}\to\Delta^0$.  Denote the resulting degenerate constructor
strip by $c^{\mathrm{deg}}:K\star\Delta^{\{0,2\}}\to\G_\T$.  Its edge $02$ is
$\id_W$.  It agrees with $b^\sharp$ on $K\star\{2\}$.
The two strips therefore define an array on
\[
 S_K=K\star\Lambda^2_2
  =\bigl(K\star\Delta^{\{0,2\}}\bigr)
    \cup_{K\star\{2\}}
    \bigl(K\star\Delta^{\{1,2\}}\bigr).
\]
The cocone at vertex $0$ is $c$ and carries its given $\HC_K$
certificate.  The vertices $0,1,2$ are respectively $W,E,W$.

\par\smallskip\noindent\textit{The underlying section and its local comparisons.}\enspace
Apply \textup{(HC-comp)} in dimension $1$ to the two cocones on
$K\star\{0\}$ and $K\star\{1\}$.  This adjoins the face
$K\star\Delta^{\{0,1\}}$.  Its edge $01$ is a substitution
$s:W\to E$, and its triangle on $k01$ is
$\xi_k:s\iota_k\simeq b_k$.  All higher simplices of this face
are supplied by the same rule instance.  The enlarged array has
shape $K\star\partial\Delta^2$.

\par\smallskip\noindent\textit{The dependent section triangle.}\enspace
Apply \textup{(HC-comp)} in dimension $2$ to this complete boundary,
distinguishing its first-vertex cocone.  It gives an array
$\Phi:K\star\Delta^2\to\G_\T$.  The triangle
$\tau:=\Phi|_{012}$ satisfies the judgmental boundary equations
\[
 d_0\tau\equiv p,\qquad d_1\tau\equiv\id_W,
 \qquad d_2\tau\equiv s.
\]
Thus $\tau$ is a term of the fixed display $p$ by
\eqref{eq:term-triangle}.  We name this term
$\mathsf{ind}_{c,p}(b^\sharp)$.  For every $k\in K_0$, the tetrahedron
$\Phi|_{k012}$ has the four specified faces
\[
 d_0(\Phi|_{k012})\equiv \tau,\qquad
 d_1(\Phi|_{k012})\equiv \theta_k,\qquad
 d_2(\Phi|_{k012})\equiv c^{\mathrm{deg}}|_{k02},\qquad
 d_3(\Phi|_{k012})\equiv \xi_k.
\]
It is the compatibility between the section equation $ps\simeq\id_W$
and the local computation $s\iota_k\simeq b_k$ over $W$.
For a $q$-simplex $\sigma$ of $K$, the cell
$\Phi|_{\sigma\star012}$ has dimension $q+3$ and supplies the
corresponding higher compatibility.  Its face on
$\sigma\star12$ is exactly the original cell of $b^\sharp$.
All the face and degeneracy equations are the constructor equations
\eqref{eq:output1}--\eqref{eq:simplicial-eq}.

Equivalently, the inclusion of this entire dependent elimination
packet has the explicit cellular factorization
\begin{equation}\label{eq:hc-elimination-factorization}
 K\star\Lambda^2_2
 \lhook\joinrel\longrightarrow K\star\partial\Delta^2
 \lhook\joinrel\longrightarrow K\star\Delta^2.
\end{equation}
The first arrow is a pushout of the dimension-$1$ comparison on
face $01$; the second is the dimension-$2$ comparison itself.
Both pushouts inherit every certificate on the input, including
$\Disp(p)$ and the given constructor certificate.  This also
describes the construction for $K=\varnothing$.

\par\smallskip\noindent\textit{Higher term boundaries.}\enspace
The higher computations can be read directly in the term convention
\eqref{eq:termspace}.  Let $\mathsf{Elim}(c,p,b^\sharp)_n$ consist of arrays
on $K\star\Delta^{n+2}$ whose apex vertices
$0,\ldots,n,n+2$ are $W$, whose vertex $n+1$ is $E$, and whose
restrictions satisfy the following two equations.  On
$K\star\Delta^{\{0,\ldots,n,n+2\}}$, the array is the degenerate
extension of $c$ obtained by collapsing all its apex vertices.
On $K\star\Delta^{\{n+1,n+2\}}$, it is the given array
$b^\sharp$.  Faces and degeneracies act on the first block
$0,\ldots,n$.  Restriction to the apex simplex gives an
$n$-dimensional term of $p$, with its complete local computations.

A compatible boundary in dimension $n\geq1$ prescribes the faces
obtained by deleting one of $0,\ldots,n$.  These faces and the fixed
all-$W$ face form the union
$K\star\Lambda^{n+2}_{n+2}$.  The fixed local face is contained in
this union.  In dimension $0$, the same source is $S_K$: its two
faces are the fixed local strip and the degenerate constructor
strip.  In every dimension, attach the missing face
$K\star\Delta^{\{0,\ldots,n+1\}}$ by
\textup{(HC-comp)} in dimension $n+1$, and then attach the full
simplex by \textup{(HC-comp)} in dimension $n+2$.  The distinguished
first cocone remains $c$ throughout.  This produces the required
$n$-simplex while retaining every prescribed face.  Hence the
simplicial set $\mathsf{Elim}(c,p,b^\sharp)$ has the right lifting property
against every boundary inclusion $\partial\Delta^n\hookrightarrow
\Delta^n$: it is a contractible Kan complex of explicit elimination
packets.  The argument gives both the computation cells and their
higher comparisons as rule applications.

\par\smallskip\noindent\textit{Relative parameters.}\enspace
Put $T_K=K\star\Delta^2$.  With inputs varying over $I$ and complete
outputs prescribed on $J_0$, the starting array has shape
\[
 (S_K\times I)\cup_{S_K\times J_0}(T_K\times J_0).
\]
Apply the relative dimension-$1$ comparison to add
$K\star\Delta^{\{0,1\}}$ over $I$, using its already prescribed
restriction over $J_0$.  The resulting array has shape
\[
 ((K\star\partial\Delta^2)\times I)
   \cup_{(K\star\partial\Delta^2)\times J_0}
          (T_K\times J_0).
\]
The relative dimension-$2$ comparison then extends it to
$T_K\times I$.  These are precisely the two pushout products of
the comparisons in~\eqref{eq:hc-elimination-factorization} with
$J_0\hookrightarrow I$, with vertexwise certificates.  Thus the
input over $I$ and the entire output over $J_0$ are retained by
the boundary equations of the two rules.  The same two-step
argument, in dimensions $n+1$ and $n+2$, gives the relative form
of every higher term comparison.

\par\smallskip\noindent\textit{A fixed outer context.}\enspace
Suppose the data are already given over $\Gamma$.  Adjoin a final
vertex $3$ carrying $\Gamma$.  The local input and the degenerate
constructor strip specify the two faces $123$ and $023$ of the
apex $3$-simplex, including all their $K$-indexed cells.  Their
union is $K\star(\Lambda^2_2\star\Delta^0)$.  Adjoin, in order,
the faces $01$, $012$, $013$, and $0123$.  The four attachments
are instances of \textup{(HC-comp)} in dimensions $1,2,2,3$,
respectively; at each step, the boundary of the indicated face
has already been specified.  Each instance distinguishes the
same certified cocone at vertex $0$.  The resulting array keeps
$\Gamma$, the motive, every context substitution, and their
given coherence cells fixed.  Its face $012$ is the section
triangle, its face $013$ is the section comparison over
$\Gamma$, and its $3$-simplex makes these comparisons compatible.
The relative forms of these four instances retain any prescribed
parameter boundary as well.\qed
\end{proof}

The dimension-$1$ and dimension-$2$ comparisons assemble the eliminator and its computation
on the shape
\begin{equation}\label{eq:dependent-HC-shape}
 (K\star\Delta^1)\star\Delta^0=K\star\Delta^2,
\end{equation}
whose three final vertices represent $W,E,W$.
Its section and computation witnesses are the constructor outputs
specified in the proof.

The categorical interpretation identifies this explicit derivation
with the dependent universal property.  In the slice over $W$, put
$d^W_k=(d_k\xrightarrow{\iota_k}W)$.  The cocone with apex
$\id_W$ is a colimit in that slice: slice colimits are computed on
total objects, and its total cocone is the given $c$.  The array
$b^\sharp$ is a cocone on $d^W$ with apex $p$.  The first comparison
in Proposition~\ref{prop:hc-derived} constructs its underlying map,
and the second constructs its section triangle and computation
compatibilities in the slice.  Consequently the entire output is
the array $K\star\Delta^1\to(\G_\T)_{/W}$ that realizes dependent
elimination.  In mapping spaces, the universal property is
\begin{equation}\label{eq:dependent-HC-sections}
 \Map_{\G_\T/W}(\id_W,p)
 \simeq\lim_{k\in K^{\op}}
           \Map_{\G_\T/W}(d_k\to W,p).
\end{equation}
The terms on the right are sections of the pullback motives, with
their coherent transition comparisons.  The raw array construction
realizes the inverse of this equivalence together with its
computation witnesses.  Rule \textup{(Stab)} certifies the reindexed
constructor diagram, and the same rule-level derivation applies
with its pullback comparisons.  The relative comparison packets
provide compatibility with reindexing and all higher substitution
comparisons.

\subsection{Deriving the Gluing Computations from Pullback Certificates}\label{app:glue-computation}
Fix a small diagram $d:K\to\G_\T$ with colimit cocone $\bar d:d\to W$, and let $P$ be the local family represented by the displays $p_k:E_k\to d_k$ and their certified cartesian transition squares. Lemma~\ref{lem:relative-closed-colimit}, with $I=\Delta^1$ and $J_0=\{1\}$, extends exactly the union of the local prism and the chosen bottom cocone. It produces an upper colimit $E\to V$, an apex map $g:V\to W$, and all mixed comparison cells. Rule (Desc) certifies its apex squares. Form
\[
 \glue_d(P)(w)=\Sigma_{v:V}\Id_W(gv,w).
\]
Write $q:G\to W$ for this family's display and $\alpha_g:V\simeq G$ for its fiber-presentation equivalence. Compose the top cocone with $\alpha_g$ and use its specified triangle over $W$. Reindexing comparison and (Tr) give the resulting certified pullback squares
\begin{equation}\label{eq:glue-local-square}
\begin{tikzcd}
 E_k\ar[r,"v_k"]\ar[d,"p_k"']&G\ar[d,"q"]\\
 d_k\ar[r,"\iota_k"']&W.
\end{tikzcd}
\end{equation}
Independently apply (Reind) to $q$ and $\iota_k$, obtaining the chosen pullback $F_k=d_k\times_WG$. Both the original square~\eqref{eq:glue-local-square} and this chosen reindexing square are certified terminal cones on the same cospan. The $n=1$ comparison rule gives a cone map
\[
 c_k:E_k\longrightarrow F_k,
\]
and a second instance with the original square last gives $\beta_k:F_k\to E_k$. Their cone boundaries contain the triangles over $d_k$ and $G$. Apply the $n=2$ comparison rule to the boundary with cone vertices $(E_k,F_k,E_k)$, distinguished last cone $E_k$, and edges $c_k,\beta_k,\id_{E_k}$. It produces the cone computation $\beta_k c_k\simeq\id_{E_k}$. The boundary with vertices $(F_k,E_k,F_k)$ similarly gives $c_k\beta_k\simeq\id_{F_k}$. Higher comparison rules extend these computations with every prescribed boundary. Thus $c_k$ is an equivalence over $d_k$, and its inverse gives the local computation of~\eqref{eq:glue-beta},
\[
 \beta_k:\glue_d(P)[\iota_k]\simeq P_k.
\]
The equivalence extends to a $J$-array by the equivalence criterion, so (E) gives its explicit certificate and coherent inverse data.

Let $j:\operatorname{sk}_0 K\hookrightarrow K$ be the inclusion of the simplicial subset generated by the vertices. The cone prism gives the cospans and their original squares over $K$. Apply the relative reindexing formation rule~\eqref{eq:relative} along $j$, prescribing at each vertex $k$ the chosen pullback square with apex $F_k$. With both cone diagrams now defined over $K$, apply the two dimension-$1$ comparisons of (Reind) relative to $j$, prescribing the cone maps $c_k$ and $\beta_k$ with their comparison cells. The resulting maps $c$ and $\beta$, together with the identity cone maps, give the boundaries for the two dimension-$2$ comparisons over $K$. Apply these comparisons relative to $j$ with the chosen inverse computations at each vertex. In each higher comparison, the preceding faces give the boundary over $K$, and the prescribed vertex fillers give the output on $\operatorname{sk}_0 K$. The boundary equation~\eqref{eq:output1} identifies the restriction of each output with those vertex data. Thus
\begin{equation}\label{eq:glue-computation-natural}
 c:P\xrightarrow{\simeq}R_d(q)
\end{equation}
is an equivalence of coherent local diagrams. On each edge, it recovers the given transition equivalence, and on each higher simplex, it recovers the corresponding coherence, since the relative comparisons extend the input prism.

For a family $B\to W$, reindex the base constructor cocone. Rule (Stab) certifies the top row, with apex $B$, as a colimit. The derived gluing construction forms another colimit of the same local total diagram and then its fiber presentation. Its total context is therefore a colimit with a specified map to $W$. The comparison from the glued family to $B$ is the counit $\epsilon_B:L_dR_dB\to B$ under fiber presentation. This gives global reconstruction
\[
 \glue_d(R_d B)\simeq B.
\]
For $P=R_dB$, the forward comparison $c$ in~\eqref{eq:glue-computation-natural} represents the unit $\eta_{R_dB}$. The triangle identity
\[
 R_d(\epsilon_B)\circ\eta_{R_dB}\simeq\id_{R_dB}
\]
therefore identifies the restriction of the counit at each $k$ with $\beta_k$, the inverse of $c_k$. The parameterized comparisons give naturality in $B$.

\section{Coherent Rule Extension over Diagrams of Equivalences}\label{app:equivalent-parameters}
A small simplicial shape $I$ lets a rule act on a diagram of substitutions or family maps. Its vertices describe the individual inputs and outputs, and its higher simplices record how these vary together. The primitive relative rules~\eqref{eq:relative} apply to every $J_0\hookrightarrow I$. The shapes $J[n]$ describe coherent equivalences between choices of presentation. They give an explicit groupoidal instance of this general parameter construction and also allow the equivalence parameter to vary inside a certified diagram.

Let $J[n]_m=\{0,\ldots,n\}^{m+1}$, with deletion and repetition as faces and degeneracies. For $E\subseteq\{0,\ldots,n\}$, write $J[E]$ for the simplicial subset with $J[E]_m=E^{m+1}$. Then $\partial J[n]$ is the union of the $J[E]$ over nonempty proper subsets $E$. For $n=0$, this union is empty. For a presentation $P$, let $P\otimes J[n]$ have underlying shape $|P|\times J[n]$, and certify every pair $(m,v)$ with $m:H_t\to P$ certified and $v:H_t\to J[n]$ arbitrary. Define its boundary presentation by
\begin{equation}\label{eq:J-boundary-presentation}
 \partial J(P\otimes J[n])
 :=
 \mathop{\operatorname{colim}}_{\varnothing\ne E\subsetneq\{0,\ldots,n\}}
       P\otimes J[E].
\end{equation}
For $n=0$, this is the empty presentation. Its underlying shape is
$|P|\times\partial J[n]$, and its certificates are exactly those inherited
from the indicated proper faces. Equivalently, a pair $(m,v)$ is certified
when $m$ is certified in $P$ and $v:H_t\to\partial J[n]$ factors through
$J[E]$ for some nonempty proper subset $E$.

\begin{proposition}[Equivalence-parameter extension]\label{prop:equivalent-parameters}
For a primitive rule $r:A\hookrightarrow B$, every derivable input on
\[
(A\otimes J[n])\amalg_{\partial J(A\otimes J[n])}
\partial J(B\otimes J[n])
\]
extends to $B\otimes J[n]$ by syntactic cells and derivable certificates. Adjoining chosen constructors for these extensions preserves the model $\infty$-categories and the syntactic geometric universe.
\end{proposition}
\begin{proof}
The vertexwise certificates in the given input specify the source of the relative rule
$r\squ(\partial J[n]\hookrightarrow J[n])$.
Apply this rule to obtain the entire underlying output array with its vertexwise certificates, retaining the prescribed boundary exactly.

For any certified $m:H_t\to B$ and any $v:H_t\to J[n]$, choose a vertex $i$ of $J[n]$. There is an explicit map $H:J\times H_t\to J[n]$. For a $q$-simplex $h$ of $H_t$, write $v(h)=(v_0,\ldots,v_q)$, and for $\tau=(\tau_0,\ldots,\tau_q)\in J_q$, set
\[
 H(\tau,h)_j=\begin{cases}i,&\tau_j=0,\\ v_j,&\tau_j=1.\end{cases}
\]
Deleting or repeating an entry commutes with this formula, so it is a simplicial map. Its endpoint restrictions are the constant map at $i$ and $v$. With first component $m$, it gives a map $J\times H_t\to B\times J[n]$ and hence the required equivalence-parameterized array in the output. Rule (Tr) transfers the vertexwise certificate to $(m,v)$. Thus every target certificate has a derivation. The small collection of such arrays and witnesses lies at a bounded syntactic stage.

In a geometric target, for each nonempty finite set $E$ of parameter vertices, this contraction identifies the $\infty$-categories of models by
\[
 \Mod_{P\otimes J[E]}(\D)
 \cong\Fun(J[E],\Mod_P(\D))\simeq\Mod_P(\D).
\]
The first isomorphism follows by currying in each simplicial degree. Certificate invariance shows that its object conditions agree. The second is evaluation at any vertex of the contractible groupoid $J[E]$. Passing to maximal subgroupoids gives precisely
\begin{equation}\label{eq:equivalent-parameter-core}
 M(P\otimes J[E],\D)
 \cong\Fun(J[E],\Mod_P(\D))^{\simeq}
 \simeq M(P,\D).
\end{equation}
The same contraction proves the equivalence of the all-parameter and vertexwise certificate conditions on every natural transformation. The resulting extension on marked mapping cores is consequently the relative trivial fibration of Lemma~\ref{lem:relative-extension}. Initiality and Corollary~\ref{cor:definitions} therefore identify the models and syntaxes of the two parameter presentations.\qed
\end{proof}

The proposition applies in particular to the reindexing comparisons of Appendix~\ref{app:typing}. Two choices for the same universal construction are related by equivalences in its contractible extension fiber. Arrays on $J[n]$ collect finitely many such equivalent choices, and the proposition extends their specified proper-face comparisons. Comparing the two pullback cones on $E\to\Gamma\xleftarrow{fg}\Theta$ gives the equivalence $A[f][g]\simeq A[fg]$ in~\eqref{eq:substitution-comparison}. The same relative lifting problem gives the comparisons for three substitutions and their pentagon boundary. The $J[n]$ parameters relate choices of these comparisons by coherent equivalences. Parameters of shape $I$ give the naturality of the comparisons as the substitutions vary. Both follow from the relative extension rule.
\section{Construction and Base Change of the Universal Families}\label{app:universal-families}

\subsection{The Theory of Indexed Boundary Presentations}

For the data of Definition~\ref{def:uf-presentation}, write $A$ for the object of colimit marks and $K_\bullet\to A$ for their indexing simplicial sets. The relative right cone in the slice over $A$ has degree-$n$ object
\[
 (K^{\triangleright_A})_n=A\amalg K_0\amalg\cdots\amalg K_n.
\]
The summand $K_j$ consists of a prefix of $j+1$ old vertices followed by apex vertices. The all-apex simplex belongs to $A$. A simplex operator restricts the old prefix, using the augmentation to $A$ when the new prefix is empty.

Form an ordinary category $D$ with objects $Q_n,R,L_J,A,K_n$. Its generators include the simplex operators on $Q$ and $K$, the augmentations to $A$, and the map $R\to Q_1$. For each $n$-simplex of $J^\triangleleft$, there is an incidence map $L_J\to Q_n$. For each summand of $(K^{\triangleright_A})_n$, there is an incidence map from that summand to $Q_n$. Impose the simplicial identities and the commutation equations for the augmentations and incidences. These are equations between unary composites. There are countably many generators and equations, since there are countably many finite simplicial shapes up to isomorphism. Diagrams $D\to\X_{\leq0}$ and their natural transformations are precisely indexed boundary presentations in $\X$ and their maps.

Start with the unmarked theory on $N(D)$ and write $Z_d$ for its sort at $d$. For each $d$, name the finite-limit comparison
\[
 Z_d\longrightarrow Z_d\times_{Z_d\times Z_d}Z_d
\]
and impose its invertibility. This is the second diagonal of $Z_d\to1$, so its invertibility expresses $0$-truncation. By Lemma~\ref{lem:support-realization}, the naming uses a small definitional extension. The resulting small theory $\Theta_{\mathrm{pres}}$ therefore classifies the indicated diagrams in set objects, including their natural transformations.

There is also a presheaf description of the base. Let $\mathcal A$ be the ordinary finite-colimit completion of $D^{\op}$, realized as the closure of the corepresentables under finite colimits in $\Fun(D,\mathbf{Set}_{\U})$ and replaced by a small skeleton. Then
\[
 \mathfrak P\simeq \Fun(\mathcal A,\Sps_\U).
\]
Indeed, Lemma~\ref{lem:lex-extension} identifies maps into $\Psh(\mathcal A^{\op})$ with finite-limit-preserving functors $\mathcal A^{\op}\to\X$. Each of their values is $0$-truncated because the second diagonal of every object in an ordinary finite-limit category is invertible. The ordinary free finite-limit completion property of $\mathcal A^{\op}$ now identifies these functors with $\Fun(D,\X_{\leq0})$. Under this description, the generic value at $d\in D$ is $a\mapsto a(d)$ by Yoneda.

\subsection{Models and the Comparison for a Variable Cocone}
Write $\operatorname{LF}_{\X}(Q)$ for the full subcategory of $(s\X)_{/Q}$ on the left families of~\eqref{eq:uf-left}, and $|E_\bullet|$ for geometric realization.

\begin{lemma}[Diagrams on an external shape]\label{lem:uf-external}
For a small simplicial set $H$, apply the constant-object functor $\Delta_{\X}$ in each degree. There is a natural equivalence
\[
 \operatorname{LF}_{\X}(\Delta_{\X}H)\simeq\Fun(H,\X).
\]
If $E_\bullet$ corresponds to $F$, then $|E_\bullet|\simeq\colim_H F$. Both statements hold in every slice of $\X$.
\end{lemma}
\begin{proof}
Disjointness and universality of coproducts identify an object over $\coprod_{H_n}1$ with an $H_n$-indexed collection. Assembling the simplicial operators gives a diagram on $(\Delta/H)^{\op}$, where $\Delta/H$ is the category of simplices. Conversely, taking the coproduct of these entries in each degree reconstructs the simplicial family. These operations apply also to maps of families. Condition~\eqref{eq:uf-left} inverts restriction to the first vertex of each simplex. By composition and two-out-of-three, this is equivalent to inverting all operators preserving that vertex. The first-vertex map $N((\Delta/H)^{\op})\to H$ is the localization at these operators, by applying the last-vertex theorem to the vertex reversal of $H$ and passing to opposites~\cite{Kerodon}. Restriction gives the stated equivalence for every simplicial set $H$.

A map $|E_\bullet|\to Y$ is a coherent family of maps from all the summands to $Y$, hence a transformation on $(\Delta/H)^{\op}$ to the constant diagram at $Y$. Full faithfulness of restriction along the localization identifies its space with the space of cocones from $F$ to $Y$. This proves the colimit formula. The same proof applies in the slice, which is again an $\infty$-topos.\qed
\end{proof}

For the generic form of the colimit condition, pull a left family over $Q$ back along $K^{\triangleright_A}\to Q$. Denote the result by $D^+$ and its restriction to $K$ by $D^-$. Write $E_a=A\times_{Q_0}E_0$ for the value at the apex $A\to Q_0$. Inclusion of the diagram and of its apex gives a span in the slice
\begin{equation}\label{eq:uf-realization}
 |D^-|\longrightarrow |D^+|\longleftarrow E_a.
\end{equation}
Indeed, degeneracies identify the constant family on $E_a$ with the restriction to the simplicial apex by~\eqref{eq:uf-left}.

\begin{lemma}[The variable right cone]\label{lem:uf-cone}
The apex map $E_a\to |D^+|$ is an equivalence. Geometric inverse images preserve the span~\eqref{eq:uf-realization}.
\end{lemma}
\begin{proof}
Work in $\mathcal Y=\X_{/A}$ and choose a left-exact localization $L:\Psh(C)\rightleftarrows\mathcal Y:j$ with $j$ fully faithful. In the presheaf category, form the right cone $W=(jK)^\triangleright$. Its degreewise coproduct formula gives $LW\simeq K^\triangleright$ in $\mathcal Y$, hence a map $W\to j(K^\triangleright)$. Pull $jD^+$ back along this map. At $c\in C$, the levels of $(jK)(c)$ are sets because $j$ and evaluation preserve limits and truncated objects. The pullback at $c$ is a left family on the ordinary right cone on $(jK)(c)$. By Lemma~\ref{lem:uf-external}, its realization is the colimit over that cone, and inclusion of the final vertex gives an equivalence onto this colimit. Realizations and equivalences of presheaves are computed objectwise. Applying $L$, which preserves pullbacks and realizations, recovers the apex map in~\eqref{eq:uf-realization}. Finally, all the operations used to form the span are coproducts, pullbacks and geometric realizations in slices. The induced inverse-image functors on slices preserve them.\qed
\end{proof}

\subsection{The Theory of Pairs and Relative Classification}
\begin{proposition}\label{prop:uf-pair}
There is a small extension $i:\Theta_{\mathrm{pres}}\to\Theta_{\mathrm{pair}}$ with models the indexed presentations equipped with a model. Its restriction functor forgets the model. This identification includes all compatible maps of pairs and is natural under geometric inverse image.
\end{proposition}
\begin{proof}
Adjoin a map of simplicial objects $E_\bullet\to Q_\bullet$ to the generators of $\Theta_{\mathrm{pres}}$. Explicitly, glue the simplicial cylinder $N(\Delta^{\op})\times\Delta^1$ to that presentation along its endpoint $1$, using the diagram of the $Q$-sorts. Keep all existing marks. Restriction from the arrow category of diagrams to its target is a categorical fibration, so the diagram category of the resulting pushout gives the homotopy pullback specifying a presentation with a map to its simplicial object $Q$. Name the comparisons~\eqref{eq:uf-left} and mark them as equivalences. Call this small intermediate theory $\Theta_0$. The sort $E_0$ takes values in arbitrary objects of the model topos.

Form the following small collection of maps for the universal $\Theta_0$-model in $\B_{\Theta_0}$.
\begin{enumerate}
\item Over $R$, restrict the left family to the marked edges. Lemma~\ref{lem:uf-external} in the slice, with $H=\Delta^1$, gives the arrow to be tested.
\item Over each $L_J$, restrict it to the marked cones. The same lemma, with $H=J^\triangleleft$, gives a cone $F$ and the comparison $F(*)\to\lim_J F|_J$ in that slice.
\item Over $A$, take the left arrow of~\eqref{eq:uf-realization}. Lemma~\ref{lem:uf-cone} identifies its target with the value at the apex.
\end{enumerate}
For the first two restrictions, the incidence families give maps from the degreewise constant indexing shapes in the corresponding slices. In the second comparison, $J$ is finite. Lemmas~\ref{lem:uf-external} and~\ref{lem:uf-cone} thus show that the three comparisons commute with geometric inverse images. Invertibility of a map in a slice is detected on its underlying arrow.

Transport the underlying arrows through the equivalence $\G_{\Theta_0}\simeq\B_{\Theta_0}$ of Theorem~\ref{thm:reconstruction}, choosing representatives with their endpoints and comparison identifications. Lemma~\ref{lem:support-realization} names this small collection in a small definitional extension. By Corollary~\ref{cor:definitions}, restriction across this naming step is an equivalence of model $\infty$-categories. Now, impose $\Eq$ on the named arrows. These axioms select the full subcategory satisfying the three tests, since their interpretations in any model are the inverse images of the universal comparisons. This gives $\Theta_{\mathrm{pair}}$ and the interpretation $i$ extending the inclusion of the presentation data.\qed
\end{proof}

For the relative assertion of Theorem~\ref{thm:uf-pres}, let $u:\X\to\mathfrak P$ classify $P$ and put $\mathfrak M_P=\X\times_{\mathfrak P}\mathfrak M$ in geometric morphisms. For $g:\mathcal Y\to\X$, the pullback universal property first identifies the \emph{space} of lifts with the fiber over $g^*P$ of the forgetful functor on model cores. Transport of simplicial families along isomorphisms of the indexed presentation identifies this fiber with $\Mod_{g^*P}(\mathcal Y)^\simeq$. The identification with $g^*P$ is part of a lift, so the maps in this fiber are taken over the fixed presentation.

To obtain the $\infty$-category of lifts with arbitrary transformations over $\id_g$, apply the same computation to $\Fun([n],\mathcal Y)$ for all $n$, using the constant $[n]$-diagram at $g$. A pointwise left-exact localization of a presheaf presentation exhibits $\Fun([n],\mathcal Y)$ as a Grothendieck $\infty$-topos, since $\Fun([n],\Psh(C))\simeq\Psh([n]^{\op}\times C)$. Since finite limits and small colimits are pointwise, for every Grothendieck $\infty$-topos $Z$, we have
\[
 \Geom(\Fun([n],\mathcal Y),Z)\simeq\Fun([n],\Geom(\mathcal Y,Z)).
\]
The presentation and model tests are likewise pointwise. Taking cores now gives in each degree the complete Segal spaces of the homotopy fiber of $\Geom(\mathcal Y,\mathfrak M_P)\to\Geom(\mathcal Y,\X)$ at $g$ and of $\Mod_{g^*P}(\mathcal Y)$. The equivalences commute with the simplex operators and hence identify these $\infty$-categories.

\begin{samepage}
\subsection{Evaluation of the Family over the Morita Space}
\begin{lemma}[Evaluation over a space of equivalences]\label{lem:uf-eval}
Let $K$ be a small space in the working universe, let $B$ be a $K$-indexed diagram of Grothendieck $\infty$-toposes in that universe with geometric equivalences as transition functors, and put $\D=\lim_{k\in K} B(k)$ in $\infty$-categories. Then $\D$ is a Grothendieck $\infty$-topos. Applying the constant-object functors pointwise defines the inverse image of a geometric morphism $p:\D\to\Sps_{/K}$. If $h_k(\ell)=\Map_K(k,\ell)$, then
\[
 \D_{/p^*h_k}\simeq B(k).
\]
These equivalences respect transport in $K$.
\end{lemma}
\end{samepage}
\begin{proof}
The category of Grothendieck $\infty$-toposes and left-exact left adjoints admits small limits, computed in $\infty$-categories~\cite{HTT}. The objects of $\D$ are coherent sections. Their limits and colimits are computed pointwise since the transitions are equivalences. Under $\Sps_{/K}\simeq\Fun(K,\Sps)$, the functor $p^*$ sends $V$ to $\ell\mapsto\Delta_{B(\ell)}V(\ell)$. It preserves finite limits and small colimits and hence has a right adjoint.

In any Grothendieck $\infty$-topos $\mathcal E$ and for any small space $S$, descent for $\Delta_{\mathcal E}S=\colim_S 1_{\mathcal E}$ gives
$\mathcal E_{/\Delta_{\mathcal E}S}\simeq\Fun(S,\mathcal E)$. Here, inversion of paths identifies the two choices of variance. Slicing the section category over a section amounts to specifying its component slice objects and their coherent transports, including such transports on maps. Applying this description at $p^*h_k$ gives
\[
 \D_{/p^*h_k}
 \simeq \lim_{\ell\in K} B(\ell)_{/\Delta_{B(\ell)}h_k(\ell)}
 \simeq \lim_{(k\to\ell)\in K_{k/}} B(\ell).
\]
In the second equivalence, an object specifies an object of $B(\ell)$ for each path $k\to\ell$, compatibly with transport along the category of elements of $h_k$, whose objects are these paths. Evaluation at the initial object $\id_k$ identifies this last limit with $B(k)$. Under straightening, $h_k$ corresponds to $*\xrightarrow{k}K$, so slicing at $p^*h_k$ computes geometric base change to $k$ by the base-change property of \'{e}tale morphisms~\cite{HTT}. All identifications are natural in the pointed representable and respect equivalences of the indexing points.\qed
\end{proof}


\begin{thebibliography}{99}
\addcontentsline{toc}{section}{\refname}
\bibitem{ABFJ}
Anel, M., Biedermann, G., Finster, E., Joyal, A.: Left-exact localizations of $\infty$-topoi I: Higher sheaves. Advances in Mathematics 400, 108268 (2022)

\bibitem{ABFJII}
Anel, M., Biedermann, G., Finster, E., Joyal, A.: Left-exact localizations of $\infty$-topoi II: Grothendieck topologies. Journal of Pure and Applied Algebra 228(3), 107472 (2024)

\bibitem{AnelLejay}
Anel, M., Lejay, D.: Exponentiable higher toposes. arXiv:1802.10425 (2018)

\bibitem{AvigadKapulkinLumsdaine}
Avigad, J., Kapulkin, K., Lumsdaine, P.L.: Homotopy limits in type theory. Mathematical Structures in Computer Science 25(5), 1040--1070 (2015)

\bibitem{BauerHaselwarterLumsdaine}
Bauer, A., Haselwarter, P.G., Lumsdaine, P.L.: A general definition of dependent type theories. arXiv:2009.05539 (2020)

\bibitem{Boulier}
Boulier, S.: Colimits in HoTT. Homotopy Type Theory (8 January 2016)

\bibitem{BuchholtzSchipp}
Buchholtz, U., Schipp von Branitz, J.: Propositional geometric type theory. HoTT/UF 2025, workshop abstract

\bibitem{Caramello}
Caramello, O.: Theories, Sites, Toposes: Relating and Studying Mathematical Theories through Topos-Theoretic ``Bridges''. Oxford University Press (2017)

\bibitem{CasacubertaEtAl}
Casacuberta, C., Guti\'errez, J.J., Mart\'inez-Carpena, D.: Sketchable infinity categories. arXiv:2511.01497 (2025)

\bibitem{ClairambaultDybjer}
Clairambault, P., Dybjer, P.: The biequivalence of locally cartesian closed categories and Martin-L\"of type theories. Mathematical Structures in Computer Science 24(6), e240606 (2014)

\bibitem{DiLibertiLobbia}
Di Liberti, I., Lobbia, G.: Sketches and classifying logoi. arXiv:2403.09264 (2024)

\bibitem{DiLibertiMeadows}
Di Liberti, I., Meadows, N.J.: Classifying infinity topoi via weighted limits. arXiv:2512.15613v2 (2026)

\bibitem{GambinoGarner}
Gambino, N., Garner, R.: The identity type weak factorisation system. Theoretical Computer Science 409(1), 94--109 (2008)

\bibitem{HartHou}
Hart, P., Hou (Favonia), K.-B.: Coslice colimits in homotopy type theory. In: 33rd EACSL Annual Conference on Computer Science Logic (CSL 2025), LIPIcs, vol.~326, pp.~46:1--46:20 (2025)

\bibitem{HeindelSobocinski}
Heindel, T., Soboci\'nski, P.: Being Van Kampen is a universal property. Logical Methods in Computer Science 7(1), article~14 (2011)

\bibitem{Isaev}
Isaev, V.: Model category of marked objects. arXiv:1610.08459 (2016)

\bibitem{Johnstone}
Johnstone, P.T.: Sketches of an Elephant: A Topos Theory Compendium, Vols. 1--2. Oxford University Press (2002)

\bibitem{KapulkinLumsdaine}
Kapulkin, K., Lumsdaine, P.L.: The homotopy theory of type theories. Advances in Mathematics 337, 1--38 (2018)

\bibitem{KapulkinSzumilo}
Kapulkin, K., Szumi{\l}o, K.: Internal languages of finitely complete $(\infty,1)$-categories. Selecta Mathematica 25, article~33 (2019)

\bibitem{LicataShulman}
Licata, D.R., Shulman, M.: Calculating the fundamental group of the circle in homotopy type theory. In: 28th Annual ACM/IEEE Symposium on Logic in Computer Science (LICS), pp.~223--232 (2013)

\bibitem{LumsdaineShulman}
Lumsdaine, P.L., Shulman, M.: Semantics of higher inductive types. Mathematical Proceedings of the Cambridge Philosophical Society 169(1), 159--208 (2020)

\bibitem{LumsdaineWarren}
Lumsdaine, P.L., Warren, M.A.: The local universes model: An overlooked coherence construction for dependent type theories. ACM Transactions on Computational Logic 16(3), 23:1--23:31 (2015)

\bibitem{LurieDAGV}
Lurie, J.: Derived Algebraic Geometry V: Structured Spaces. arXiv:0905.0459 (2009)

\bibitem{HTT}
Lurie, J.: Higher Topos Theory. Annals of Mathematics Studies, vol. 170. Princeton University Press (2009). Updated edition

\bibitem{Kerodon}
Lurie, J.: \href{https://kerodon.net/}{Kerodon}

\bibitem{LynchMyers}
Lynch, O.: An invitation to geometric type theory. Joint work with D.J.~Myers. CT 2026, abstract

\bibitem{Macpherson}
Macpherson, A.W.: Adjoining colimits. arXiv:2111.12117 (2021)

\bibitem{MakkaiReyes}
Makkai, M., Reyes, G.E.: First Order Categorical Logic: Model-Theoretical Methods in the Theory of Topoi and Related Categories. Lecture Notes in Mathematics, vol. 611. Springer (1977)

\bibitem{MaruyamaHigherTripos}
Maruyama, Y.: Higher tripos theory. arXiv:2610.27163 (2026)

\bibitem{MaruyamaNasu}
Maruyama, Y., Nasu, R.: Measuring univalence and its failure in homotopy type theory via synthetic cohomology. In: 9th International Conference on Applied Category Theory (ACT 2026), EPTCS

\bibitem{MollerScherer}
M{\o}ller, J.M., Scherer, J.: Can one classify finite Postnikov pieces? Bulletin of the London Mathematical Society 42(4), 661--672 (2010)

\bibitem{NguyenUemura}
Nguyen, H.K., Uemura, T.: $\infty$-type theories. Higher Structures 9(1), 179--226 (2025)

\bibitem{Nikolaus}
Nikolaus, T.: Algebraic models for higher categories. Indagationes Mathematicae 21(1--2), 52--75 (2011)

\bibitem{Puppe}
Puppe, V.: A remark on ``homotopy fibrations''. Manuscripta Mathematica 12, 113--120 (1974)

\bibitem{RaptisSchappi}
Raptis, G., Sch\"appi, D.: Flat functors in higher topos theory. arXiv:2208.13897 (2022)

\bibitem{RezkSheaves}
Rezk, C.: Fibrations and homotopy colimits of simplicial sheaves. arXiv:math/9811038 (1998)

\bibitem{RezkToposes}
Rezk, C.: Toposes and homotopy toposes. Lecture notes (2010)

\bibitem{RezkLeeds}
Rezk, C.: Lectures on Higher Topos Theory (Leeds, 2019). Revised 2021

\bibitem{Rijke}
Rijke, E.: Classifying Types. Ph.D. thesis, Carnegie Mellon University (2018). arXiv:1906.09435

\bibitem{RijkeIntro}
Rijke, E.: Introduction to Homotopy Type Theory. Cambridge University Press (2025)

\bibitem{Riley}
Riley, M.: Geometric Type Theory, Done Two Ways. Topos Institute (15 June 2026)

\bibitem{ShulmanDescent}
Shulman, M.: $(\infty,1)$-toposes via descent. IAS Univalent Foundations discussion (22 February 2013)

\bibitem{ShulmanUniverses}
Shulman, M.: All $(\infty,1)$-toposes have strict univalent universes. arXiv:1904.07004 (2019)

\bibitem{Sojakova}
Sojakova, K.: Higher inductive types as homotopy-initial algebras. In: 42nd ACM SIGPLAN-SIGACT Symposium on Principles of Programming Languages (POPL), pp.~31--42 (2015)

\bibitem{StenzelSheaves}
Stenzel, R.: Higher geometric sheaf theories. arXiv:2205.08646v2 (2024)

\bibitem{StenzelSites}
Stenzel, R.: Notions of $(\infty,1)$-sites and related formal structures. arXiv:2306.06619 (2023)

\bibitem{ToenVezzosi}
To\"en, B., Vezzosi, G.: Homotopical algebraic geometry I: Topos theory. Advances in Mathematics 193(2), 257--372 (2005)

\bibitem{UemuraFramework}
Uemura, T.: A general framework for the semantics of type theory. Mathematical Structures in Computer Science 33(3), 134--179 (2023)

\bibitem{UemuraHIT}
Uemura, T.: Higher inductive types in $(\infty,1)$-categories. arXiv:2410.17615 (2024)

\bibitem{UemuraDiagrams}
Uemura, T.: Homotopy type theory as a language for diagrams of $\infty$-logoses. Logical Methods in Computer Science 22(1), 25:1--25:39 (2026)

\bibitem{HoTTBook}
The Univalent Foundations Program: Homotopy Type Theory: Univalent Foundations of Mathematics. Institute for Advanced Study, Princeton (2013)

\bibitem{Vickers}
Vickers, S.: Arithmetic universes and classifying toposes. Cahiers de Topologie et G\'eom\'etrie Diff\'erentielle Cat\'egoriques 58, 213--248 (2017)
\end{thebibliography}
\end{document}